\RequirePackage{snapshot}

\documentclass{article}
\usepackage[utf8]{inputenc}
\usepackage[a4paper, total={6in, 8in}]{geometry}
\usepackage{setspace}

\usepackage{hyperref}
\usepackage{amssymb}
\usepackage{fullpage}
\usepackage{enumerate}
\usepackage{amsthm}
\usepackage{amsmath,amssymb}
\usepackage{mathtools}
\usepackage{graphics,graphicx}
\usepackage{subcaption}
\usepackage{booktabs}
\usepackage{multirow}
\usepackage{multicol}
\usepackage{fancyhdr}
\usepackage{textcomp}
\usepackage{setspace}
\usepackage{amsfonts}
\usepackage{upgreek}
\usepackage{dsfont}
\usepackage{natbib}
\usepackage{hyperref}
\usepackage{tikz}
\usepackage{array}
\usepackage{float}
\usepackage{dsfont}
\usepackage{booktabs} 
\usepackage[ruled]{algorithm2e} 

\SetAlFnt{\small}
\SetAlCapFnt{\small}
\SetAlCapNameFnt{\small}
\SetAlCapHSkip{0pt}
\IncMargin{-\parindent}
\usepackage{arydshln}
\usepackage{makecell}
\usepackage{svg}
\usepackage{threeparttable}

\newtheorem{lemma}{Lemma}

\title{An Empirical Analysis of Optimal Nonlinear Pricing in Business-to-Business Markets\thanks{We thank Saman Ghili for his advice on optimization methods and literature. We also thank Jason Abaluck, Tat Chan, Sam Goldberg, Yufeng Huang,  Yewon Kim, Natalia Kyui, Tesary Lin, Nikhil Malik, Dan Miller, K. Sudhir, Raphael Thomadsen, Caio Waisman, Wenting Yu, Jidong Zhou, and various conference and seminar participants for their helpful comments. Ghili acknowledges financial support from the Yale Center for Customer Insights. We thank Wanxi Zhou for outstanding research assistance. \href{https://sites.google.com/view/soheil-ghili/nonlinearpricing}{Click here for the most current version}. All errors are our own.}}
\author{Soheil Ghili\thanks{Yale University. Email: soheil.ghili@yale.edu}, Russ Yoon\thanks{Massachusetts Institute of Technology. Email: ykyoon@mit.edu}}

\date{\today}

\begin{document}
\pagenumbering{gobble} 
\maketitle

\onehalfspacing

\begin{abstract}
    In continuous-choice settings, consumers decide not only on whether to purchase a product, but also on how much to purchase. Thus,  firms optimize a full price schedule rather than a single price point. This paper provides a  methodology to empirically estimate the optimal schedule under multi-dimensional consumer heterogeneity with a focus on B2B applications. We apply our method to novel data from an educational-services firm that contains purchase-size information not only for deals that materialized, but also for potential deals that eventually failed. We show that this data, combined with identifying assumptions, helps infer how price sensitivity varies with ``customer size''. Using our estimated model, we show that the optimal second-degree price discrimination (i.e., optimal nonlinear tariff) improves the firm's profit upon linear pricing by at least 8.2\%. That said, this second-degree price discrimination scheme only recovers 7.1\% of the gap between the profitability of linear pricing and that of infeasible first degree price discrimination. We also conduct several further simulation analyses (i) empirically quantifying the magnitude by which incentive-compatibility constraints impact the optimal pricing and profits, (ii) comparing the role of demand- v.s. cost-side factors in shaping the optimal price schedule, and (iii) studying the implications of fixed fees for the optimal contract and profitability.
\end{abstract}

\newpage

\setcounter{page}{1} 
\pagenumbering{arabic} 

\section{Introduction}

``Continuous choice'' products are increasingly common. For such products (unlike in discrete choice environments), each consumer decides not only on whether to purchase, but also on how much. Traditional B2C examples of such products are cell-phone plans and utility. More recently, many B2B products are of this form. Some major examples are SaaS products, cloud services (Azure, AWS, Google Cloud, ect.), and B2B online pay (e.e., PayPal, Amazon Pay, etc). In continuous choice settings, the firm's pricing problem goes beyond finding an optimal price point; it entails optimizing a full pricing schedule. The objective of this paper is to provide an empirical framework for optimizing a nonlinear price schedule, with a focus on B2B applications. To this end, we (i) develop a structural model of nonlinear pricing motivated by key insights from the mechanism design literature, and (ii) provide and implement an estimation procedure that is cognizant of what data sources are typically scarce in B2B cases and what data sources are available instead.

Our model is one of the continuous-choice demand models with multi-dimensional heterogeneity across customers. It is built to capture a key insight from the economic theory literature on multi-dimensional screening. More specifically, we develop  a demand model that allows for flexibility in the joint distribution of a customer's ``size of use'' (i.e., how many units of the product the customer needs) and her price sensitivity. The correlation across customers between these two quantities (or more broadly, the joint distribution of them) is critical for determining the shape of the optimal nonlinear pricing scheme. To illustrate, if ``larger'' customers are more price sensitive, then  flatter tariffs are more profitable whereas if ``smaller'' ones have a higher sensitivity to price, then steeper schedules are recommended. This is closely related to  results from recent multi-dimensional-screening/2nd-dgree-disrimination literature    on how the optimal tariff is impacted by the shape of the joint distribution between price sensitivity and taste/need for quality/quantity \citep{anderson2009price,haghpanah2019pure,ghili2022characterization,yang2021costly}.

Ideally, empirical estimation of the relationship between customer size-of-use (which we henceforth refer to as ``size'') and customer price sensitivity requires exogenous price changes to which a statistically representative subset of customers are exposed. Such variation would be useful because it would allow the researcher to compare reactions to the price change across customers with different pre-price-change purchase sizes. Nevertheless, such price variations are typically unavailable. This is especially common for most B2B products which tend to have moderate data sizes (hundred or thousands of customers in total) but large revenue per customer, which can render varying the prices risky.\footnote{A prime example of such data issues is cloud computing where the total number of clients is rather small, price schedules are fixed for years, and firms are unwilling/unable to run pricing experiments.}

To provide a broadly applicable solution to this problem, we turn to novel data which, although not yet used in academic work to our knowledge, is increasingly collected and maintained in B2B settings. Specifically, we leverage a dataset from a B2B firm that sells educational workshops and that records not only deals that succeeded, but also sale efforts that started but did not lead to a transaction. Crucially, for unsuccessful sale efforts, the data records, among other things, the \textit{would-be number of workshops that the potential customer was considering buying}.   

Combined with the right identifying assumptions, such a dataset allows to identify the key object of interest in estimation: the joint distribution over customer size and price sensitivity. What makes it possible to flexibly recover this distribution is, roughly, the variation across deal sizes in deal success rate. To informally illustrate, if a significantly higher fraction of larger potential deals fail relative to smaller potential deals, then large customers are on average more price sensitive than smaller customers. More complex joint distributions may also be recovered if, for instance, medium-sized potential deals are more likely to be unsuccessful compared to both larger and smaller ones. The key feature of our data (i.e., observing the intended sizes of eventually unsuccessful deals) gives us access to this necessary variation in deal success rates across sizes, thereby enabling us to estimate the joint distribution of interest.

To our knowledge, data on intended sizes of unsuccessful deals is commonly  recorded in B2B settings.\footnote{Some firms, such as LifeLabs Inc. where our data comes from, directly record the intended purchase size data. Other firms use software such as Salesforce that have designated features (e.e., ``Pipeline Management'') that records properties of deals along the negotiation journey, including but not limited to intended sizes.} As a result, we believe our estimation method may be applied beyond the specific context of our analysis to other major B2B applications where direct price variation is limited but granular data on failed deals is available.

We use our estimation and price-schedule optimization methodology to simulate the optimal nonlinear schedule (i.e., optimal 2nd degree price discrimination) and assess its effects on profit and welfare. We find that the optimal schedule lowers the per-unit prices for larger deal sizes, relative to those for smaller deals. The cost-side reason for this is that a portion of production costs is incurred per customer rather than per unit. The demand side reasons are more complex and will be discussed in detail later in the paper; but a simplified intuition is that a higher portion of larger deals fail relative to smaller ones, even though the observed pricing strategy of the firm already involved some volume discounting. To give a sense of the magnitude of our results, we find that it is best to offer about 29\% lower per-unit fees to customers purchasing more than 100 units relative to those who purchase less than five. The discount is around 6\% for deals of sizes 10-19 relative to those less than five. The optimal nonlinear pricing schedule  delivers at least a 8.2\% higher profit relative to optimal linear pricing. In addition, we find that the optimal second-degree schedule covers 7.1\% of the profitability-gap between  optimal linear pricing and optimal first-degree price discrimination. This stands in  contrast to third-degree price discrimination which has been shown to closely approximate the profitability of first degree discrimination (see \cite{dube2017scalable}). Also, we find that the optimal second degree price discrimination leads to an almost 10.2\% increase in consumer welfare relative to linear pricing (consumer welfare for first-degree discrimination is, by construction, zero). In addition, nonlinear pricing (i.e., second degree price discrimination) increases total social welfare relative to linear pricing by about 8.9\%. In addition to these counterfactuals, we also carry out a segment-by-segment analysis of how different pricing schemes affect profit and welfare, as well as an analysis of the implications of adding a fixed fee to the pricing structure.

We conduct additional simulation analyses that both produce useful insights and demonstrate the ability of our framework to generate similar insights in other settings: (i)  We study the role of ``incentive compatibility constraints''. More specifically, we quantify the profit from a price schedule in which the per-unit price for each deal-size range is optimized only considering the per-unit willingnesses to pay by customers whose sizes of use are in that range. We show that such ``separately optimized'' price schedules lead  to significant losses if they charge for a deal-size range a substantially different marginal price compared to its adjacent ranges. Such losses arise from the fact that customers belonging to a size-group that faces a high marginal price are incentivized to adjust their purchase sizes to take advantage of substantially lower per-unit prices offered to other size groups. We show that the ``globally optimal'' contract that accounts for such incentives  tends to moderate the variation of the per-unit prices across sizes. (ii) We empirically compare the role of cost side factors to those of demand side ones in determining the shape of the optimal price schedule. We show that the nonlinearities in the optimal schedule are more heavily shaped by the heterogeneity in demand than by nonlinearity in the cost function. We then document the  interaction between demand- and cost-side factors. In addition, we dive deeper into the role of costs. Among other things, we show that ``customer-level'' fixed costs, which are a phenomenon specific to continuous-choice settings, have key implications for optimal pricing--unlike traditional firm-level fixed costs. 


The rest of this paper is organized as follows. Section \ref{sec: literature} reviews the related literature. Section \ref{sec: data and setting} describes the data and setting and provides some summary statistics. Section \ref{sec: model} presents the model. Section \ref{sec: estimation} discusses the estimation procedure, identification, and estimation results. Section \ref{sec: optimal schedule} presents the optimal nonlinear price schedule and discusses in detail how it compares to a linear schedule and to first-degree price discrimination. Section \ref{sec: further CF analysis} conducts further counterfactual analysis. \ref{sec: discussion}  discusses the general applicability of our method beyond our specific setting, and points to some caveats and avenues for future research. Section \ref{sec: conclusion} concludes.

\section{Related Literature}\label{sec: literature}

This paper sits at the intersection of two strands of the literature on theory and empirics of nonlinear pricing. On the theoretical end, there exists a large literature on ``screening,'' with one of its major applications/interpretations being nonlinear pricing. Many papers in this domain, such as seminal work by  \cite{mussa1978monopoly} and \cite{maskin1984monopoly}, focus on uni-dimensional consumer type spaces. The literature on multi-dimensional screening (e.g., \cite{wilson1993nonlinear,armstrong1996multiproduct,laffont1987optimal,rochet2002nonlinear,rochet2003economics,carroll2017robustness}) relaxes this assumption. The present paper relates to this literature by directly analyzing optimal screening and second-degree price discrimination in an empirical setting. We empirically optimize a full price schedule for a firm that, when deciding the charge amount for a given quantity/quality level $q$, tries not only to maximize the profit made from those customers who purchase at $q$, but also to incentivize as many customers as possible to purchase more profitable quantities $q'$.

Though it may not be clear in the first glance, our emphasis on correlation between size and value makes this paper related to the literature on bundling of products with non-additive values. In that literature, there are results stating that bundling a set of products is recommended if less price-sensitive consumers tend to perceive a lower degree of complementarity among products (see \cite{anderson2009price,haghpanah2019pure,ghili2022characterization} among others). This has parallels to our intuition that flatter contracts are optimal when less-price-sensitive customers tend to be the smaller ones.

On the empirical side, there have been a number of studies on nonlinear tariffs. Most commonly studied applications  have been electricity and cell-phone plans. Some of these papers examine a one-dimensional type space \citep{luo2018structural,aryal2020empirical,bodoh2023stress}\footnote{\cite{aryal2020empirical} examine a two-dimensional heterogeneity model. But they study two products; and preferences over each of the products has been captured using a single dimension of heterogeneity.} whereas others study mutli-dimensional type spaces (e.g., \cite{narayanan2007role,iyengar2008conjoint,iyengar2012conjoint,nevo2016usage,reiss2005household,mcmanus2007nonlinear}).

The present paper belongs to the latter group. Our key contribution in this paper is that (i) we focus on  flexibly capturing the joint distribution between ``customer size'' and price sensitivity (as motivated by the mechanism design literature), and that (ii) we estimate this joint distribution by developing a method that is focused on B2B applications. Specifically, our estimation method is cognizant of the limited price variation in those applications, and instead leverages detailed information about the purchase process that is commonly recorded and maintained by B2B firms about sizes of deals that eventually succeeded \textit{and those that ultimately failed}. Although customer-size has been captured in multiple empirical papers studying nonlinear pricing (e.g., \cite{iyengar2009nonlinear,howell2016price,liu2022non}), to our knowledge, they do not focus on a flexible joint distribution between size and price sensitivity. The only empirical paper that captures such a distribution is \cite{reiss2005household} which studies nonlinear pricing in the electricity market and examines heterogeneity across consumers in the number and types of electric home-appliances they own. \cite{reiss2005household} do not study the optimal price schedule and. Additionally, their measures of customer size is specific to their application and may not generalize to most B2B settings which are our focus.

Our work is also related to the empirical literature that quantifies the efficiency of price discrimination strategies. For example, \cite{dube2017scalable} quantify that a sufficiently fine-grained third-degree price discrimination strategy can replicate the profitability of first degree price discrimination in the context of their study. Our paper is complementary in that it quantifies the same effect for second degree price discrimination, and is one of the few to do so, especially in the context of nonlinear pricing (for other examples of empirical quantification of the effects of second-degree discrimination--or more generally, screening mechanisms--see \cite{hendel2013intertemporal,goldberg2021designing,leslie2004price,verboven2002quality,iyengar2009nonlinear,kadiyali1996empirical,draganska2006consumer} where the latter two papers consider competing sellers. For surveys of this literature, see \cite{chan2009structural} and \cite{lambrecht2012price}. We show that at least in our context, second degree price discrimination recovers only a small portion of the profitability gap between first degree price discrimination and no discrimination at all. 

Finally, this paper is related to the literature on continuous-choice demand models (see, among others,  \cite{dube2004multiple,hendel1999estimating,kim2002modeling,chan2006estimating,song2007discrete}). The key objective of this literature is to examine environments where multiple brands are available and consumers might choose to purchase a positive amount from each brand and form a mixed basket. Consequently, demand models in this literature are more general than ours on the fronts related to this objective, such as the inclusion of multiple products in the utility function. Our paper, on the other hand, seeks to optimize a nonlinear tariff for second-degree price discrimination. As such, we move away from the ``multiplicatively separable'' utility functions assumed in this literature to one that allows for heterogeneity in the two key dimensions of size of use and price sensitivity. We then focus on leveraging  data to identify the joint distribution on these two dimensions, and use our estimated model to optimize the nonlinear tariff.

\section{Data, Setting, and Descriptive Statistics}\label{sec: data and setting}

\begin{figure}[H]
    \centering
    \includegraphics[height=2.2in]{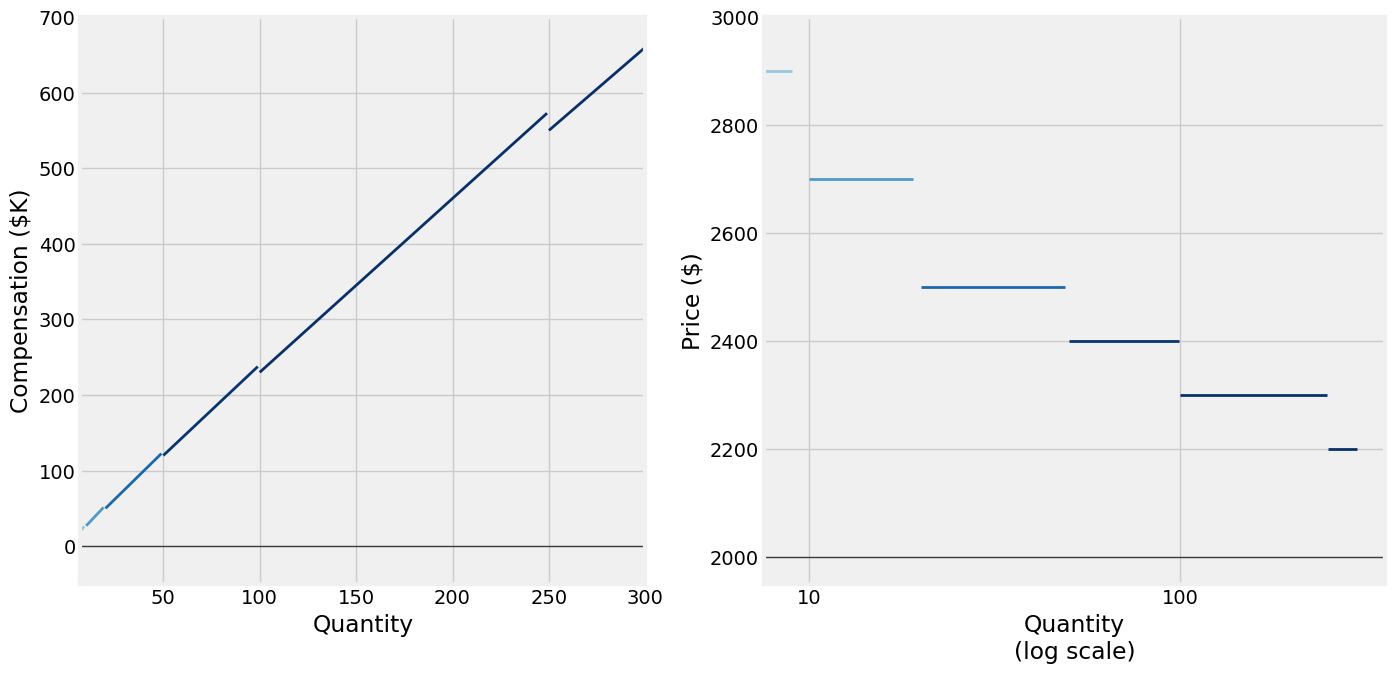}
    \caption{LifeLabs Price Schedule. The left panel shows the total charge as a function of quantity while the right panel depicts the marginal price. Note that the schedule is approximately concave. }
        \label{fig:pricesched}
\end{figure}

We study pricing by LifeLabs Learning, a New-York-based HR-Services company offering workshops to employees of its business clients. Lifelabs serves customers within and outside of the United States. The workshops are on leadership and other business-related skills.  The time window of our data from the company encompasses 2020 and 2021.  As of 2021, LifeLabs did not directly approach potential customers and its marketing activities were mainly based on word of mouth. When a potential client reaches out to LifeLabs, a conversation about a potential deal begins. The most important aspects of each deal in our data are quantity (the number of workshops to be delivered by LifeLabs to the company) and the total price. The price is determined based on a pre-set schedule as a function of quantity. Figure \ref{fig:pricesched} depicts LifeLabs' current price schedule.

In addition to the price schedule, our data consists of demand- and cost-side information, which we turn to subsequently.

\textbf{Demand side data:} For each potential  deal, aside from the quantity and the total price, we also observe whether the deal eventually succeeded. In other words, our data provides information on  price and quantity not only for actual transactions, but also  for  potential ones that did not actualize. As we will discuss later on, this is key to identifying how per-workshop valuation across firms correlates with the sizes of their needs. 

\begin{figure}[H]
    \centering
    \includegraphics[width=0.7\textwidth]{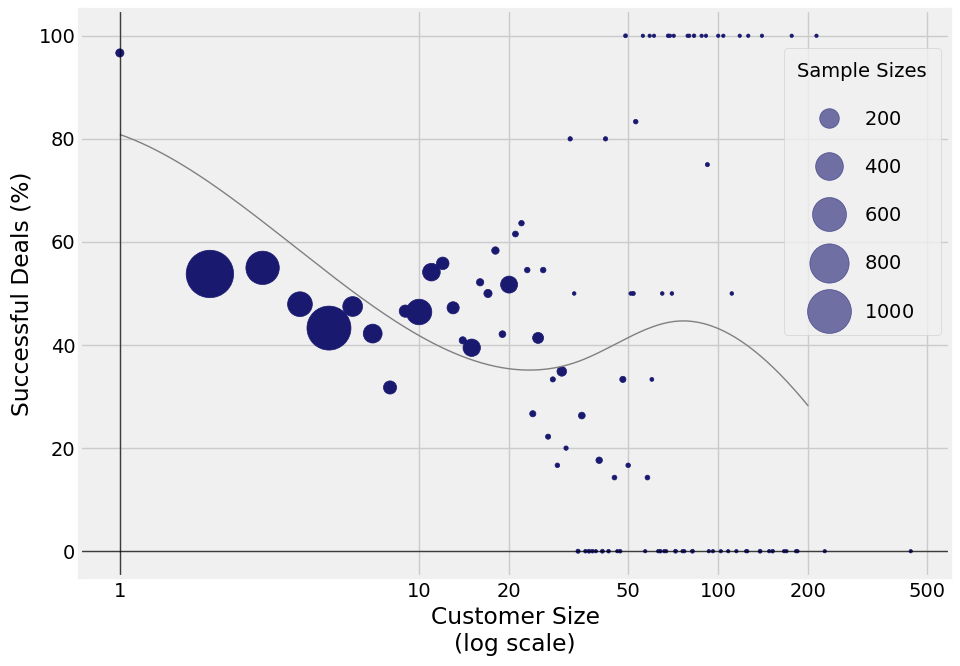}
    \caption{Deal Success Rate v.s. Deal Size}
    \label{fig:successVSsize}
\end{figure}

Figure \ref{fig:successVSsize} plots deal success rate against deal size. As can be seen from this figure, a higher percentage of deals fail as we look at larger sizes; and this is in spite of the fact that for those deal sizes the per-workshop price is cheaper. This is suggestive that LifeLabs might want to further sharpen its volume discount policy in order to increase profitability. Of course a structural analysis encompassing both demand- and cost-side data would be necessary before one can (i) determine  whether such a strategy is indeed recommended to LifeLabs and (ii) quantify the extent of it.

\begin{figure}[H]
    \centering
    \includegraphics[width=.9\textwidth]{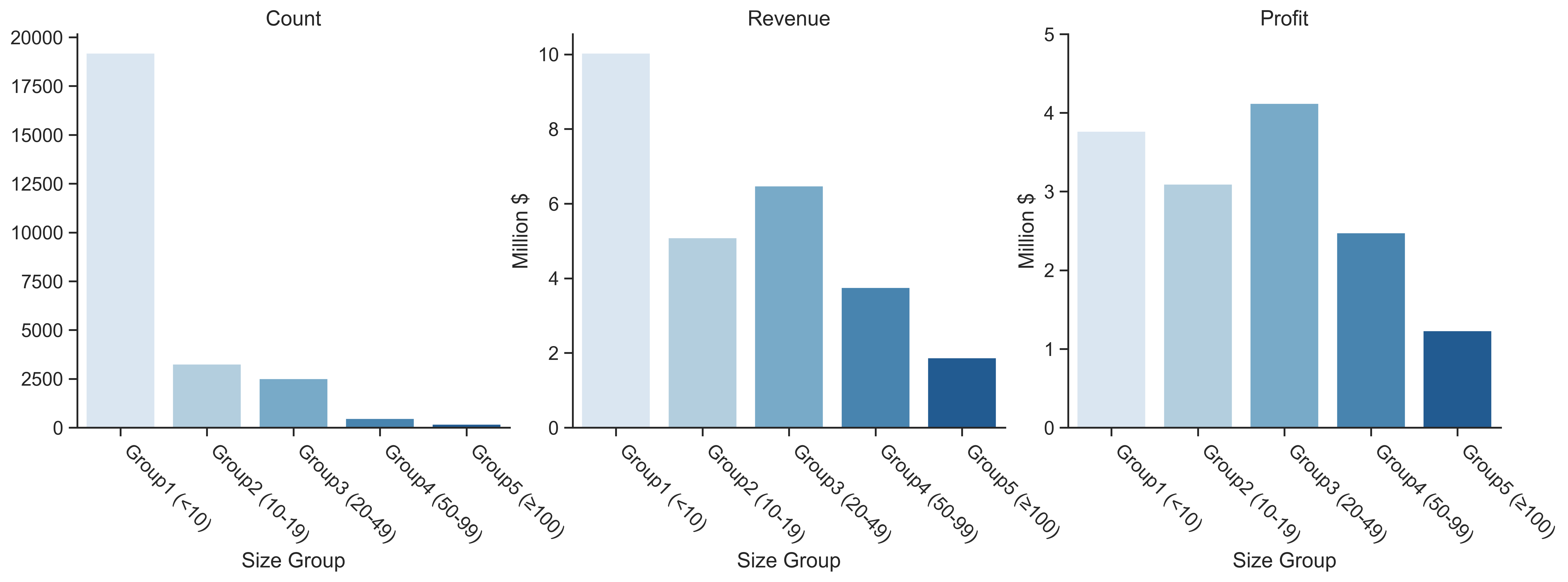}
    \caption{Larger deals are infrequent but meaningful for bottomline revenue and profit.}
    \label{fig: size and rev hist}
\end{figure}

Figure \ref{fig: size and rev hist} presents three histograms. The left panel shows the counts of deals  of different size groups. The middle and right panels respectively depict the total revenue and total profit (in \$M) from each such size group. As these panels together depict, larger deals are substantially less frequent than smaller ones. Nevertheless, they contribute meaningfully to the total firm revenue and especially profit. This is because each large deal contributes a more substantial revenue compared to a small deal; and the contrast is even sharper (due to the role of costs) when we compare the profits instead of the revenues.

\begin{figure}[H]
\begin{subfigure}{0.55\textwidth}
\centering
\includegraphics[width=\textwidth]{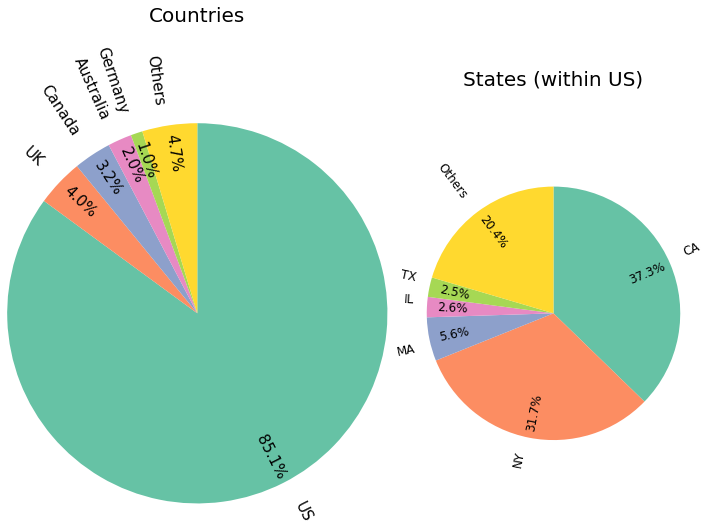}
\caption{By countries and By states}
\label{fig:subim1}
\end{subfigure}
\begin{subfigure}{0.4\textwidth}
\centering
\includegraphics[width=\textwidth]{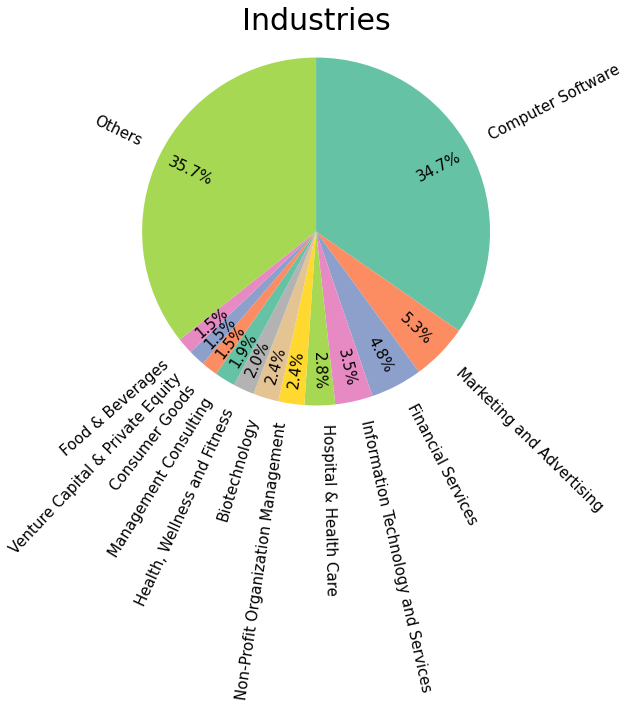}
\caption{By industries}
\label{fig:subim1}
\end{subfigure}
\begin{center}
\begin{subfigure}{\textwidth}
\centering
\includegraphics[width=\textwidth]{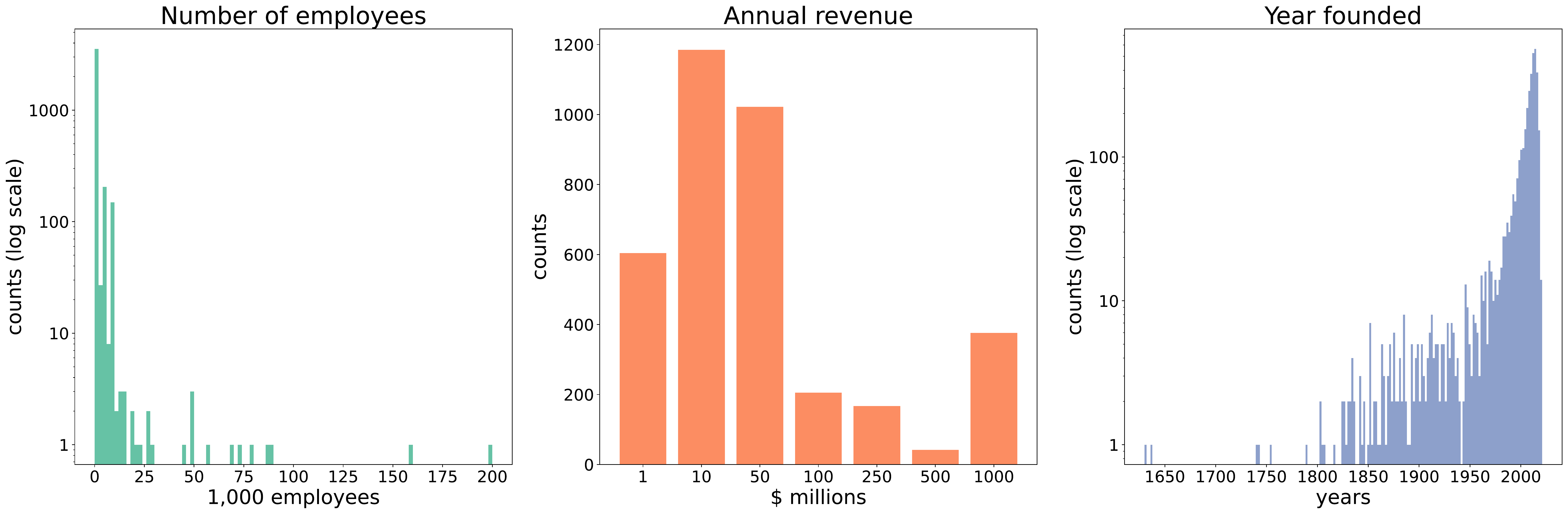} 
\caption{Histograms}
\label{fig:subim1}
\end{subfigure}
\end{center}
\caption{Summary stats on customer characteristics}
\label{fig:summarystats}
\end{figure}

In addition to the deals data, we also possess a client-level dataset which provides information on different client characteristics. Among those are number of employees, a coarse measure of annual revenue ($<\$1M$, $\$1-10M$, $\$10-50M$, $\$50-100M$,  $\$100-1,000M$) geographical location (country if abroad, city if in the U.S. or Canada), industry, year founded, and some behavioral characteristics that we may not publicly disclose. Figure \ref{fig:summarystats} provide summary statistics on deals and customer characteristics. Customers come from various industries with ``Computer Software'' being the most common. Although most customers are U.S. based, there is a non-trivial international demand. Finally, customers are mostly companies with fewer than 10,000 employees, \$50M or less in annual revenue, and founded after 1950.

Customer characteristics data is also helpful in providing assurance that our data on unsuccessful deals is meaningful (such assurance is important because one might be concerned that the information recorded for deals that eventually did not happen might be too noisy to be useful). To this end, we briefly report some summaries that examines how (intended) deal size and deal success status  fall into clear patterns as they relate to some observables. As Table \ref{tab: suggestive evidence size}  shows, and as expected, the percentage of deals that are of size 10 or larger increases with company size. Same holds for deals of size 20 or larger and 50 or larger. Crucially, this pattern holds not only for the sizes of successful deals but also for the intended sizes of unsuccessful ones. For instance, only 1\% of successful deals are of size 20 or larger for clients with 1-100 employees. This share increases to 7\% when we consider clients with 101-1000 employees, and to 10\% for those with 1000+ employees. The same shares for unsuccessful deals are 1\%, 12\%, and 14\%, an overall similar magnitude and pattern when compared with the successful deals.\footnote{Note that the patterns are not exactly the same. And they need not be. In fact our identification of how price sensitivity varies with size comes from the differences between these two patterns, as will be explained later in the paper. Nevertheless, there is sufficient similarity to strongly suggest our data on unsuccessful deals is meaningful.} The fact that this observable has such a clear relationship not only with sizes of successful deals but also with those of unsuccessful ones is suggestive that the  size  information recorded for deals that eventually fail is meaningful.

\begin{table}[H]
\centering \footnotesize
\begin{tabular}{@{}rrcccc@{}}
\toprule
 \multicolumn{1}{l}{\textbf{Number of Employees}} & \thead{\textbf{Count}}  & \textbf{\% Size over 10} & \textbf{over 20} & \textbf{over 50} \\
\midrule
\multicolumn{1}{l}{\textit{\textbf{Unsucessful Deals}}} \\[-1ex]
 1-100 & 989 &  9 & 1 & 0 \\
 101-1000 & 1,206 & 27 & 12 & 2 \\
 1000+ & 369 & 22 & 14 & 6 \\\midrule
 \multicolumn{1}{l}{\textit{\textbf{Successful Deals}}} \\[-1ex]
 1-100 & 753 & 7 & 1 & 0 \\
 101-1000 & 1,235  & 18 & 7 & 1 \\
 1000+ & 383  & 26 & 10 & 3 \\
\bottomrule
\end{tabular}
\caption{Evidence in suggesting that size data for unsuccessful deals is meaningful. }
\label{tab: suggestive evidence size}
\end{table}

Table \ref{tab: suggestive evidence success rate}
briefly compares deal success rates between customers belonging to the ``Computer Software'' industry category and those belonging to ``Marketing and Advertising''. The table shows that a potential deal with the firm from the former group is 53\% (18pp) more likely to be successful relative to the latter (although  not shown in the table, this gap is statistically significant). Again, this relationship between deal success status and observables is suggestive that data on unsuccessful deals were systematically collected.

\begin{table}[H]
\centering \footnotesize
\begin{tabular}{@{}rrcc@{}}
\toprule
 \multicolumn{1}{l}{\textbf{Industry}} & \thead{\textbf{Count}}  & \textbf{Deal Success Rate}\\
\midrule
 Computer Software & 1,892 &  0.53 \\
 Marketing and Advertising & 218  & 0.35\\
 All Other & 2,885 & 0.46 \\
\bottomrule
\end{tabular}
\caption{Deal success rate by industry. There is a meaningful difference between ``Computer Software'' and ``Marketing and Advertising''. }
\label{tab: suggestive evidence success rate}
\end{table}

The final table we present is based on a behavioral characteristic of the customers, the nature of which we may not disclose. We term it ``behavioral feature 1''. As Table \ref{tab: suggestive evidence behavioral feature} shows, customers with low level of behavioral feature 1 are much less likely to purchase and more likely to be of smaller sizes. This, again, is suggestive of the meaningfulness of our data on unsuccessful deals.

\begin{table}[H]
\centering \footnotesize
\begin{tabular}{@{}rrcccc@{}}
\toprule
 \multicolumn{1}{l}{\textbf{Behavioral feature 1 level}} & \thead{\textbf{Count}} & \textbf{Deal Success Rate} & \textbf{\% Size over 10} & \textbf{over 20} & \textbf{over 50} \\
\midrule

 Low & 3,442 & 0.35 & 15 & 6 & 1 \\
 High & 1,553 & 0.76 & 24 & 10 & 3 \\
\bottomrule

\end{tabular}
\caption{Customers with low level of behavioral feature 1 are much less likely to purchase and more likely to be of smaller sizes.}
\label{tab: suggestive evidence behavioral feature}
\end{table}

\textbf{Costs Data.} In addition to purchase data, we were provided with unique and detailed costs data. Using this data, we are able to obtain measures of per-workshop cost to the firm and a ``per-customer fixed cost.''\footnote{Details about how these elements of cost are estimted will be provided in section \ref{subsub: cost}.} The former is the standard marginal cost. The latter is incurred for every customer purchasing a positive amount, but does not change with the amount purchased. This type of fixed cost is specific to continuous choice settings. While to our knowledge this component of cost has not been examined in the continuous-choice literature, we empirically show that it is relevant to optimal nonlinear pricing.

\section{Model}\label{sec: model}

We seek to construct a model that allows us to capture the notion central to our problem: a flexible joint distribution over customers' ``size'' (i.e., how many units of the product they need) on the one hand and their willingness to pay per unit on the other. Below, we describe different ingredients of this model.

\textbf{Consumer Preferences.} Each potential consumer is modeled using a value function $V_i(\cdot)$ where $V_i(q)$ is the willingness to pay for $q$ units of the product.  Our model of $V_i(\cdot)$ has two parameters: ``size'' $\Bar{q}_i$ captures how many workshops customer $i$ needs at most, and ``valuation'' $v_i$ denotes the customer's willingness to pay for each workshop. Formally, customer $i$'s willingness to pay for $q$ workshops is given by:

\begin{equation}\label{eq: WTP}
    V_i(q)=v_i\times\min(q,\Bar{q}_i)
\end{equation}

In words, the customer values each workshop at $v_i$ until it has received $\Bar{q}_i$ ones, at which point it will no longer value additional workshops. This formulation has been used before in the literature to model demand for similar products to what we are considering, such as cloud computing services \citep{devanur2020optimal}. 

\textbf{Market.} A ``market'' is a collection of individual potential customers $i$ each described using the two parameters $\Bar{q}_i$ and $v_i$. Formally, one can model it using a scalar $I$ which represents the total number of potential customers as well as a joint distribution $f(\cdot,\cdot)$ over $\Bar{q}_i$ and $v_i$. 

\begin{figure}[H]
    \centering
    \includegraphics[height=3.5in]{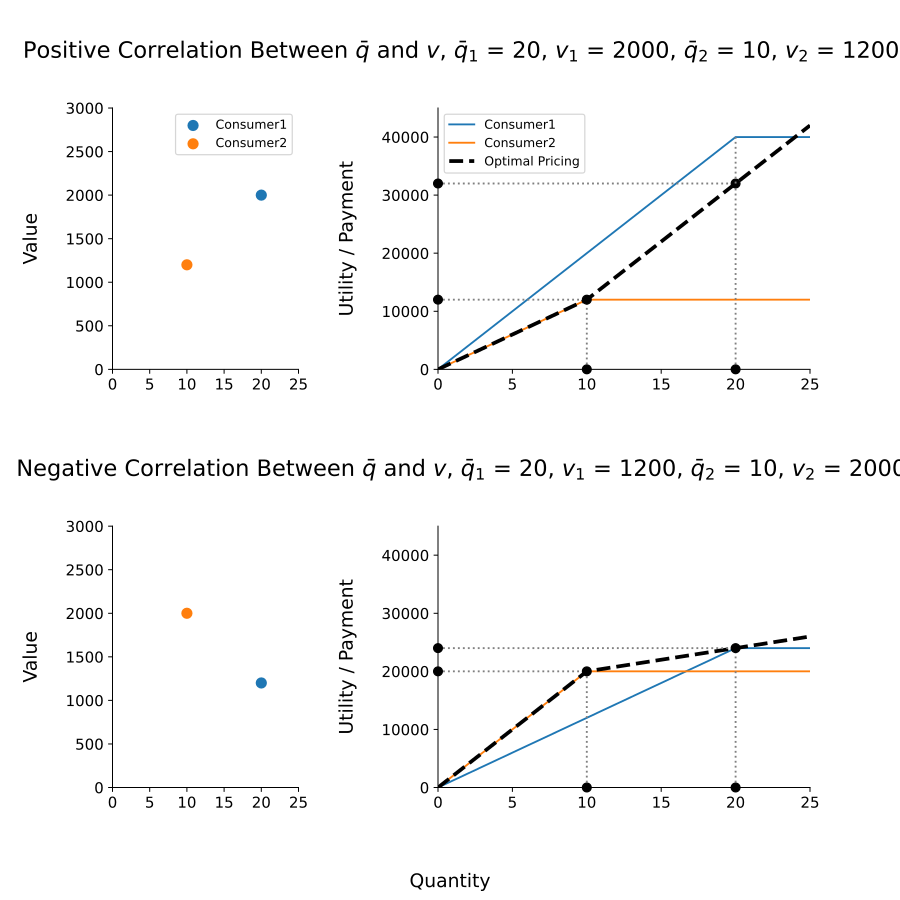}
    \caption{A two-customer illustration for the role of the size-value relationship on the shape of optimal pricing strategy. A positive correlation between the two (top panels) yields a steeper optimal tariff whereas a negative correlation (bottom panels) implies a flatter one.}
    \label{fig: illustration}
\end{figure}

In spite of its parsimony, our formulation for value functions $V_i(\cdot)$ has the advantage that it captures exactly the two dimensions (size and value) along which consumer heterogeneity, as captured by $f(\cdot,\cdot)$ is of first-order importance for nonlinear pricing. Figure \ref{fig: illustration} should help illustrate this matter. This figure uses two simple examples to show how our parsimonious value function captures a key economic force. Each example describes a market with only two customers $i=1,2$. As the figure depicts, when the customer with a larger size $\Bar{q}_i$ has a lower valuation $v_i$, a concave (i.e., flattening) contract is optimal. On the other hand, when the larger customer has a higher per-unit valuation, then a convex (steepening) schedule is the best (for a robustness analysis to smooth--as opposed to piece-wise linear--value functions, see appendix \ref{appendix: smoothness}). To sum up, Figure \ref{fig: illustration}  suggests that the relation between $\bar{q}_i$ and $v_i$ across customers $i$, formally captured by joint distribution $f(\cdot,\cdot)$,  is critical for optimal nonlinear pricing. The importance of the relationship between a ``baseline willingness-to-pay'' and ``preference for quality/quantity'' is congruent with theoretical research on price discrimination, market design, and multi-dimensional screening. Instances of such  papers are \cite{anderson2009price,haghpanah2019pure,ghili2022characterization,yang2021costly}.\footnote{This theoretical emphasis on the importance of the relationship between these two dimensions stands in contrast to a large body of the multi-dimensional screening literature, such as \cite{rochet2002nonlinear}, that assumes independence across dimensions.}

Note that  either scenario described in Figure \ref{fig: illustration} is empirically plausible. On the one hand, a customer with larger need for the product may have a higher willingness-to-pay per unit, perhaps because this customer is  more resourceful (e.g., in B2B, it is a larger firm), or perhaps because this product a more essential to the customer. On the other hand, a larger user of a product may have better information about/access to alternative options, lowering its willingness-to-pay per unit for the current option. This empirical ambiguity about $f(\cdot,\cdot)$ ex-ante, alongside its theoretical importance for the shape of the optimal schedule, strongly motivates the need for its empirical estimation. For examples of other empirical papers emphasizing the importance of estimating joint distribution of multi-dimensional heterogeneity for market design, see \cite{nevo2016usage,derdenger2013dynamic}; and for a paper emphasizing the importance of the flexibility of similar joint-distribution estimations, see \cite{goldberg2021designing}.

As a result, unlike with the form of the value functions $V_i(\cdot)$, we will not be parsimonious about the modeling of $f(\cdot,\cdot)$. We will, rather, estimate this object flexibly. This flexibility, as will be demonstrated in next sections, will allow us to consider cases that are relevant to the design of the optimal pricing schedule but are more complex than a simple ``positive or negative correlation'' between size $\Bar{q}$ and value $v$. An example is a scenario in which ``mid-size'' customers have on average lower valuations $v_i$ compared to both smaller and larger customers.

As a final note, observe that Figure \ref{fig: illustration} also demonstrates how our model relaxes a restrictive assumption in the literature: As the bottom-right panel shows, $V_1(\cdot)-V_2(\cdot)$ is non-monotonic in $q$ and crosses zero twice. This relaxes the ``single-crossing'' assumption  in much of the theory  on nonlinear pricing (e.g., \cite{maskin1984monopoly,mussa1978monopoly} and a long line of uni-dimensional-screening literature following them) and some of the empirical studies (e.g., \cite{luo2018structural,mcmanus2007nonlinear}).

\textbf{Nonlinear price schedule.} Denote the nonlinear tariff using function $P(q)$. This simply means any customer who purchases $q$ units of the product will have to pay the total amount of $P(q)$ dollars. 

\textbf{Timeline.} We assume a simple timeline. In stage 1, all $I$ potential customers learn about their $\bar{q}_i$ but not $v_i$. In stage 2, all potential customers engage in \textit{costless} exchange with the seller to learn about the product. At the end of this stage, each customer $i$ learns its $v_i$ and the firm learns $\bar{q}_i$. In stage 3, potential customers make decisions on how many units of the product (if any) to purchase. 
 
\begin{figure}
    \centering
    \includegraphics[scale=.5]{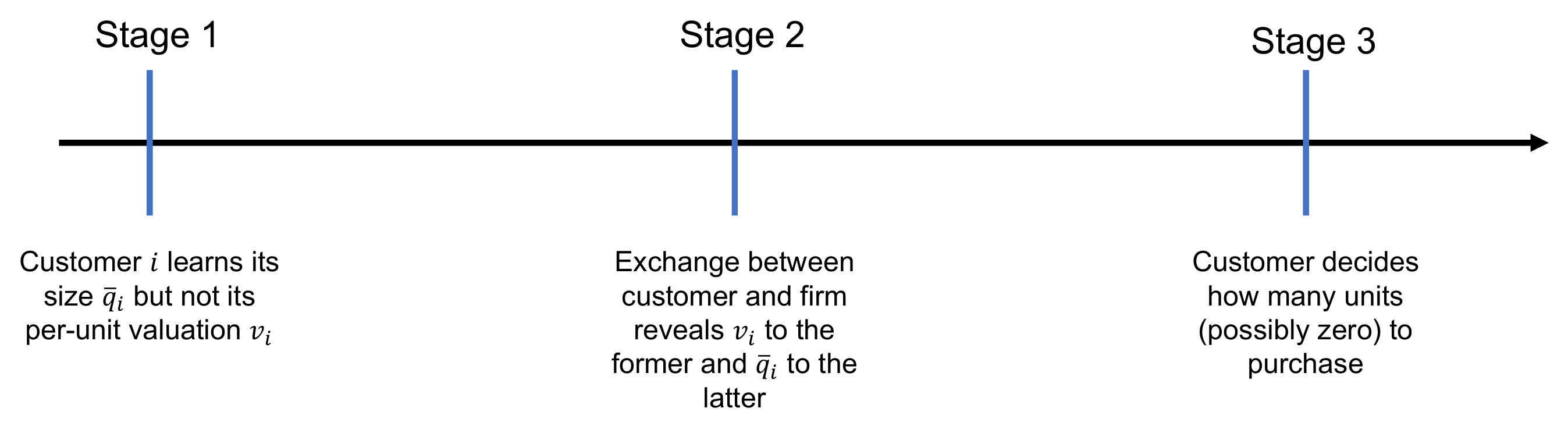}
    \caption{A simple timeline that provides a micro-foundation for why potential customers $i$ that end up not purchasing take the time to communicate their $\bar{q}_i$ to the seller.}
    \label{fig: timeline}
\end{figure}

Figure \ref{fig: timeline} depicts the model timeline graphically. Note that the sequential revelation of $\bar{q}_i$ and $v_i$ to the buyer does not have a central role in how the purchase decisions are made: the buyer has already learned both of these parameters before stage 3 where it makes the purchase decision. The purpose of setting up this timeline is, hence, \textit{not} to model incomplete or asymmetric information. But rather, the purpose is to construct the most succinct possible micro-foundation that can explain why those customers $i$ who do not end up purchasing take the time to communicate their $\bar{q}_i$ to the seller. The simple answer provided by this model is that it is only at the end of this very communication that they learn their $v_i$ is too low to justify any purchase. As a result, Figure \ref{fig: timeline} should be considered more a ``data timeline'' than a ``model timeline.''

To sum up, the importance of the timeline in Figure \ref{fig: timeline} is that it provides a framework that allows to interpret and structurally analyze datasets that record deal-size information for unsuccesful deals. This simple setup is also helpful in clarifying the directions in which the model can be extended, a discussion of which is provided in section \ref{sec: discussion}.

\textbf{Customer purchase decisions and firm profit.} We now turn to quantifying the purchase decisions. Each consumer's net value for $q$ workshops will be given by: $V_i(q)-P(q)$. Thus, customer $i$'s purchase decision $q_i$ will solve the following optimization problem:

\begin{equation}\label{eq: customer decision}
q_i=q^*(P|v_i,\Bar{q}_i)\coloneqq\arg\max_{q\geq 0} V_i(q)-P(q) 
\end{equation}

The seller's expected profit under pricing strategy $P(\cdot)$ is given by:

\begin{equation}\label{eq: profit}
\pi(P)=N\times\int_{v,\Bar{q}} P\big( q^*(P|v,\Bar{q}) \big)-c_1\times \textbf{1}_{q^*(P|v,\Bar{q})>0}-c_2\times \big(q^*(P|v,\Bar{q})\big) f(v,\Bar{q})dvd\Bar{q}
\end{equation}

In words, the expected profit is given by expected revenue net of expected costs, integrated over customer types. Note that in this continuous-choice setting, the cost function is more complex than it would be under discrete choice. In particular, the cost has two components: $c_1$ is the fixed customer-level component of costs which would be incurred once any customer decides to buy a positive amount.  Such fixed costs are specific to continuous choice environments and, unlike ``sunk'' firm-level\footnote{Here, by ``firm'', we mean the seller. Even though customers are also firms in this B2B context, we refer to them simply as customers.} fixed costs, can shape a firm's optimal pricing. The marginal cost $c_2$  would be incurred with every additional workshop provided to a customer. 

The firm's problem is to find the price schedule $P(\cdot)$ to maximize the expected profit:

\begin{equation}\label{eq: optimal tariff}
  P^*=\arg\max_{P(\cdot)}\pi(P)  
\end{equation}

With the model fully specified, we next turn to the estimation procedure and identification.

\section{Estimation}\label{sec: estimation}

The object we seek to estimate is the joint probability distribution $f(\cdot, \cdot)$ over values $v$ and sizes $\Bar{q}$. We do this in two steps. We start by estimating the marginal distribution $f_{\Bar{Q}}(\cdot)$ for size $\Bar{q}$, and then move on to estimate the conditional distribution $f_{V|\Bar{Q}}(\cdot)$ over values $v$.

For the first step, we take advantage of a feature of a model, alongside an approximation.

\begin{lemma}\label{claim: concavity and estimation of q bar}
If price schedule $P(\cdot)$ is strictly increasing and concave, then for any customer type $(v,\Bar{q})$, we have $q^*(P|v,\Bar{q})\in\{0,\Bar{q}\}$.
\end{lemma}

The proof of this lemma is relegated to the appendix. This lemma simply says that under a weakly concave price schedule, any customer will either buy no workshops or exactly as much as its size $\Bar{q}$. 

Due to gradually decreasing marginal prices (as depicted by figure \ref{fig:pricesched}), we treat the observed price schedule in our data as  approximately concave. As a result, we assume, for each customer $i$ who purchased $q_i$ units, that $q_i=\Bar{q}_i$. Also, for those who did not purchase, we assume the $q_i$ size in our data which they were considering purchasing was indeed their size $\Bar{q}_i$. To sum up, we estimate the marginal distribution $f_{\Bar{Q}}(\cdot)$ by equating it with the distribution of observed $q^*$ amounts (across both successful and unsuccessful deals).

In the appendix, we perform a robustness check to our concavity approximation by removing the few data points for which the observed $q_i$ cannot be equal to $\Bar{q}_i$ (due to the discontinuities in the schedule). The results change only negligibly. Also note that the marginal distribution $f_{\Bar{Q}}(\cdot)$ can still be estimated without a concave (or approximately concave) price schedule, though in that case some parametric assumptions may be necessary.

With our estimation for $f_{\Bar{Q}}(\cdot)$ at hand, we next turn to estimating $f_{V|\Bar{Q}}(\cdot)$. We use the following model:

\begin{equation}\label{eq: value regression}
v_{it}=\beta\times X_i+\alpha_t+\gamma_{\Tilde{q}}+\epsilon_{it}
\end{equation}

In this equation, $i$ denotes the customer and $t$ represents the year. Also, $X_i$ captures observable customer characteristics, $\beta$ is a vector of coefficients determining the weights of different customer characteristics,  and $\alpha_t$  represents yearly fixed effects. Additionally, in order to directly relate $\Bar{q}$ to $v$ beyond what could be explained by observables, we allow ``size-group'' fixed effects $\gamma_{\Tilde{q}}$. Here, $\Tilde{q}$ (which is simplified notation for $\Tilde{q}_{it}$) indicates   which of the three intervals $[1,20),[20,50)$ or $[50,\infty)$ customer size $\Bar{q}_{it}$ falls into.\footnote{See our identification argument for why we chose three bins. Conditional on working with three bins, we chose this specific partition in order to, roughly speaking, strike a good balance between (i) the data points in each bin being close in size to each other, and (ii) each bin being of a non-trivial sample size. See appendix \ref{appendix: other specs} for a sensitivity analysis on the choice of more bins.} These size-group fixed effects have a critical role in our model's ability to capture the relationship between $\Bar{q}_i$ and $v_i$ across customers $i$. Finally, $\epsilon_{it} \overset{iid}{\sim} logistic(0,\sigma)$. That is, $\epsilon$ is the mean-zero error term with a logistic distribution. Note that we do not normalize the standard deviation of $\epsilon_{it}$ to 1. This is because the price coefficient in the customers' net value formula has already been normalized to -1.\footnote{The reason why we choose this less standard normalization is that we would like our $v_i$ values to be measured in dollar terms.}

We estimate the model in equation \ref{eq: value regression} using a MLE approach. The likelihood function is as follows:

\begin{equation}\label{eq: likelihood}
    \mathcal{L}(\beta,\alpha,\gamma,\sigma)=\Pi_{it}\text{Prob}\big(s_{it}=\textbf{1}_{q^*(P|v_{it},\Bar{q}_{it})>0}\big)
\end{equation}

where $s_{it}$ is the observable binary variable denoting whether deal $it$ was successful. This likelihood function gives the probability that the predicted deal success matches with the observed one. 

We do not estimate this likelihood function on the original dataset. Rather, we augment the dataset in the following way. We create a copy of the original dataset and make the following modifications to it for each row $it$: first, we set $p_{it}=0$. Second, we set $s_{it}=1$. Then, we concatenate the original dataset with this modified copy. The purpose of this data-augmentation step is to bring in, as an extra moment, the assumption that almost all potential customers in our record would have purchased if the prices were sufficiently small. As we will argue, this is key to identifying price effects in our model.

\textbf{Identification.} Overall, the entire variation in the data identifies all of the variables of interest. But an informal description of our intuition for what mainly identifies what would still be useful. The  vector $\beta$ is identified by the variation in firm characteristics. Year fixed effects are identified by the variation in deal success across years for similarly looking firms negotiating similarly sized deals.  Size bucket fixed effects are identified by the differential rates of deal success across different deal sizes. The standard deviation $\sigma$ is identified by choosing the value that, for each size group, would (i) match the predicted demand levels by the model under zero price to the observed volume of potential demand, and (ii) match the predicted demand levels by the model under the observed price schedule to the observed realized demand. If $\sigma$ is too large, then predicted deal success rates across all customer sizes and all prices will be close to 0.5, which is punished by the likelihood function. For $\sigma$ values that are too small to also be punished, some model assumptions help. One key assumption that assists our identification is that we do not interact the size-group fixed effects with year fixed effects. As a result, any change in relative deal-success rates across size groups from 2020 to 2021 could only be explained by $\sigma$. Another aspect of the model that would punish $\sigma$ levels that are too small is the fact that our size-group fixed effects are coarser than the number of different marginal prices (3 v.s. 5) in the observed schedule. Thus, within some of these size bucket there is some variation in price. If $\sigma$ is too small, the price variation within the size bucket will lead to a prediction that the subset of the size bucket facing the lower marginal price will purchase almost certainly and the other subset with a vanishingly small probability. This would not match the data and gets punished. Note that although our model in equation \ref{eq: value regression} includes both of these assumptions (no interaction between size-group and year fixed effects, and the size fixed effects being coarser than the price schedule), either one of them is sufficient to ensure identification. For robustness of the analysis to having five instead of three size fixed effects, see Appendix \ref{appendix: other specs}.

We finish this section by a focused discussion of how price effects are identified. Note that although we have normalized the price coefficient in the model to -1, notions such as price elasticity are still meaningful and are governed by a combination multiple parameters. The key price variation in our setting that helps identify the price effects (and do so by size group) arises from two things (i) observed ``intended'' sizes for unsuccessful deals and (ii) an identifying assumption. More specifically, our estimation procedure interprets the total number of potential customers (i.e., customers regardless of deal-success status) for each size-group as the demand-volume to match under negligible prices. The estimation procedure also attempts to match observed demand volumes (this time deals that indeed successfully closed) for each size group at observed prices. This effectively creates demand variation (from the volume of all potential deals to that of successful deals) in response to a price variation (from zero to the observed prices) that is used for identification. As we discussed before, this idea was incorporated into the estimation by augmenting the original dataset with a copy of it in which we set $p_{it}=0$ and $s_{it}=1$.


The above argument is equivalent to interpreting the percent success rate for each deal size group as the \textit{``semi elasticity of demand''} for customers of that size when price per-unit is moved from zero to the observed price for that size. It is this interpretation that identifies price effects by size-group in our model. Of course, we have only two price points (zero and the observed level) for each size group. As a result, the identification of the demand function requires parametric assumptions. This is one reason why we have the formulation in equation \ref{eq: value regression} for values $v_i$, unlike the fully non-parametric estimation that we had for the distribution of $\Bar{q}_i$.

\subsection{Estimation Results}

There are both cost parameters and demand parameters in our model. This section describes how we estimate the demand-side parameters using the procedure described before and how we directly calibrate the cost-side ones based on data from the company.

\subsubsection{Demand Side Parameters}

As mentioned before, the objective here is to estimate the joint distribution $f(\cdot,\cdot)$ over $\bar{q}$ and $v$ as flexibly as possible. We estimate $f_{\bar{Q}}(\cdot)$ directly off the data by taking the empirical distribution. We then estimate $f_{V|\bar{Q}}(\cdot)$ by finding the parameter values $(\beta,\alpha,\gamma,\sigma)$ in equation  \ref{eq: value regression} for $v_{it}$ that maximize the likelihood function in equation \ref{eq: likelihood}. The MLE results are presented in  Table \ref{tab:mleestimates}. As for what company characteristics to include in $X$ from equation \ref{eq: value regression}, we chose  a number of features that seemed to have the most predictive power on whether a deal would happen, conditional on $\alpha$ and $\gamma$. These features were: the age of the customer (as a firm), two industry group indices (``computer software'' and ``marketing and advertising''), and two behavioral features which we term ``feature 1'' and ``feature 2''.\footnote{We cannot disclose the nature of feature 1 and feature 2 due to non-disclosure agreement. For a robustness analysis to the inclusion of more or fewere behavioral features, see Appendix \ref{appendix: other specs}.} Features such as number of employees, revenue, location, or some other industry categories industry were not highly predictive once size fixed effects $\gamma$ are included in our regression. 

\begin{table}[H]
\begin{center}
{
\begin{tabular}{ l  c c  }
\toprule
\textit{\textbf{Coefficient}} &  &  \textit{\textbf{Estimate}}\\
\midrule
\textit{Intercept} & $\beta_{0}$ & 2260.56 \\[-0.3ex]
&& (56.25)\\[1ex]
\textit{Log Feature 1} & $\beta_{1}$ & 133.79 \\[-0.3ex]
&& (9.10)\\[1ex]
\textit{Feature 2} & $\beta_{2}$ & 686.64\\[-0.3ex]
&& (29.67)\\[1ex]
\textit{Computer Software} & $\beta_{cs}$ & 39.18 \\[-0.3ex]
&& (28.71)\\[1ex]
\textit{Marketing and Advertisement} & $\beta_{ma}$ & -224.29 \\[-0.3ex]
&& (66.90)\\[1ex]
\textit{Log Firm Age} & $\beta_{age}$ & -40.99 \\[-0.3ex]
&& (16.43)\\[1ex]
\textit{Time} & $\alpha_{2021}$ & 70.2\\[-0.3ex]
&& (25.61)\\[1ex]
\textit{Mid Size} & $\gamma_{medium}$ & -657.40\\[-0.3ex]
&& (44.05)\\[1ex]
\textit{Large Size} & $\gamma_{big}$ & -835.73 \\[-0.3ex]
&& (67.53)\\[1ex]
\textit{Scale} & $\sigma$ & 385.44\\[-0.3ex]
&& (8.46)\\[1ex]
\midrule
\textit{Negative} & & \\
\textit{Log-Likelihood}& &2648.59\\
\bottomrule
\end{tabular}}
\end{center}
\footnotesize
\emph{Notes:} Regression results for equation \ref{eq: value regression}. Constrained optimization with bound on positive scale parameter(or precision parameter) was implemented using \textit{L-BFGS-B} method with stopping tolerance of $10^{-15}$.
\caption{Maximum Likelihood Estimates for the parameters describing $f_{V|\bar{Q}}(\cdot)$ according to equation \ref{eq: value regression}. Bootstrapped standard errors are shown in parentheses.}
\label{tab:mleestimates}
\end{table}

 

As the results from Table \ref{tab:mleestimates} suggest, both mid-size and large customers have, on average and all else equal\footnote{Of course all is is not equal; larger and smaller customers may differ from each other systematically on other characteristics.}, smaller per-unit willingnesses to pay--respectively by \$657 and \$854 per unit--for the company's product relative to smaller customers (again, recall that by ``size'' we do not mean the size of the customer as a firm. We mean the size of their need for the product the company sells to them). We also see that clients in 2021 valued the product more than those in 2020 did. Younger clients seem to place a higher value on LifeLabs' services. Clients categorized as ``computer software'' tend to have a higher valuation for the product relative to the average industry, while clients from ``Marketing and advertisement'' tend to value it less. These results are directionally congruent with the data patterns presented in section \ref{sec: data and setting}.

\begin{figure}[H]
    \centering
    \includegraphics[scale=.5]{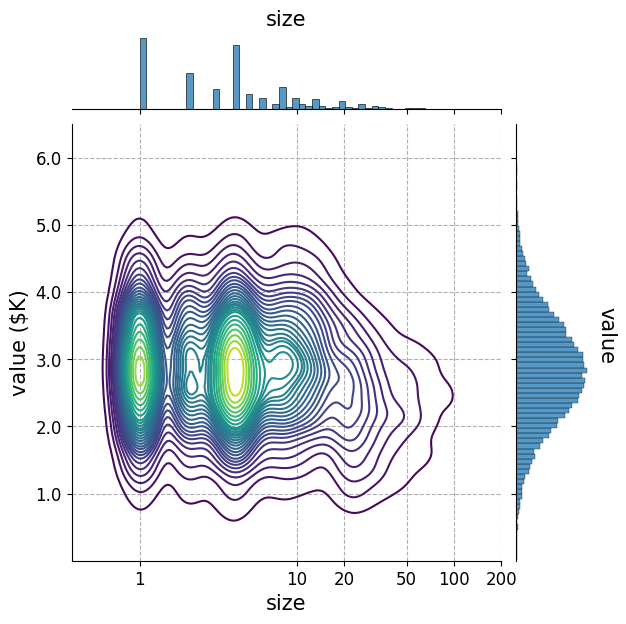} 
    \footnotesize
\\ \linespread{1}\small
\textit{Notes:}  \textit{Size} used in visualizing the distribution are log transformed.

\caption{Estimated Joint Distribution over $\widehat{v}_{it}$ and $ln\,\bar q$}
    \label{fig: joint dist}
\end{figure}

With both $f_{V|\bar{Q}}(\cdot)$ and $f_{\bar{Q}}(\cdot)$ in hand, the joint distribution $f(\cdot,\cdot)$ is recovered. Figure \ref{fig: joint dist} visually presents this joint distribution. The  figure confirms the suggestive evidence in  Figure \ref{fig:successVSsize} that the per-unit willingness to pay seems to be smaller for customers with larger needs. As a result, it seems natural to expect the company to want to offer lower per-unit prices for larger deals. The question is ``By how much," which our estimates of the demand parameters, as well as cost estimates provided below, help us quantify.

\textbf{Robustness to Demand Specification.} For a robustness analysis to the specification used in equation \ref{eq: value regression}, see Appendix \ref{appendix: other specs}. To summarize the finding in that appendix: the results are robust both on the front of the estimated joint  distribution $f(\cdot,\cdot)$ and on the front of the implied optimal price schedule.

\subsubsection{Cost Parameters}\label{subsub: cost}

Detailed cost-side data, alongside conversations with the company, allows us to obtain estimates of $c_{1}$ and $c_{2}$ for all deals. Based on internal documents from LifeLabs, the total cost $c_{1}+c_{2}\times q_{it}$ for successful deals has the following three components:

\begin{enumerate}
    \item Setup costs: This is an element of costs that is fixed per deal. It is evaluated at $\$1,253$ per deal  by LifeLabs estimates.

    \item SNC (Service and Consulting Costs): This element of costs pertains to supporting the customer once a deal is purchased. Internal estimates put SNC in three tiers and have the following average estimate for the tiers:

    \begin{itemize}
        \item Low: $\$1,868$ per deal
        \item Medium: $\$4,670$ per deal
        \item High: $\$9,340$ per deal
    \end{itemize}

    Two additional notes about the SNC cost element: (i) which of the three groups low, medium, or high a deal is categorized under is decided by a ``client administrator'' at LifeLabs based on client's observable and unobservable characteristics. This classification need \textit{not} coincide with our small/medium/large classification for deal sizes. (ii) Based on LifeLabs' internal estimates, overall,  65\% of SNC costs is fixed and 35\% is variable.
    \item Per-workshop Cost (e.g., workshop leader wages, etc)
    \begin{itemize}
        \item $\$601$ per workshop
    \end{itemize}
\end{enumerate}

We use this information to calibrate a cost model consisting of two parameters  $c_{1}$ and $c_{2}$. The fixed per-deal cost parameter $c_1$ is measured in dollars per deal. The variable cost parameter $c_2$ is measured in dollars per workshop. Formal measurements are carried out as follows:

Fixed cost parameter $c_1$ consists of the setup cost and 65\% of SNC.

\begin{equation}\label{eq: fixed cost calibration}
    c_{1t}= 1253 + 0.65 \times \frac{\sum_{i\,\text{s.t. } d_{it}=1}{SNC}_{it}}{\sum_{i\,\text{s.t. } d_{it}=1}1}
\end{equation}

and variable cost parameter $c_2$ consist of the per-workshop cost and 35\% of SNC:

\begin{equation}\label{eq: var cost calibration}
    c_{2t} = 601 + 0.35 \times \frac{\sum_{i\,\text{s.t. } d_{it}=1}{SNC}_{it}}{\sum_{i\,\text{s.t. } d_{it}=1}\bar{q}_{it}}
\end{equation}

where ${SNC}_{it}$ takes one of the three values 1868, 4670, or 9340, as described above.\footnote{The $SNC_{it}$ levels taking only one of three values may not be realistic even though it is based on the best estimates the company has. As a result, one can consider a robustness analysis in which the $SNC$ values are ``smoothed out'' based on a regression model before being used as an input into equations \ref{eq: fixed cost calibration} and \ref{eq: var cost calibration}. In appendix \ref{apx: alternative cost model}, we carry out this analysis and find that the results are indeed robust.}

\begin{figure}[H]
    \centering
    \includegraphics[height=2in]{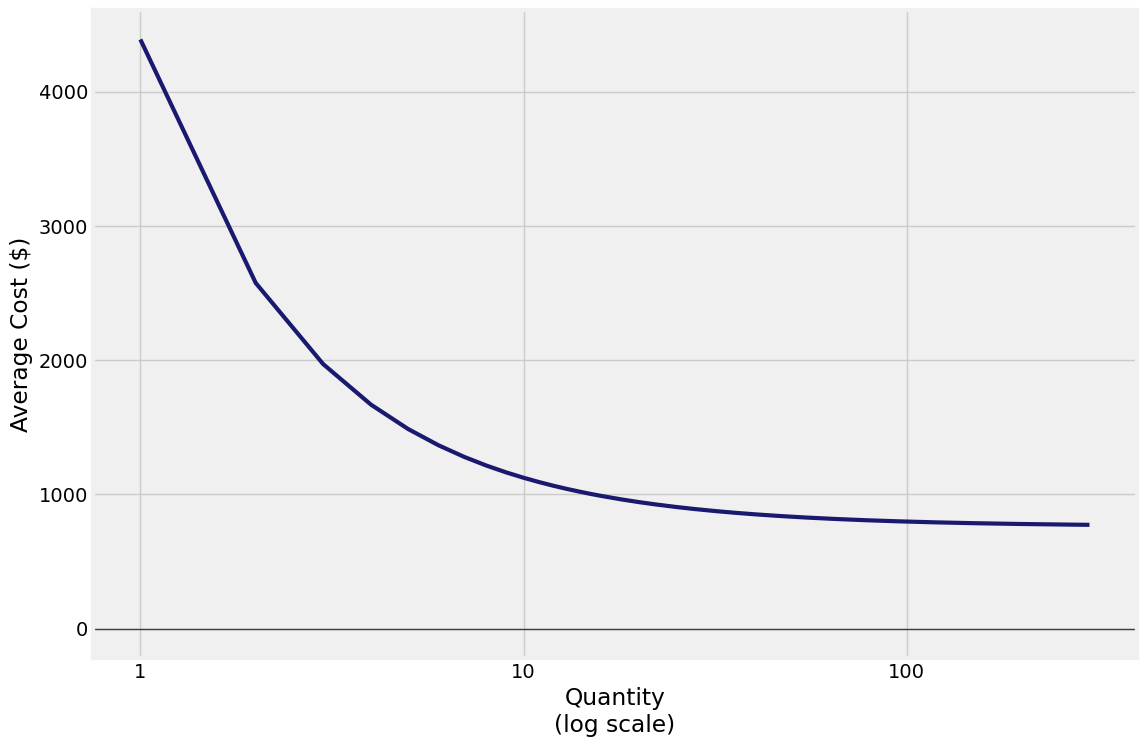}
    \caption{Average Cost Curve}
    \label{fig:avgcost}
\end{figure}

Based on this analysis, we arrive at the following estimates: $\hat c_{1,2021} = $ \$3,630/customer and  $\hat c_{2,2021} = $ \$760/workshop. Figure \ref{fig:avgcost} plots the average cost (i.e., $\frac{c_{1,2021}}{q}+c_{2,2021}$) as a function of quantity $q$.\footnote{We only estimate the cost parameters for 2021 because this is the year we choose to conduct our counterfactual analyses for.} This figure shows that, similar to demand side parameters, our cost side analysis suggests that the company's optimal nonlinear tariff will lower the per-unit price for larger deals. 

\subsection{Model Fit}\label{subsec: model fit}

\begin{figure}[H]
    \centering
    \includegraphics[width=\linewidth]{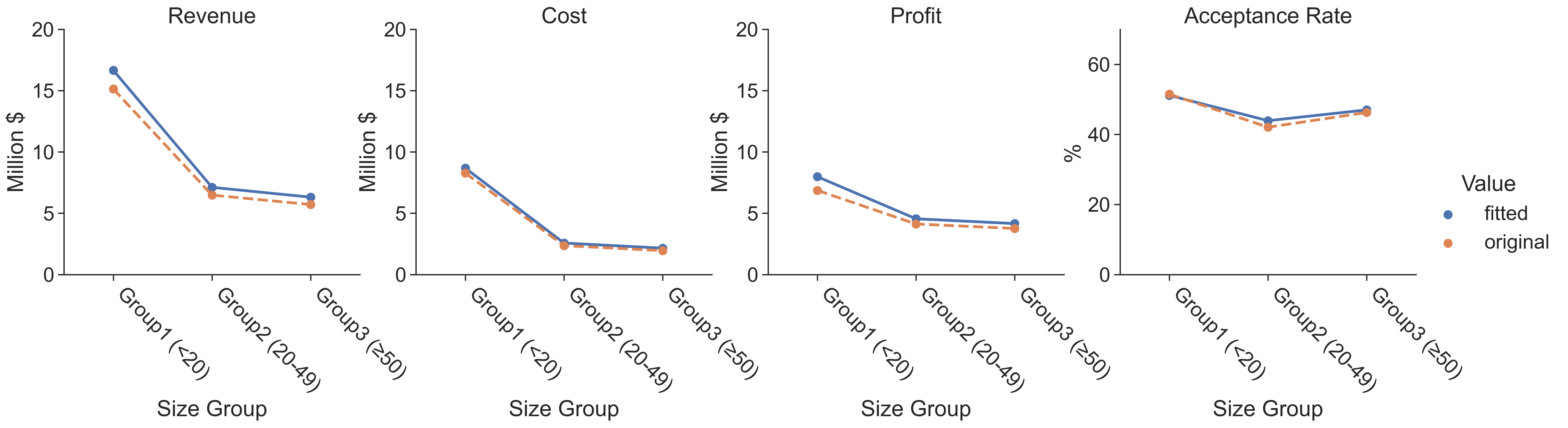}
    \caption{Model Prediction VS Original Data}
    \label{fig:modelfit3}
\end{figure}

Figure \ref{fig:modelfit3} describes goodness of fit for our estimated model. It compares multiple quantities measured directly on the data to their counterparts generated by the model. In particular, it examines deal purchase rate, total revenue, total cost, and total profit (revenue net of costs) by three size groups. As can be seen from the figure, our model fits the data quite closely. 

We now turn to quantifying the exact shape of this optimal nonlinear contract.

\section{Optimal Nonlinear Pricing Scheme}\label{sec: optimal schedule}

With an estimated model of demand and costs in hand that closely fits the data, we are now ready to solve the optimization problem in equation \ref{eq: optimal tariff} and arrive at the profit-maximizing schedule. Focusing on the market in 2021, in this section we discuss three topics.  First, we present the main results on the optimal schedule. Next, we move to discussing this optimal schedule's profitability, and consumer- and social-welfare implications. Finally, we study the implications of charging fixed fees.

We need to start with a parameterization for the price schedule $P(\cdot)$. The main parameterization we work with is similar to LifeLabs' current pricing strategy: it consists of a few linear segments where the continuation of each of them would pass through the origin. Formally, consider intervals $I_k=[q_k,q_{k+1})$ where $k=1,2,...$ and $q_1=0$. We allow price schedule $P$ to take the form $P(q)=p_k\times q, \forall q\in I_k$, where $p_k$ values are constants. In our application, we consider $I_1$ through $I_5$ to be, respectively, $[0,10),[10,20),[20,50),[50,100),[100,\infty)$. Observe this parameterization restricts the space of all possible schedules from an infinite dimensional object to a 5-dimensional one. As a result, with some abuse of notation, we sometimes refer to the function $P(\cdot)$ as a vector $P=(p_1,...,p_5)$. 

Note that we could in principle restrict this space in alternative ways. For instance, we could consider a piecewise linear and continuous schedule using the five intervals above. We in fact do examine this continuous alternative initially (see next subsection). But we chose the approach described in the previous paragraph as our main specification for two reasons: first, it is the structure that the firm is using; and second, as we will see shortly, our estimates for the optimal through-origin schedule is substantially more precise (smaller confideince intervals) relative to the optimal continuous schedule in our simulations. With the parameterization structure in place, we next turn to the empirical analysis, skipping the details of the optimization method we develop. For more details on our  method, how it compares to alternative optimization algorithms, and our recommendations for algorithm choice in similar situations, see appendix \ref{appendix: optimization method}.

\subsection{Optimal Price Schedule}

\begin{figure}[H]
\begin{center}
\begin{subfigure}{0.475\textwidth}
    \includegraphics[width=\textwidth]{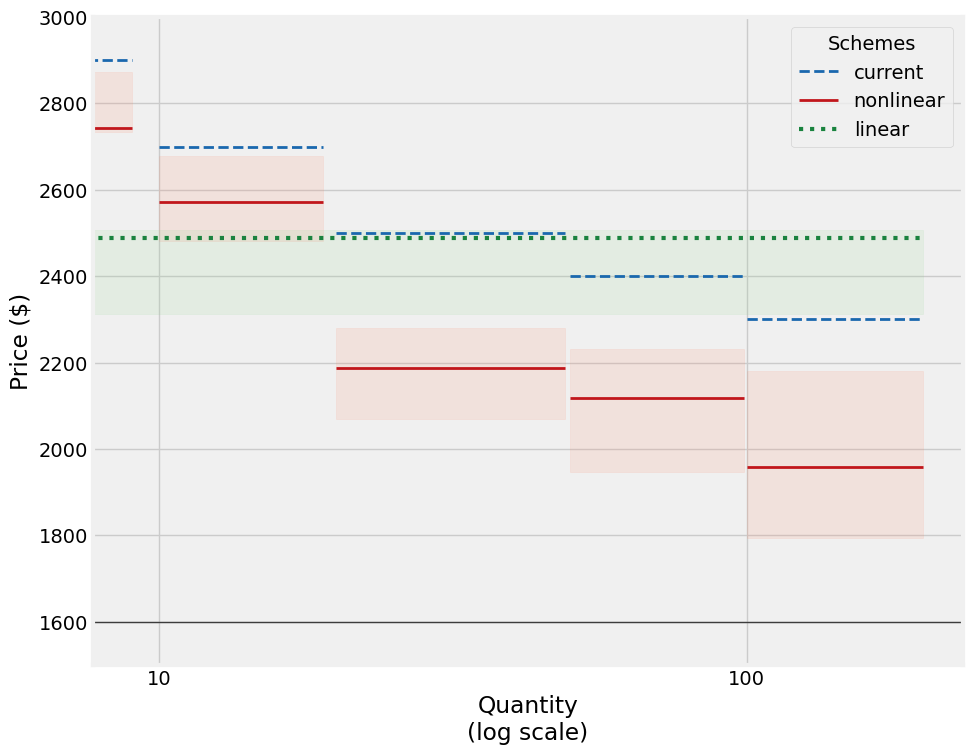}
    \caption{Marginal Price}
    \label{fig:marginalprice-all} 
\end{subfigure}
\begin{subfigure}{0.475\textwidth}
    \includegraphics[width=\textwidth]{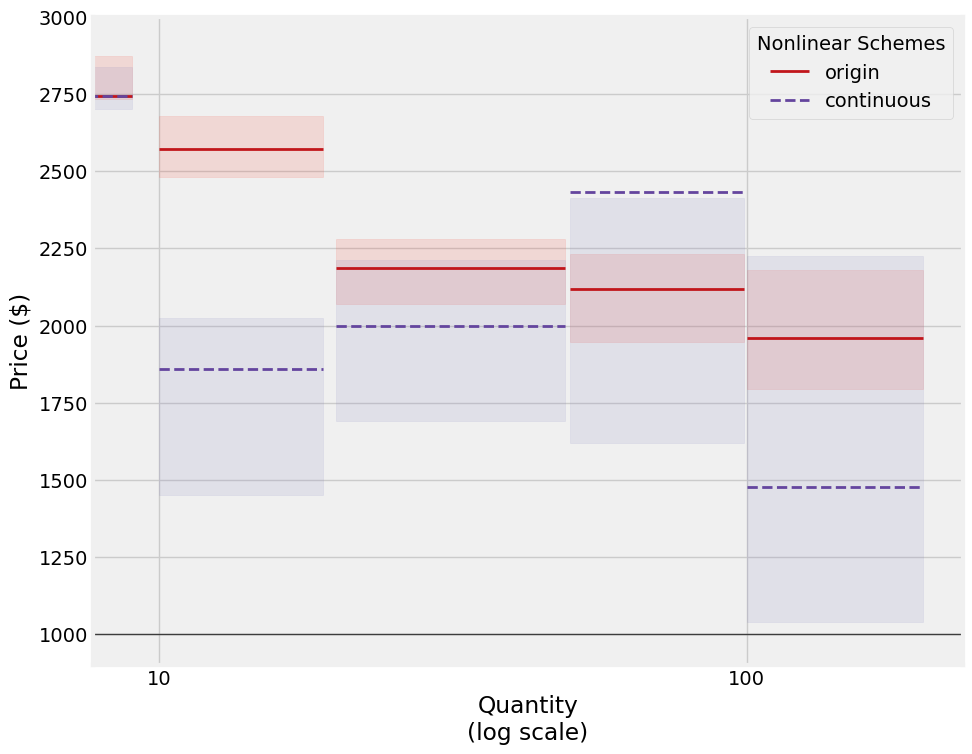}
    \caption{Marginal Price: nonlinear price schemes}
    \label{fig:marginalprice-nonlin} 
\end{subfigure}
\end{center}
\linespread{1}
\footnotesize\emph{Notes:}  Panel (a) shows the marginal prices for \textit{current}, \textit{(via-origin) nonlinear}, and \textit{linear} price schedules. (b) is the comparison of marginal prices between \textit{continuous} and \textit{via-origin} nonlinear price schedules. The results also exhibit 95\% confidence intervals using 500 bootstrapped samples of the data. Note the horizontal axes are on the log scale.
    \caption{Pricing Schemes}
    \label{fig:schemes}
\end{figure}

Figure \ref{fig:schemes} compares multiple pricing schemes: (i) the current price schedule by LifeLabs, (ii) the optimal linear price schedule, and (iii) the optimal nonliner price schedule. Notably, as mentioned in the previous subsection, we solve for the optimal nonlinear schedule in two ways: first we find the optimal schedule among those that charge fixed per-unit prices within each of the five intervals mentioned above. Second, we find the optimal schedule among those that charge a fixed per-unit \textit{incremental} rate within each of these five intervals. In other words, the first type of nonlinear schedule consists of five segments where the continuation of each of them passes through the origin; whereas the second type consists of five segments that are connected together and form a continuous function.

Panel (a) provides a  comparison across different contracts by plotting marginal prices as a function of deal size (for all contracts except the continuous one.) As can be seen from this panel, the optimal nonlinear schedule charges more than the optimal linear one for small deals and less for larger deals. The optimal difference between the per-unit prices for largest and smallest deal sizes is more than \$700. The panel also shows the bootstrapped 95\% confidence intervals. As expected, the confidence intervals are wider for larger deal sizes due to fewer data points.

Panel (b) attempts to provide a comparison between marginal prices of the optimal continuous schedule and those of the optimal schedule with segments that would pass through origin.\footnote{Though some of the marginal prices are far from each other, this is in fact an indication of overall similarity. To illustrate, in the optimal continuous schedule, the segment for deals of sizes 10 to 19 would have a higher starting point relative to a similarly sloped segment for those sizes in a schedule with segments passing through origin (this is because the former simply starts form the end of the segment for smaller deals). As a result, a lower slope in this case for the continuous schedule compensates for the higher starting point.} Next section shows that in terms of profitability, the two schedules indeed behave similarly. 

\subsection{Profit and Welfare Analysis}

Table \ref{tab:profitandwelfare} shows how different pricing schemes fare against one another with respect to a number of measures. In addition to the current pricing by the firm, three other  pricing schemes are examined. The first scheme is first-degree price discrimination in which the firm tailors pricing toward each individual customer. In this regime, the firm will sell to any customer $i$ with $v_i\times \Bar{q}_i\geq c_1+c_2\times \Bar{q}_i$, charging exactly $v_i\times\Bar{q}_i$. The second regime is linear pricing, meaning the firm is only allowed to charge tariffs in the form $P(q)\equiv p\times q$, choosing $p$ optimally. Finally, the third tariff is the optimal nonlinear pricing scheme $P^*(\cdot)$ which was described shortly before. Both the oprimal linear price and the optimal nonlinear schedule were shown in figure \ref{fig:schemes}. The three measures on which the above pricing schemes are compared against each other are (i) firm profit, (ii) consumer surplus, and (iii) social surplus.


\begin{table}[H]
\begin{center}
{\small
\begin{tabular}{ l r l r l r l r l }
\toprule
 &  &  &  &  & \textit{Consumer} &  & \textit{Social} & \\
\hspace{5mm}\textit{Scheme} & \textit{Revenue} & change & \textit{Profit} & change & \textit{Welfare} & change & \textit{Welfare} & change \\
 & \textit{(\$M)} & (\%) & \textit{(\$M)} & (\%) & \textit{(\$M)} & (\%) & \textit{(\$M)} & (\%)\\
 \midrule
\hspace{4mm}\textit{current}  
& 30.08 & \multicolumn{1}{c}{-} & 16.68 & \multicolumn{1}{c}{-} & \ 7.35 & \multicolumn{1}{c}{-} & 24.03 & \multicolumn{1}{c}{-} \\
\cdashline{2-9}[1pt/1pt]
\textit{1$^\text{st}$ degree} 
& 58.55 & +94.65\% & 34.92 & +109.35\% &	\ 0 & $-$100.00\% & 34.92 & +45.32\%\\
\hspace{6mm}\textit{linear} 
& 32.39 & +\ 7.68\% & 16.21 & $-$\ \ 2.82\% &\ 9.33& +\ 26.94\% & 25.54 & +\ 6.28\% \\[-1.5ex]
& \scalebox{.6}[.7]{(30.30, 35.75)} & 
& \scalebox{.6}[.7]{(14.39, 17.72)} & 
& \scalebox{.6}[.7]{(8.91, 11.88)} &
& \scalebox{.6}[.7]{(23.53, 29.20)}\\
\textit{nonlinear}\\
\makecell[r]{\hspace{5mm}continuous} 
& 33.62 & +11.77\% & 17.55& +\ \ 5.22\% & 9.8 & +\ 33.33\%	& 27.35 & +13.82\%\\[-1.5ex]
& \scalebox{.6}[.7]{(31.22, 36.17)} & 
& \scalebox{.6}[.7]{(15.84, 19.01)} & 
& \scalebox{.6}[.7]{(9.07, 11.26)} &
& \scalebox{.6}[.7]{(25.19, 30.25)}\\
\makecell[r]{origin} 
& 33.92 & +12.77\% & 17.54 & +\ \ 5.16\% & 10.28 & +\ 39.86\% & 27.81 & +15.77\%\\[-1.5ex]
& \scalebox{.6}[.7]{(31.06, 35.98)} & 
& \scalebox{.6}[.7]{(15.88, 19.04)} & 
& \scalebox{.6}[.7]{(9.07, 11.23)} &
& \scalebox{.6}[.7]{(25.11, 30.18)}\\
\bottomrule
\end{tabular}}
\end{center}
\footnotesize
\emph{Notes:} The values in parentheses represent the bootstrapped 95\% confidence intervals. For the bootstrapping process, 1,000 samples were used for the \textit{linear} scheme, while 500 samples were used for both the \textit{nonlinear-continuous} scheme and the \textit{nonlinear-origin} scheme.
\caption{Profit and Welfare Analysis}
\label{tab:profitandwelfare}
\end{table}

As can be seen from the table, the optimal linear pricing strategy delivers almost as much profit as the current nonlinear pricing scheme used by the firm. The optimal nonlinear pricing scheme improves profitability by about 5.2\%.\footnote{The profitability has been computed both for the optimal nonlinear schedule with segments passing through origin and for the optimal continuous schedule. As the table shows, they are almost identical. Given these similar performances, and due to  the fact that the through-origin schedule is the observed price structure and the fact that it is substantially more precisely estimated, we work with optimal nonlinear schedules with segments passing through origin.} Note that this 5.2\% is an underestimate because it is carried out on profit levels before taking into account firm-level fixed costs (e.g., facility rent/depreciation, full-time employee salaries and benefit, etc.). To get a sense of the magnitude by which firm-level fixed costs could impact these estimates, note that a 10 or 15 \$M/y fixed cost would, respectively, increase the 5.2\% performance to 12.9 or 51.2\%.\footnote{We did not ask LifeLabs for an estimate of firm-level fixed costs given that the number would have no bearing on the optimal pricing strategy.}

Another question that table \ref{tab:profitandwelfare} helps answer is about the comparison between first- and second-degree price discrimination. If first degree price discrimination were feasible, it would more than double the profit relative to the current strategy. This means that optimal second degree price discrimination recovers only about 7.1\% of the profitability gap between linear pricing on the one hand and first degree price discrimination on the other.

Aside from profit levels, Table \ref{tab:profitandwelfare} also compares different pricing schemes with regards to their effects on consumer- and social-surplus. Compared to linear pricing, nonlinear pricing increases consumer welfare by 10.2\%, hurting some consumers and benefiting others.\footnote{Note that this empirical result is not general. In theory, the welfare effects of price discrimination could go either way. See \cite{schmalensee1981output} or \cite{varian1985price} for instance.} On the front of social surplus, nonlinear pricing  outperforms linear pricing by about 8.9\%.

\begin{figure}[H]
    \centering
    \makebox[\linewidth]{
    \includegraphics[width=.9\linewidth]{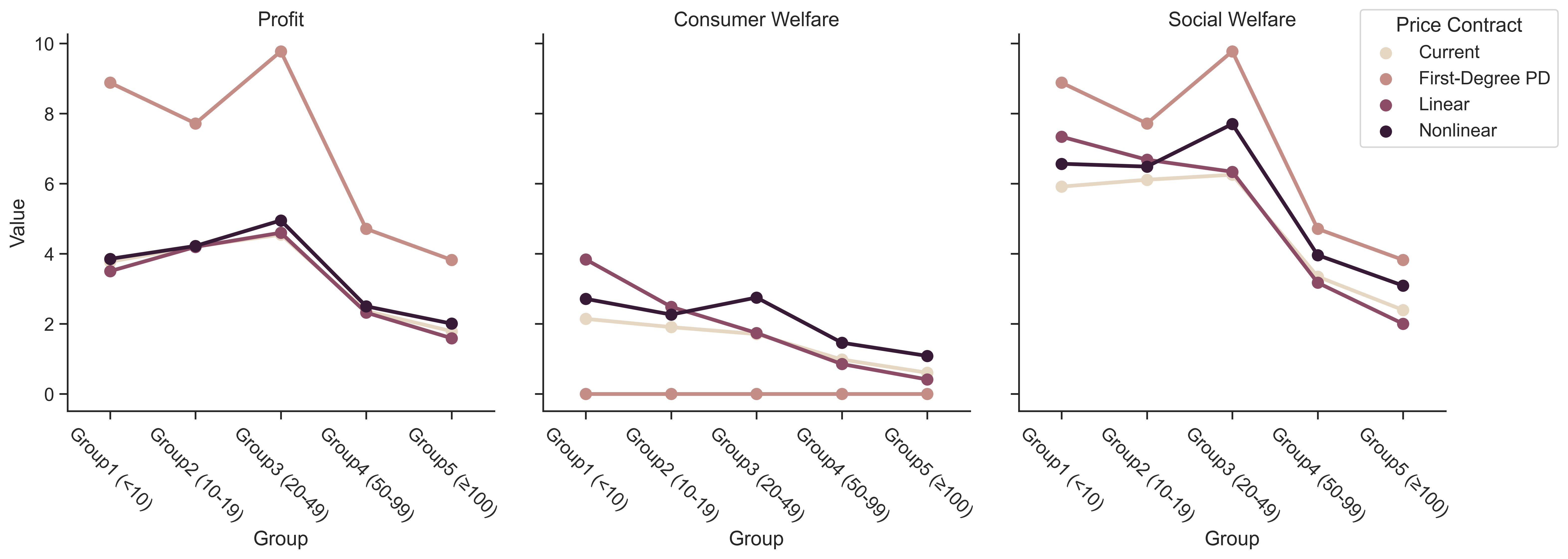}}
    \caption{Visualization of Profit and Welfare Analysis by Groups}
    \label{fig: profit and welfare by groups}
\end{figure}

\textbf{Segment-by-segment analysis.}  In addition to the aggregate analysis presented by Table \ref{tab:profitandwelfare}, it is worth conducting a segment-by-segment analysis of how different pricing policies compare against one another. Specifically, we partition customers into five groups based on the customer size variable $\bar{q}_i$.\footnote{One could, in principle, run a similar analysis with other segmentation methods as well.} Figure \ref{fig: profit and welfare by groups} presents such segment-based results. There are two general lessons from this segment-based analysis.

\begin{table}[H]
    \centering
    {\small
    \begin{tabular}{lcccccc}
    \toprule
    {} &  \makecell{Group1 \\ ($<$10)} &  \makecell{Group2 \\ (10-19)} &  \makecell{Group3 \\ (20-49)} &  \makecell{Group4 \\ (50-99)} &  \makecell{Group5 \\ ($\geq$100)} &  \makecell{Total} \\
   \midrule
   \multicolumn{7}{c}{\it Profit} \\
    Current & 3.77 & 4.21 & 4.55 & 2.37 & 1.79 & 16.68 \\
    First-Degree PD & 8.88 & 7.72 & 9.78 & 4.71 & 3.83 & 34.92 \\
    Linear & 3.50 & 4.20 & 4.60 & 2.32 & 1.59 & 16.21 \\
    Nonlinear & 3.85 & 4.22 & 4.95 & 2.50 & 2.01 & 17.54 \\
    \multicolumn{7}{c}{\it Consumer Welfare} \\
    Current & 2.14 & 1.91 & 1.71 & 0.98 & 0.61 & 7.35 \\
    First-Degree PD & 0.00 & 0.00 & 0.00 & 0.00 & 0.00 & 0.00 \\
    Linear & 3.84 & 2.49 & 1.74 & 0.85 & 0.41 & 9.33 \\
    Nonlinear & 2.71 & 2.26 & 2.75 & 1.46 & 1.09 & 10.28 \\
   \multicolumn{7}{c}{\it Social Welfare} \\
    Current & 5.92 & 6.11 & 6.26 & 3.35 & 2.40 & 24.03 \\
    First-Degree PD & 8.88 & 7.72 & 9.78 & 4.71 & 3.83 & 34.92 \\
    Linear & 7.34 & 6.68 & 6.34 & 3.17 & 2.00 & 25.54 \\
    Nonlinear & 6.57 & 6.49 & 7.70 & 3.96 & 3.09 & 27.81 \\
    \bottomrule
    \end{tabular}}
    \caption{Profit and Welfare Analysis by Groups}
    \label{tab: profit and welfare by groups}
\end{table}
First, and unsurprisingly, 1st-degree price discrimination dominates all other methods both on the fronts of firm profit and  social welfare.  Its high performance on social welfare is because this approach, by construction, allows a transaction to take place if and only if it is socially efficient (i.e., if $v_i \Bar{q}_i\geq c_1+c_2\Bar{q}_i$). Given the wide gap between the profitability of this method and other ones, it is natural to expect that the firm would benefit from a customer-by-customer negotiation approach as opposed to posted pricing; because even though this approach is unlikely to replicate the exact profit from 1st-degree price discrimination, replicating even a fraction of it would outperform nonlinear pricing.\footnote{Note that there are other hurdles in the way of adopting a negotiation strategy that may discourage a firm from going that route. In particular (and according to our interviews with the firm),  in industries such as HR services, there is a considerable chance that the outcome of the negotiation with one client is subsequently learned by other clients and acts as a barrier to negotiating desirable terms with them.} 

The second lesson from the segment-based analysis concerns the comparison between linear and nonlinear schedules. Nonlinear pricing generates lower consumer welfare for small-size customers than does linear pricing. This is reversed for larger customers. This is because, by finding the per-unit price that is, loosely speaking, ``optimal on average'', linear pricing benefits segments that received higher prices under nonlinear pricing and hurts those who received lower prices under the nonlinear schedule. With respect to profitability, nonlinear pricing fully dominates linear pricing given that nonlinear pricing allows the firm to tailor the charges toward respective size groups.\footnote{As we will see later, however, this intuition is incomplete. Optimal nonlinear pricing entails more than choosing the right price for each segment. As emphasized in a long line of theoretical research on screening, nonlinear pricing also requires careful attention to whether customers in each segment respond to relative prices by purchasing a different amount that was ``meant for them.'' }

\subsection{Fixed Fees}

Up to this point in the analysis, we optimized a nonlinear price schedule that maintained the original pricing architecture adopted by the firm: charging five different marginal prices for different deal-size intervals $I_k\in\{[1,10),[10,20),[20,50),[50,100),[100,\infty)\}$. In this section, we investigate an important possible modification to this structure: the addition of a fixed fee. We examine amending a fixed fee both to a linear contract (thereby forming a two-part tariff) and to the nonlinear, 5-dimensional, schedule. Table \ref{tab: fixed fees} shows the optimal linear and nonlinear contracts with and without (optimally chosen) fixed fees. It also presents revenues and profits generated from these four contract types.


\begin{table}[H]
\begin{center}
{\footnotesize
\begin{tabular}{ r c l c c}
\Xhline{1pt}
\thead{Scheme}&  \thead{Fixed Fee (\$)}  & \thead{Marginal Price(s) (\$/unit)} & \thead{Revenue (\$M)} & \thead{Profit (\$M)} \\
\hline 
 Linear & - & [2489] & 32.39 & 16.21 \\
 Linear + Fixed Fee & 2758 & [2102] & 32.91 & 17.83 \\
 Nonlinear & - & [2744 2572 2188 2117 1959] & 33.92 & 17.54 \\
 Nonlinear + Fixed Fee & 1736 & [2402 2441 2109 2097 1946]  & 32.92 & 18.07 \\
\Xhline{1pt}
\end{tabular}}
\end{center}
\caption{Optimal linear and nonlinear price schedules, with and without fixed fees}
\label{tab: fixed fees}
\end{table}

There are two main lessons to learn from  Table \ref{tab: fixed fees}. One is about profitability and the other has to do with the shapes of the optimal contracts.

On the front of profitability, adding a fixed fee appears to be a powerful tool. For instance, adding a fixed fee to a linear contract and forming a two-part tariff yields more than \$1.5M/y extra profitability. It slightly outperforms a full nonlinear schedule that does not have a fixed fee. A main reason for this observation is that a fixed fee effectively introduces diminishing average prices, mimicking the advantage of nonlinear contracts. To empirically verify that introducing effective nonlineariry is indeed at work here, note that the added profitability by a fixed fee shrinks from 17.83-16.21=1.62 \$M/y to 18.07-17.54=0.53 \$M/y when  the baseline contract is nonlinear as opposed to linear. The fixed fee also helps ``partially cover'' the per-customer fixed cost $c_1$=\$3,630/customer. This helps ensure that fewer individual contracts can run a loss; and those that do so run a smaller loss.  A suggestive piece of evidence that fixed fees are helping by ``shedding'' unprofitable customers is that even though they boost the total profit, their effect on the total revenue is less substantial or even net-negative. This latter comparison also suggests that if a firm has a focus on growth (and, hence, seeks to maximize revenue rather than profit), fixed fees may not be as strongly recommended.

Regarding the shape of the optimal contracts, it is worth using our empirical results to relate to the existing literature on the use of fixed fees. To ease this comparison, we focus mainly on the second row of Table \ref{tab: fixed fees} which examines the optimal two-part tariff. The literature on optimal two-part tariffs with homogeneous customers (see for instance \cite{jeuland1983managing}) has long established that, under certain conditions, the optimal pricing policy involves charging a ``variable fee'' that is equal to marginal cost and then using the fixed fee to capture all of the surplus from the consumer.\footnote{Of course, even with homogeneous customers, there are conditions under which the optimal two part tariff deviates from this standard structure. Examples are buyer risk aversion \citep{rey1986logic,ghili2018risk} to competition among buyers when they are part of the supply chain (\cite{rey2008economics}), and non-specifiable  contracts (\cite{iyer2003bargaining}) } It has also been established, however, that under heterogeneous preferences by consumers, the optimal two-part tariff may charge a variable fee different from--typically above--marginal cost (see \cite{oi1971} for an early example). This is what our empirical analysis suggests as well: the optimal variable fee is \$2,102/unit, which is almost three times as high as the marginal cost $c_2$=\$760/unit. To gain intuition for why this happens, first consider a hypothetical scenario under which the seller knows each individual buyer's value function and is able to charge an individualized two-part tariff to it. In this scenario, we have separate pricing problems in each of which consumer-homogeneity is restored (because we have only one customer in each). Thus, it is optimal to offer each consumer a two-part tariff in which the variable fee is \$760 and the fixed fee captures the entire remaining surplus of $(v_i-760)\times\Bar{q}_i$.\footnote{Of course, if $(v_i-760)\times\Bar{q}_i<c_1$, the seller will not sell to customer $i$. This can be implemented by setting the individualized fixed fee to $\max\{c_1,(v_i-760)\times\Bar{q}_i\}$} Now observe that even though our analysis found a negative relationship between $\Bar{q}_i$ and $v_i$, the relation between $\Bar{q}_i$ and $(v_i-760)\times \Bar{q}_i$ is positive. This means that if the seller were able to offer individualized two-part tariffs, it  would charge fixed fees that would tend to increase in $\Bar{q}_i$. The question that arises at this point is: given that the seller is in reality only allowed to offer a common two-part tariff to all customers, how can it attempt to mimic the aforementioned scenario in which larger customers pay larger fixed fees? The answer is: by leveraging the variable fee which helps fine-tune how closely the total pay is tied to size $\Bar{q}_i$. This is why we observe an optimal variable fee so significantly above marginal cost $c_2$=\$760/unit.

Note that the above intuition also helps to understand why the optimal fixed fee (either when coupled with a linear price or when coupled with a nonlinear schedule) is set to only partially cover  the per-customer fixed cost $c_1$=\$3,630/customer. This is because a fixed fee as high as \$3,630 would have put a downward pressure on the variable fee and curtailed the ability of the seller to price discriminate across customers of different sizes. As a result, we observe that the seller accepts the possibility of running a loss on smallest customers  in order to better price discriminate, even though doubling the fixed fee would have guaranteed no loss from any single customer.

\section{Further Simulation Analyses}\label{sec: further CF analysis}

In this section, we carry out additional simulations that serve two purposes. First they shed further light on our methodological framework and its features. Second, they yield substantive insights that could be of value independent of the methodology. We carry out two studies.  We start by examining how optimal nonlinear pricing would respond to various changes in the demand conditions, and we study the role of incentive compatibility constraints. We then dive deeper into the role of costs in shaping the optimal contract, both on their own and in comparison to demand-side factors.

\subsection{Analysis of Demand-Side Factors}\label{subsec: demand and incentives again}

This analysis delves deeper into the role of demand-side factors in shaping the optimal contract. Our objective is to convey two key  messages. First, it is the entire shape of the joint distribution $f(\cdot,\cdot)$ that matters for optimal pricing, as opposed to a simple correlation between size $\bar{q}$ and value $v$ across customers $i$. Second, due to incentive compatibility issues, sellers need to \textit{jointly} optimize all marginal prices $p_1,...,p_5$ which comprise the nonlinear schedule, as opposed to doing so separately. To this end, the rest of this section has two parts. We start by formalizing what we mean by ``separately optimizing'' prices $p_1,...,p_5$ due to ignorance of incentive compatibility constraints. Next, we quantify how the optimal price schedules, profits, and the importance of incentive compatibility constraints respond to  changes in demand conditions.

\subsubsection*{The Role of Incentive Compatibility Constraints}
As we explained in section \ref{sec: optimal schedule}, we jointly optimize the five per-unit prices $p_1,...,p_5$. A key question is: can we optimize these prices separately instead? That is, can we take each price $p_k$ for $k=1,...,5$ and optimize it \textit{only for the market of customers within the relevant range $I_k$ of sizes}? Formally, define the ``local profit function'' $\pi_k(p)$ for $p\in\mathbb{R}$ to mean the profit to the firm if (i)  the set of potential customer consisted only of those with $\Bar{q}_i\in I_k$ and (ii) the firm charged the linear price schedule of $P(q)=p\times q$:

\begin{equation}\label{eq: profit local}
\pi_k(p)=N_k\times\int_{v,\Bar{q}} p\times\big( q^*(p|v,\Bar{q}) \big)-c_1\times \textbf{1}_{q^*(p|v,\Bar{q})>0}-c_2\times \big(q^*(p|v,\Bar{q})\big) f(v,\Bar{q}|\Bar{q}\in I_k)dvd\Bar{q}
\end{equation}

where $N_k$ is the total count of all potential customers $i$ with $\Bar{q}_i\in I_k$; and with a slight abuse of notation, $q^*(p|v,\Bar{q})$ is the amount purchased by customer with size $\Bar{q}$ and value $v$ under the linear price schedule of $P(q)\equiv p\times q$. 

Denote by $\tilde{p}_k$ the separately optimized price for size group $I_k$. That is:

\begin{equation}\label{eq: locally optimal price}
\tilde{p}_k=\arg\max_{p\in\mathbb{R}} \pi_k(p)
\end{equation}

We can now formally state what it means to ``optimize prices $p_1,...,p_5$ separately". It means charging the schedule $\tilde{P}=(\tilde{p}_k)_{k=1,...,5}$ instead of $P^*=(p^*_k)_{k=1,...,5}$. Likewise, a formal way of asking ``what would be the consequences of optimizing prices separately" would be to ask ``how closely does $\pi(\tilde{P})$ approximate $ \pi(P^*)$?''  This is an important question both conceptually (are different size groups effectively ``separate'' markets?) and computationally (can we avoid the difficult joint optimization problem and  optimize one-dimensional objects instead?).

By the definition of $P^*$ as the optimal schedule, it has to be that $\pi(\tilde{P})\leq \pi(P^*)$. The reason why this inequality may be strict is ``incentive compatibility'' constraints that are ignored when prices are separately optimized. To illustrate, optimizing separately ignores the fact that if $\tilde{p}_3$ is substantially smaller than $\tilde{p}_4$, then some customers $i$ of size-group $4$  (i.e., $\Bar{q}_i\in I_4$) who would purchase $\Bar{q}_i$ under $\tilde{p}_4$ absent other options,  might take the opportunity presented by the wide gap between $\tilde{p}_3$ and $\tilde{p}_4$ and reduce their purchase sizes. Similarly, if $\tilde{p}_3$ is substantially lower than $\tilde{p}_2$, customers of size group 2 might respond by increasing their purchase sizes, paying an overall lower total price, and imposing a higher cost of production to the firm. Put differently, the schedule $\tilde{P}$ naively attempts to separately sell to different size groups at their respective optimal prices. This, if feasible, would yield a total profit of $\Sigma_k \pi_k(\tilde{p}_k)$, but is \textit{not incentive compatible} and ends up delivering a lower profit. The schedule $P^*$, however, optimizes in an incentive-compatibility-aware fashion.

As a long line of research on mechanism design suggests, it is theoretically conceivable that ignoring incentive compatibility results in strictly positive losses: $\pi(P^*)-\pi(\tilde{P})>0$. The empirical question is how substantial is the loss? This, among other demand-related ones,  is a question we turn to next.




\subsubsection*{Simulation and Analysis of Different Demand Conditions:}

\begin{figure}[H]
\begin{center}
\begin{subfigure}{0.3\linewidth}
    \includegraphics[width=\linewidth]{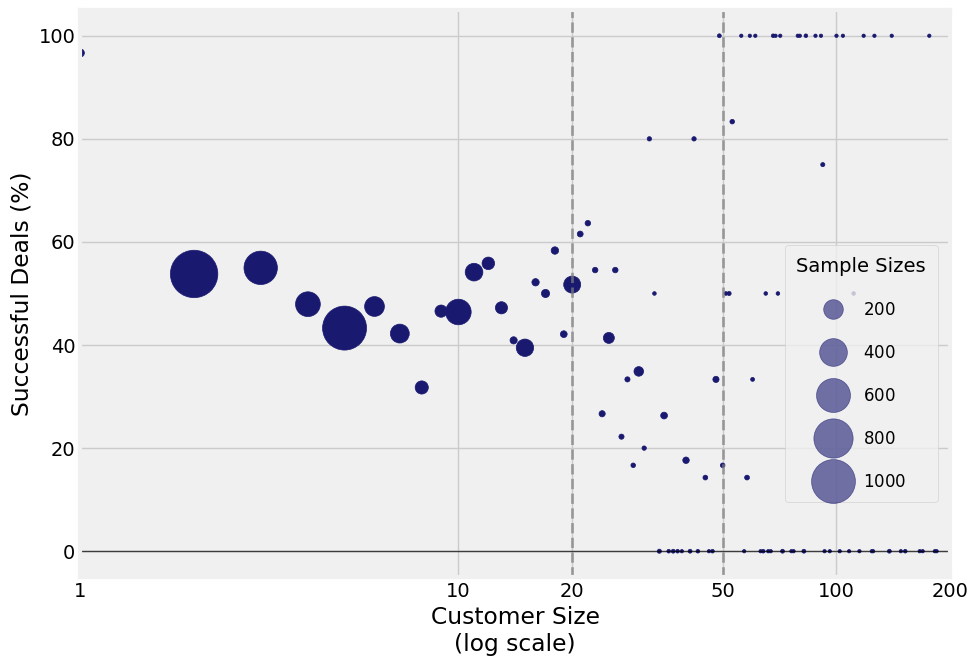}
    \caption{Original Data}
    \label{fig:(f)corr} 
\end{subfigure}
\begin{subfigure}{0.3\linewidth}
    \includegraphics[width=\linewidth]{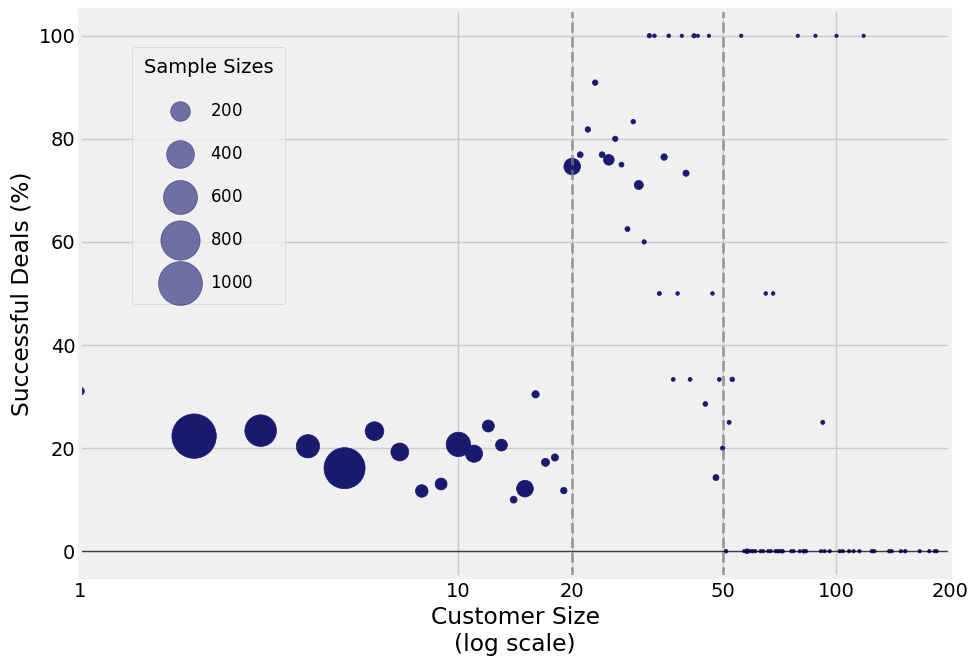}
    \caption{Counterfactual Data (case 1)}
\end{subfigure}
\begin{subfigure}{0.3\linewidth}
    \includegraphics[width=\linewidth]{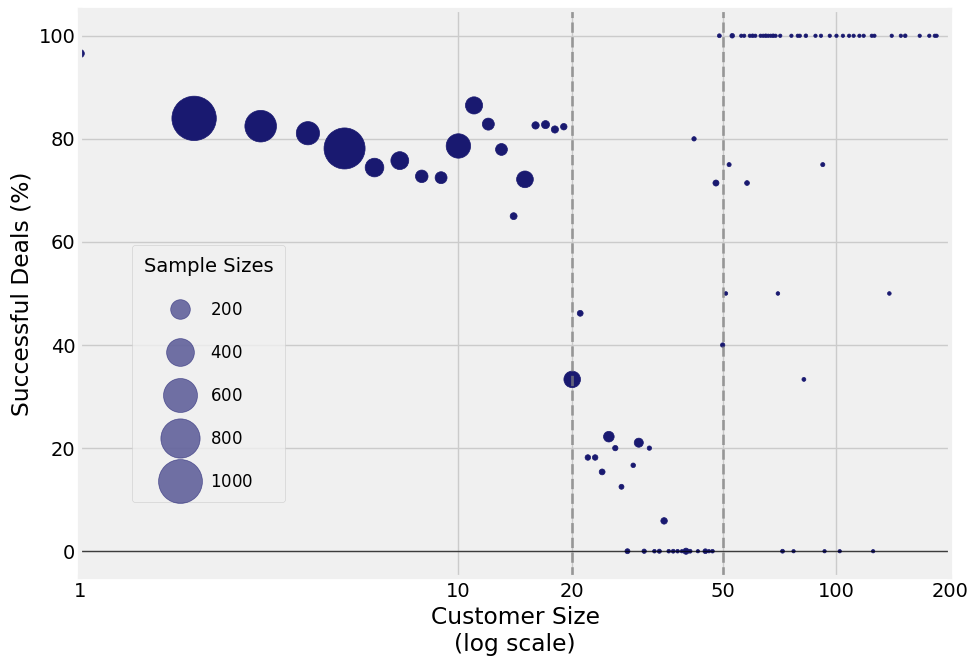}
    \caption{Counterfactual Data (case 2)}
\end{subfigure}
\begin{subfigure}{0.3\linewidth}
    \includegraphics[width=\linewidth]{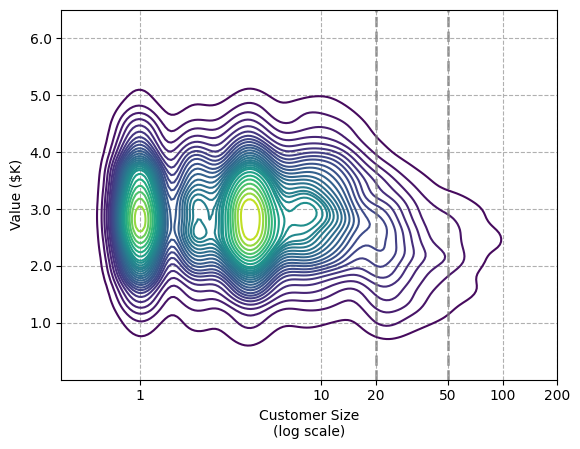}
    \caption{Estimated Demand (original)}
\end{subfigure}
\begin{subfigure}{0.3\linewidth}
    \includegraphics[width=\linewidth]{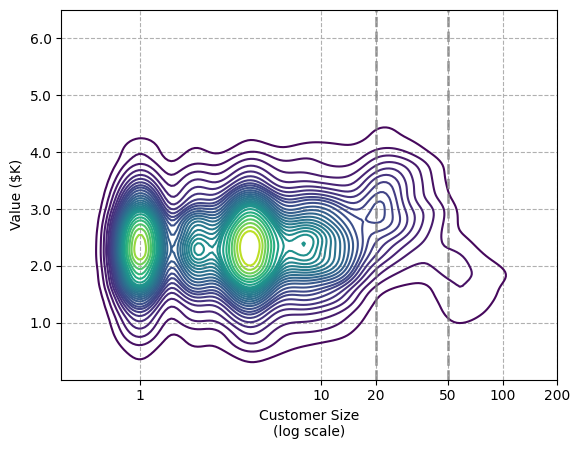}
    \caption{Estimated Demand (case 1)}
\end{subfigure}
\begin{subfigure}{0.3\linewidth}
    \includegraphics[width=\linewidth]{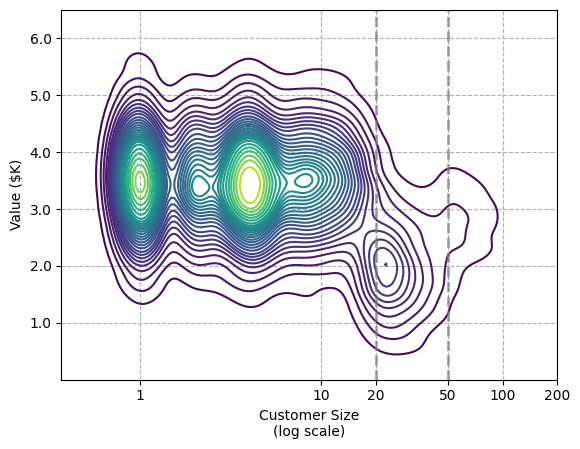}
    \caption{Estimated Demand (case 2)}
\end{subfigure}
\begin{subfigure}{0.3\linewidth}
    \includegraphics[width=\linewidth]{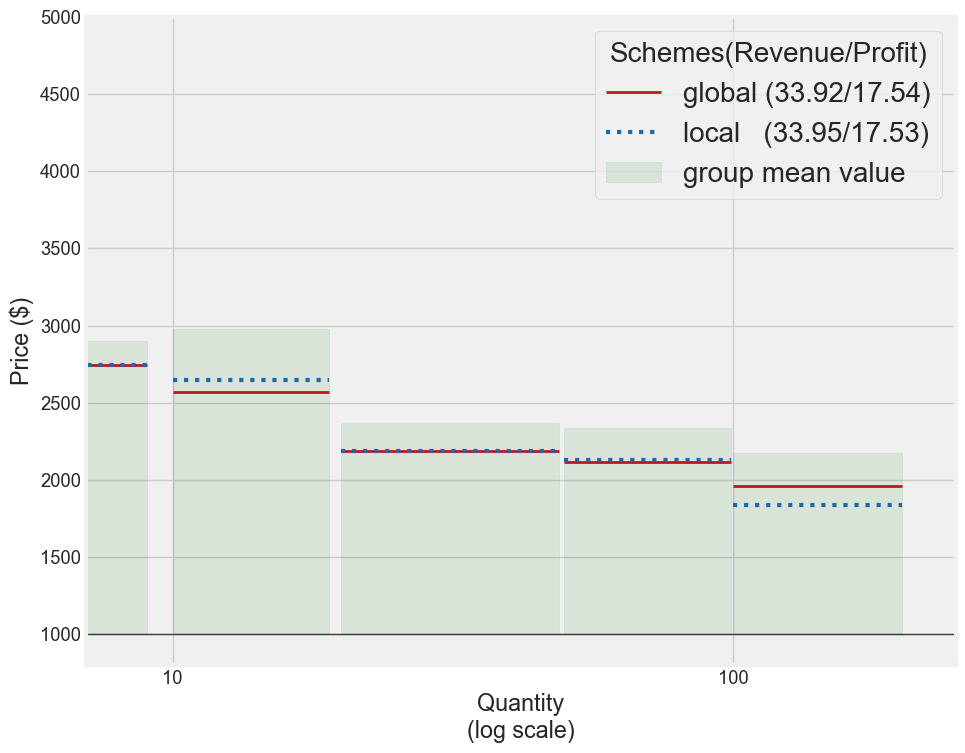}
    \caption{Optimal Pricing (original)}
\end{subfigure}
\begin{subfigure}{0.3\linewidth}
    \includegraphics[width=\linewidth]{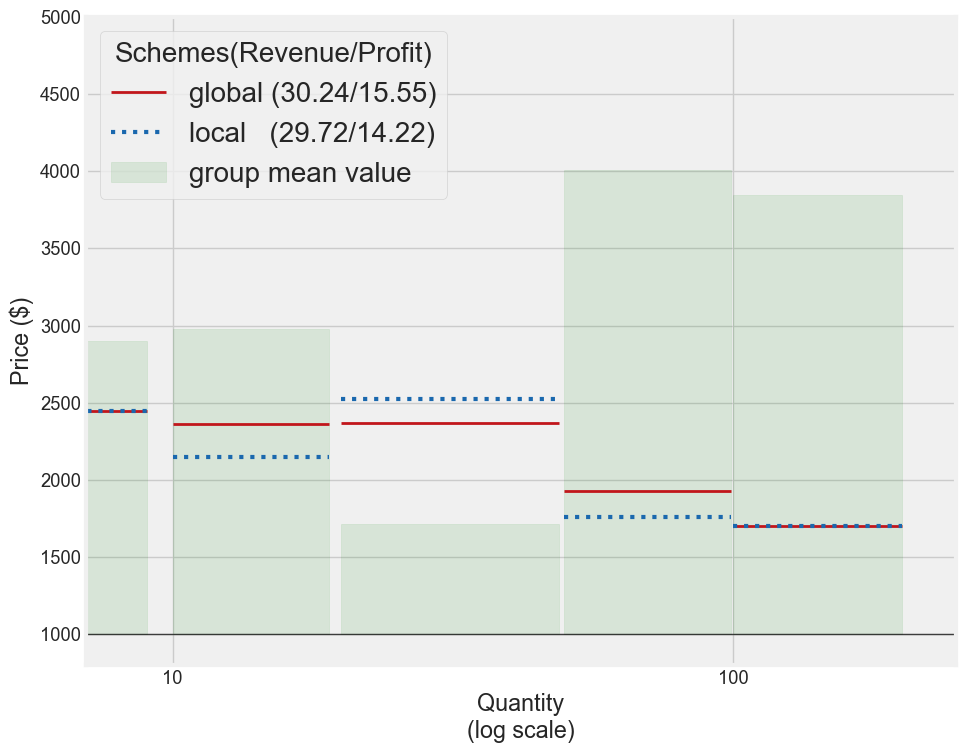}
    \caption{Optimal Pricing (case 1)}
\end{subfigure}
\begin{subfigure}{0.3\linewidth}
    \includegraphics[width=\linewidth]{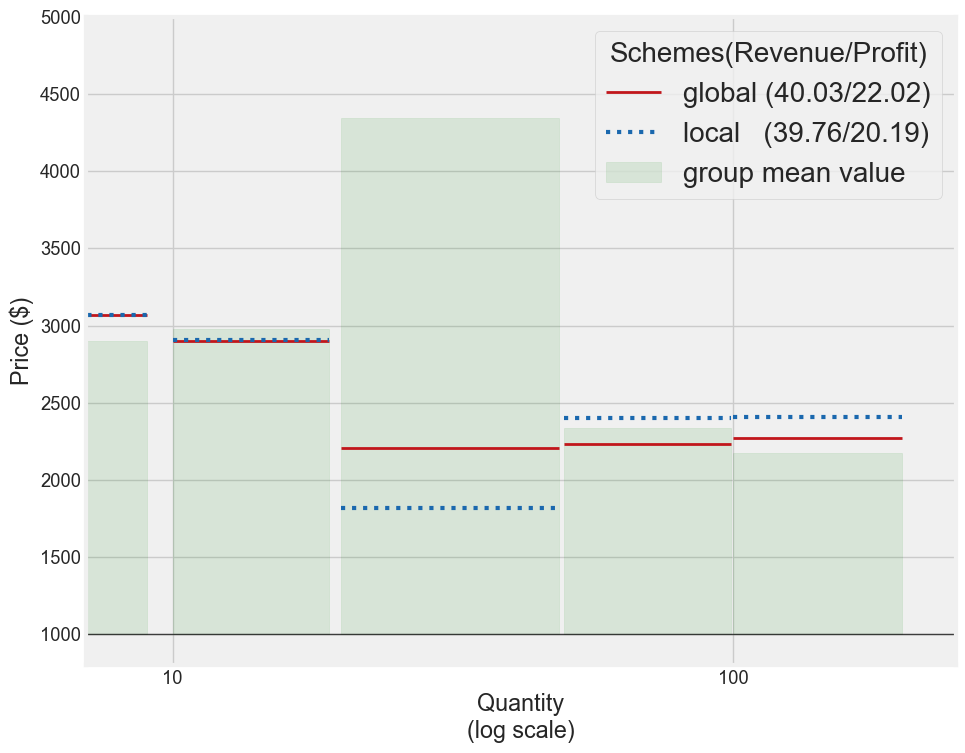}
    \caption{Optimal Pricing (case 2)}
\end{subfigure}
\end{center}
    \caption{Analysis of optimal pricing under three scenarios: Left column: original data; middle column: counterfactual data with higher acceptance rates for mid-size deals compared to small and large ones; right column: counterfactual data with lower acceptance rates for mid-size deals compared to small and large ones. For details of counterfactual data simulation, refer to appendix \ref{appendix: data simulation}. }
    \label{fig: incentive compatibility demand CF}
\end{figure}

With a formalization of individually v.s. jointly optimized prices at hand, we now turn to counterfactuals examining the role of demand side factors. We find it more illustrative to start the counterfactual analysis in this section from the data rather than from (the later stage of) the estimated model. In the first counterfactual, we modify our original dataset so that mid-size deals ($q\in[20,50)$, equivalent to $I_3$) have a meaningfully higher acceptance rate than small ($q<20$, equivalent to $I_1\cup I_2$) and large ($q\geq 50$, equivalent to $I_4\cup I_5$) ones. In a second counterfactual, we do the opposite. For each counterfactual, we re-estimate the model (i.e., the joint distribution $f(\cdot,\cdot)$), and re-compute both the jointly and individually optimized price schedules. The two right columns of Figure \ref{fig: incentive compatibility demand CF} depict these two counterfactuals. We also show the original case on the left for comparison. We now turn to analyzing these results based on the two major lessons mentioned above.

\textbf{Importance of the shape of $f(\cdot, \cdot)$  beyond simple correlation:} As can be seen from panels (d), (e), and (f) of Figure \ref{fig: incentive compatibility demand CF}, the flexibility of the shape of $f(\cdot,\cdot)$ in our model allows us to capture the nature of the relationship between deal accept/reject outcome and deal size in a comprehensive manner. All of the estimated distributions in Figure \ref{fig: incentive compatibility demand CF} have flexible forms and multiple local peaks. Such flexibility is  not feasible to capture with more restrictive models of $f(\cdot,\cdot)$ or with simple measures such as correlation. See Table \ref{tab: correlation by scenario} for illustration: the correlation between $v_i$ and $\Bar{q}_i$ across $i$ in the original model, left column of Figure \ref{fig: incentive compatibility demand CF}, and right column of Figure \ref{fig: incentive compatibility demand CF}, are -0.26 and -0.41 respectively. Although these correlations are fairly similar to one another, the shapes of the estimated distribution and the optimal contracts are meaningfully different. There are at least two reasons why such simple correlational measures cannot replace a flexible structure for $f(\cdot,\cdot)$.

\begin{table}[H]
    \centering
    \begin{tabular}{c r r r}
         \textbf{Correlation} & \textbf{Left col of Fig \ref{fig: incentive compatibility demand CF}} & \textbf{Mid col of Fig \ref{fig: incentive compatibility demand CF}} & \textbf{Right col of Fig \ref{fig: incentive compatibility demand CF}} \\\toprule
    $s_{it}$ \& $\bar{q}_{it}$ & 
    $-0.06^{*\!*\!*}$ & 
    $0.09^{*\!*\!*}$ & 
    $-0.14^{*\!*\!*}$\\\midrule
    $\hat{v}_{it}$ \& $\bar{q}_{it}$ & $-0.26^{*\!*\!*}$ & $-0.01$\hspace{4mm} & $-0.41^{*\!*\!*}$ \\\bottomrule
    \end{tabular}
    \caption{Correlation between deal success and size,  (as well as value and size) for each of the three scenarios of Figure \ref{fig: incentive compatibility demand CF}}
    \label{tab: correlation by scenario}
\end{table}

First, as can be seen from Figure \ref{fig: incentive compatibility demand CF} the relationship between $v$ and $\Bar{q}$ need not be monotone. And non-monotone relationships that are fundamentally different from one another (such as U-shaped and inverse U-shaped) may look similar when  assessed using a linear model (e.g., correlation).

Second, and more crucially, the importance of one sample in the estimation procedure might differ from its importance with respect to the counterfactual policy analysis of interest. For instance, consider the smaller local peaks on the right end of the distribution $f(\cdot,\cdot)$ in panels (e)  and (f) of Figure \ref{fig: incentive compatibility demand CF}. They  belong to mid- and large-size deals.  Note that these peaks are substantially lower compared to the corresponding peak(s) for small-size deals. This is simply because $f(\cdot,\cdot)$ is a probability distribution and there are far fewer mid- and large-size samples in the data compared to smaller ones. As a result, if we impose a restrictive model of $f(\cdot,\cdot)$ that does not adequately separate the estimation of $v$ across sizes, the weight of these fewer samples will be dwarfed in the estimation procedure by substantially more frequent small-size deals. This, in turn, would bias our estimation of average $v$ for these less frequent deal sizes. Such a bias would be detrimental to our counterfactual analysis. This is because as depicted by Figure \ref{fig: size and rev hist} earlier in the paper, larger deals, although much less frequent, have a meaningful role in shaping the revenue and profitability.

In sum, we find the analysis summarized by Figure \ref{fig: incentive compatibility demand CF} to be in support of our modeling choices on where to be parsimonious (shape of $V_i(\cdot)$) and where to be flexible (shape of $f(\cdot,\cdot)$).

\textbf{Role of incentive compatibility constraints:} The bottom panels of Figure \ref{fig: incentive compatibility demand CF} depict not only the ``jointly optimized'' schedule $P^*$ for each demand scenario, but also the ``individually optimized'' schedule $\Tilde{P}$. There are two broad lessons to learn from this figure. 

The first lesson is that if the deal acceptance rate (and hence the estimated average $v_i$) is substantially heterogeneous across size groups, then the profitability gap $\pi(P^*)-\pi(\tilde{P})$ between the individually and jointly optimized schedules may be wide. This is because under such conditions, the individually optimized marginal prices end up being far from each other, exacerbating the loss from the lack of incentive compatibility. To see this, note that $\frac{\pi(P^*)-\pi(\Tilde{P})}{\pi(P^*)}$ is less than $1\%$ in the original data (i.e., left column of figure), whereas it is 9.4\% for the demand scenario in the middle column and around 9.1\% in the right-most column (the profit and revenue levels are posted on the top left corners of bottom panels in Figure \ref{fig: incentive compatibility demand CF}). Observe that, similar to our previous analyses, these percentages are gross of firm-level fixed costs. If the firm-level fixed cost between 10 and 15 \$M/y, the former profitability gap would range from 31.5\% or more and the latter from  18.0\% to 35.3\%.

The second important lesson goes beyond profitability, and focuses on how the shapes of the two contracts compare to one another.  In general, relative to the individually optimized contract $\Tilde{P}$, the  jointly optimized contract $P^*$ seems to \textit{moderate} the price variation by size. Take the second column of the figure for illustration. Here, $p^*_3$ is substantially smaller than $\Tilde{p}_3$, even though mid-size customers (i.e., those with $\Bar{q}_i\in I_3$) have much larger average valuations $v_i$ relative to other sizes. Also, $P^*$ charges higher prices than $\Tilde{P}$ in adjacent size ranges $I_2$ and $I_4$. Conceptually similar arguments (but in part in the opposite direction) hold for the right column of the figure.

Table \ref{tab: global vs local by size segment} should shed more light on why such a moderating behavior is optimal. In this table, we compare the profits generated by  $P^*$ and $\Tilde{P}$ from different size-groups. In particular, we break down $\pi(P^*)-\pi(\Tilde{P})$ for customers that are small ($\Bar{q}_i<20$), mid-size ($\Bar{q}_i\in[20,50)$), and large ($\Bar{q}_i>50$). 


\renewcommand\theadfont{\bfseries}
\begin{table}[H]
    \centering
    \small
    \begin{tabular}{llc}
         \textbf{Scenario} & \thead{Customer\\Size Segment} & \thead{Profit Difference\\from the Segment}  \\
          \toprule
    & small: $\Bar{q}_i<20$ & $-$\$0.14M \\
\textbf{Mid col of Fig \ref{fig: incentive compatibility demand CF}}  
    & medium: $\Bar{q}_i\in[20,50)$ & +\$1.51M  \\
    & large: $\Bar{q}_i>50$ & $-$\$0.04M \\\hline
    & small: $\Bar{q}_i<20$& +\$0.83M \\
\textbf{Right col of Fig \ref{fig: incentive compatibility demand CF}} 
    & medium: $\Bar{q}_i\in[20,50)$& $-$\$0.36M \\
    & large: $\Bar{q}_i>50$& +\$1.36M \\
 \bottomrule
    \end{tabular}
    \caption{jointly  vs individually optimized prices: performance comparison segment  by segment. A positive sign in the right-most column means the jointly optimized schedule delivers a higher profit from the respective segment relative to the individually optimized schedule.}
    \label{tab: global vs local by size segment}
\end{table}

As the table shows, for the demand system in the middle column of Figure \ref{fig: incentive compatibility demand CF}, $P^*$  delivers \$1.51M/y more profit from medium-size customers, relative to $\Tilde{P}$. This happens because in the vector $P^*$, the element $p^*_3$ is only moderately larger than the other elements, which stands in contrast to the wider gap between $\Tilde{p}_3$ and other elements of $\Tilde{P}$. This prevents many mid-size customers from ``flocking'' to cheaper quantities, thereby helping to boost the profitability from those customers. Also helpful toward this objective is the fact that $p^*_2$ and $p^*_4$ are elevated (relative to their $\Tilde{p}$ counterparts). Such elevation in $p^*_2$ and $p^*_4$ either further prevents mid-size customers from adjusting purchase sizes, or helps make a higher profit off of those mid-size customers who do adjust nonetheless. Of course this means that the tuning of $p^*_2$ and $p^*_4$ is done in part with the ``global'' goal of taming mid-size customers' behavior; which means those prices are ``locally'' sub-optimal. Thus, it should not be surprising that $P^*$ delivers less profit than $\Tilde{P}$ from small and large customers, by the amounts of \$0.14M/y and \$0.04M/y respectively.

Similarly, as Table \ref{tab: global vs local by size segment} shows, $P^*$ makes \$0.36M/y less profit than $\Tilde{P}$ from mid-size customers under the demand system in the right-most column of Figure \ref{fig: incentive compatibility demand CF}. This is because jointly optimized $P^*$ charges a substantially higher per-unit price for mid-size deals than mid-size customers are willing to pay. Though this leads to losses from those customers, it prevents smaller and larger customers from taking advantage of low rates in the mid-range. Alongside this, $P^*$ also charges lower prices outside of mid-size deals to further discourage customers of other sizes from moving. As a result, the profitability from those sizes, which is compromised under $\Tilde{P}$ due to not accounting for incentive compatibility issues (i.e., large and small customers taking advantage of mid-size rates), is partially protected under $P^*$. As the table shows, these improvements are \$0.83M/y and \$1.36M/y respectively for small and large customers.

To recap, compared to the individually optimized schedule, the jointly optimized one seems to charge prices that vary less significantly with size.

\subsection{Analysis of Cost-Side Factors}

In this section, we examine the role of cost-side factors. We start by comparing the role of those factors against that of demand-side factors in shaping the optimal nonlinear contracts. We then study how the optimal contract responds to various cost-side scenarios.

\subsubsection{Cost- v.s. demand-side factors: a comparative analysis}\label{subsub: cost v.s. demand}
The fact that our optimal price schedule involves lower per-unit prices for larger deals seems consistent with both demand-side  cost-side estimates. On the demand side, as Table \ref{tab:mleestimates} suggests, customers with medium and large sizes $\Bar{q}_i$ tend to have lower valuations $v_i$. On the cost-side, as Figure \ref{fig:avgcost} depicts, average cost to sell deals of larger sizes is substantially lower than that for smaller deals. An empirical question is, hence, which of these two factors is a more important reason behind the shape of our optimal price schedule?

\begin{figure}[H]
\begin{center}
    \begin{subfigure}{0.475\textwidth}
        \includegraphics[width=\linewidth]{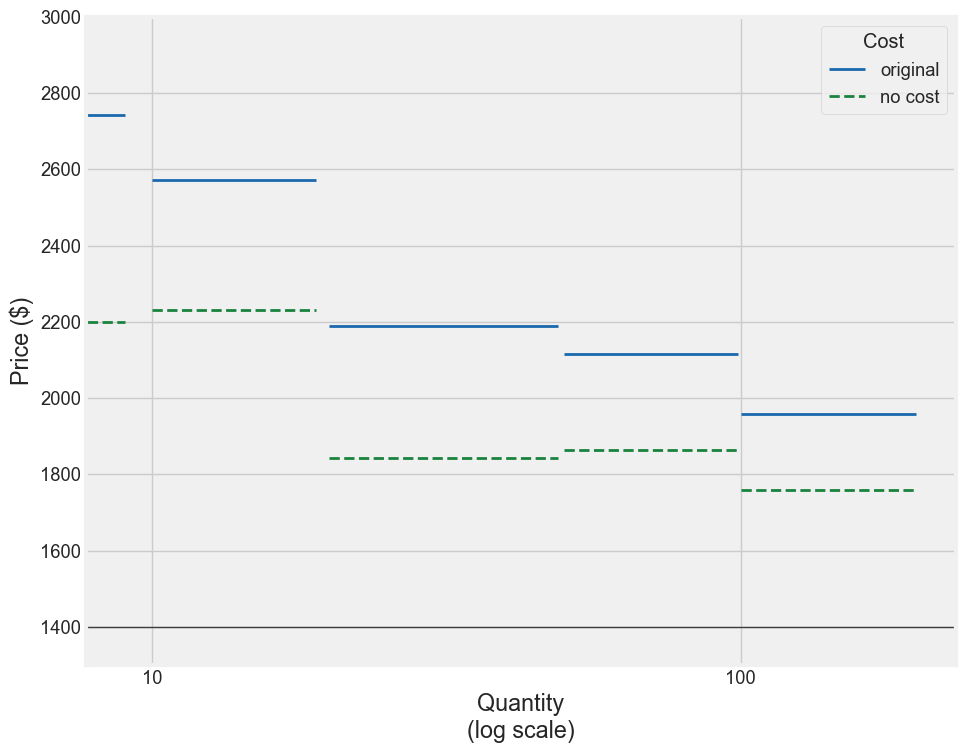}
    \end{subfigure}
    \begin{subfigure}{0.475\textwidth}
        \includegraphics[width=\linewidth]{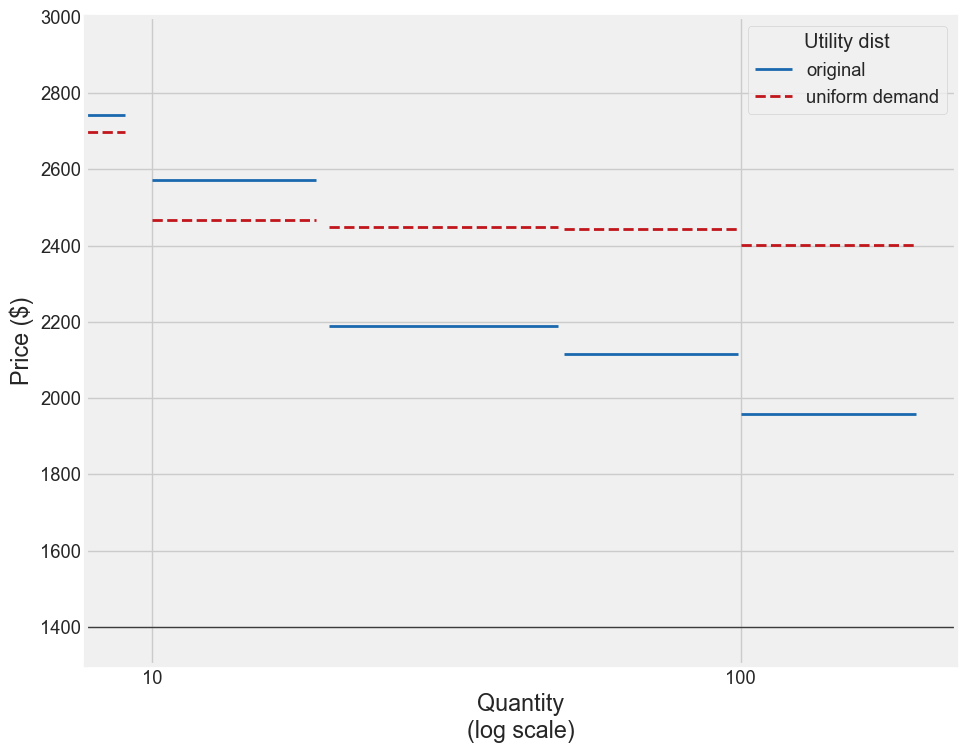}
    \end{subfigure}
\end{center}
    \caption{Counterfactual optimal price schedule under no cost (left panel) and uniform mean values (right panel). The left panel helps isolate the role of demand in shaping the optimal nonlinear contract while the right panel isolates the role of costs. }
    \label{fig: cost vs dmnd}
\end{figure}

Figure \ref{fig: cost vs dmnd} helps answer this question. On the left panel,  we compare the optimal price schedule (solid blue lines) to the optimal schedule under the counterfactual scenario in which the firm faces  no costs (i.e., $c_1=0,c_2=0$). This allows to ``shut off'' the role of costs in causing nonlinearity in $P^*$ and focus only on the role of demand. The right panel considers a counterfactual scenario in which the valuations $v_i$ are homogenized across customers, up to the idiosyncratic error term. This homogenization is formally operationalized by  by using $\Bar{v}_{it}:=\mathbb{E}_{i'}[\beta\times X_{i'}+\alpha_{t}+\gamma_{\Tilde{q}}]+\epsilon_{it}$ in the counterfactual simulation instead of original $v_{it}$.\footnote{Note that this demand homogenization takes place only on $i$ rather than on $it$. Time index $t$ is fixed at 2021, because this is the year our counterfactual analysis is focused on.} In this counterfactual, hence, values $v_i$ and sizes $\Bar{q}_i$ are independently distributed across potential customers. This allows to shut off the role of demand in shaping the nonlinearity of $P^*$ and focus only on cost factors.

As Figure \ref{fig: cost vs dmnd} shows, the downward trend of  per-unit prices in deal size is qualitatively preserved when we shut off either factor. Thus both demand and cost factors seem to have a role in the shape of the optimal contract. That said, on the front of magnitude, demand-side factors seem to have a substantially larger role relative to cost-side ones. To see this, observe that the counterfactual optimal schedule on the left panel of Figure \ref{fig: cost vs dmnd} has a much wider range of prices compared to the one on the right. In fact, the price range across sizes is almost the same between the original optimal and the no-cost counterfactual price schedules on the left panel of the figure. This seems puzzling: if costs do cause diminishing marginal prices (as seen in the right panel of Figure \ref{fig: cost vs dmnd}), then why does the range of marginal prices not shrink in a counterfactual with no costs? For an intuitive understanding of why this is the case, note that costs affect the shape of the optimal schedule via two channels. The first channel is what was  discussed so far: deal sizes with higher average costs tend to have higher per-unit prices in the optimal contract. The second channel has to do directly with the existence of costs rather than heterogeneity in them: when there are marginal costs, the optimal price schedule is in part guided by how to cover those costs; and this leaves the seller with less freedom to shape the price schedule based on demand-side factors. To sum up: when costs are assumed away (green dashed lines in the left panel), not only will their direct effect on prices will be gone, but so will their moderating effect on the role of demand.


We finish this subsection by emphasizing another insight from Figure \ref{fig: cost vs dmnd}. As mentioned earlier in the paper, an important feature of our empirical methodology is a flexible joint distribution $f(\cdot,\cdot)$ between size $\bar{q}_i$ and per-unit value $v_i$ that can allow for different correlation patterns. The right panel of Figure \ref{fig: cost vs dmnd} further demonstrates the importance of this flexibility. the ``uniform demand'' simulation effectively imposes independence between $v_i$ and $\bar{q}_i$. As the figure shows, the imposition of independence leads to a much ``less nonlinear'' price schedule, further highlighting the importance of capturing a flexible co-variation between the two quantities.

\subsubsection{Detailed analysis of cost-Side factors}

\begin{figure}[H]
\centering
\begin{subfigure}{0.475\textwidth}
\includegraphics[width=\linewidth]{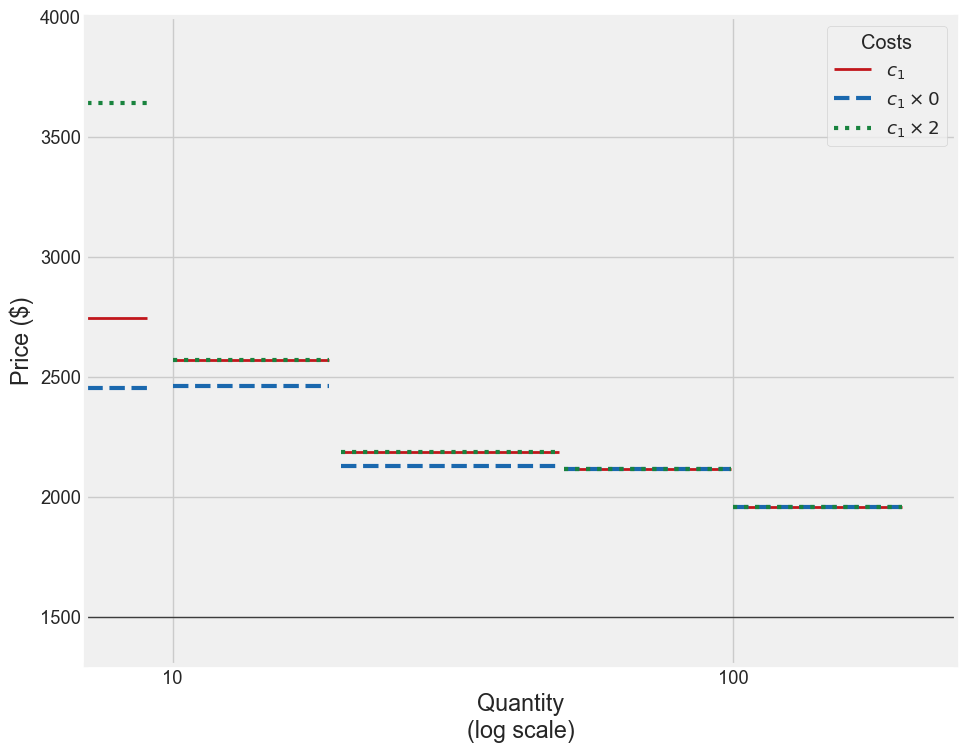}
\caption{Different $c_1$s}
\label{fig:subim1}
\end{subfigure}
\begin{subfigure}{0.475\textwidth}
\centering
\includegraphics[width=\linewidth]{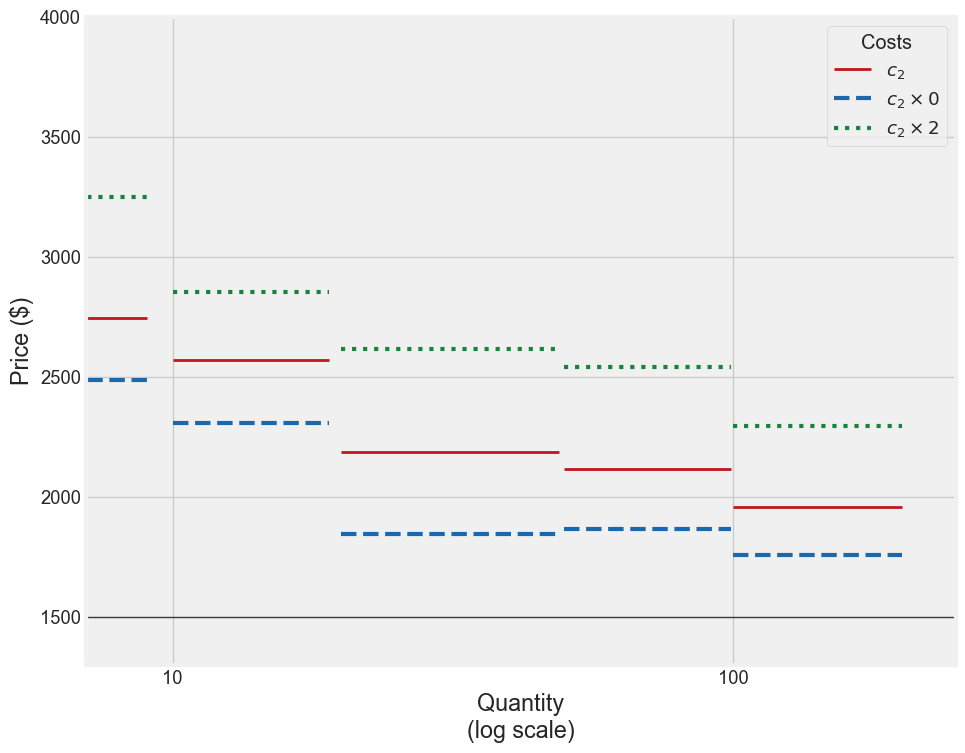}
\caption{Different $c_2$s}
\label{fig:subim2}
\end{subfigure}
\caption{Optimal price Schemes under Different Cost Parameters. Left panel compares default value for $c_1$, doubling $c_1$, and $c_1=0$. Right panel does the same for $c_2$}
\label{fig:pricebycost}
\end{figure}

In this section, we examine the way in which cost parameters $c_1$ and $c_2$ shape the optimal price schedule. Recall that $\hat c_{1,2021}=$\$3,630/customer and $\hat c_{2,2021}=$\$760/workshop. Figure \ref{fig:pricebycost} depicts the optimal contracts as these parameters change. Panel (a) shows the optimal contract for a range of $c_1$ values while panel (b) does the same for $c_2$.

As evident by panel (a) of this figure, changes in $c_1$ mostly impact the marginal price for smaller size deals. This is because the larger the deal size, the smaller the customer-level fixed cost will be as a fraction of the total cost associated with the customer. Observe that at $c_1=0$, the optimal contract qualitatively preserves its shape. This, as discussed before, indicates that demand side factors (i.e., the shape of the joint distribution $f(\cdot,\cdot)$) have a critical role in shaping the optimal price schedule. Finally, it is worth noting that the observed impact of $c_1$ on optimal pricing is seemingly at odds with the traditional wisdom that fixed costs should have no bearing on pricing. This is because $c_1$ is the customer-level fixed cost and, unlike firm-level fixed costs,  is not a ``sunk''. This notion of cost is not well-defined in the discrete choice settings where each customer consumes zero or one unit.\footnote{In discrete choice, $c_1$ would always get incurred with one unit of  $c_2$. Hence, one could think of $c_1+c_2$ as the variable cost.} But in continuous-choice environments, as we show here, having the right estimate of $c_1$ is critical for designing the right pricing strategy.

The effect of $c_2$ on the optimal pricing schedule is shown in panel (b) of the figure. Price increases across the board as the per-unit cost increases, and the pass-through is more or less uniform and around $\frac{1}{3}$. Observe that the direction in which $c_2$ affects the shape of the optimal schedule is in contrast to that of $c_1$. The larger the marginal cost $c_2$, the closer the optimal schedule is to a linear contract (the opposite was the case for $c_1$). Per our discussion in subsection \ref{subsub: cost v.s. demand}, this happens because larger values of $c_2$ restrain the ability of the seller to shape the price schedule based on demand.

\subsection*{Additional counterfactual analyses on 3rd-degree price discrimination}

Appendix \ref{apx: 3rd degree} carries out  a  number of counterfactual analyses in which we study the impact of 3rd-degree price discrimination as well as a combination of 2nd and 3rd degree discrimination strategies.

\section{Discussion}\label{sec: discussion}

This section provides discussions on broad applications of our framework, as well as caveats and future research. It also lists additional analyses and robustness checks which are relegated to the appendix.

\subsection{General Applicability}

Here, we briefly discuss the applicability of our framework both on the front of the data and on the front of the modeling and assumptions. 

\begin{figure}[H]
    \begin{center}
        \includegraphics[width=0.5\linewidth]{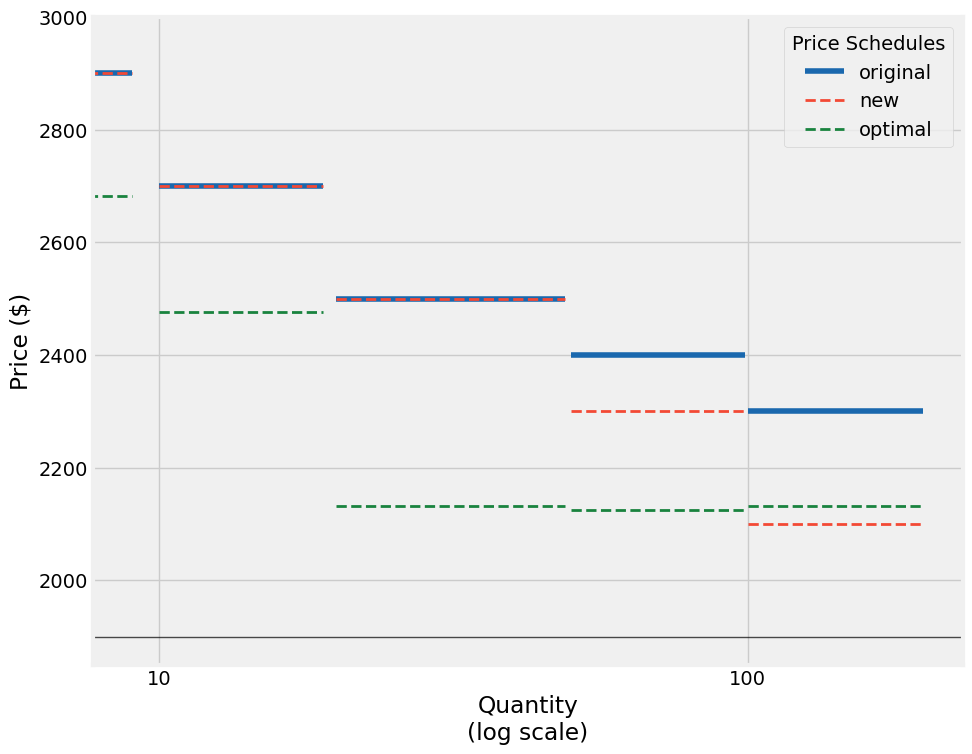}
    \end{center}
   \footnotesize{\textit{Note:} the optimal (recommended) schedule in this figure is slightly different from what we present in the rest of the paper. The reason is that the recommendation was made to the firm at an earlier stage of the research, and the marginal prices in the figure depict our default specification at that time.}   
    \caption{Comparison among: Lifelabs' original pricing, optimal schedule $P^*$, and Lifelabs' adjusted prices based on the recommendation of $P^*$.}
    \label{fig: lifelab's reaction}
\end{figure}

On the data front, empirical framework is applicable in any context in which either pricing experiments are feasible, or “intended size of use” for failed deals is directly recorded or can be estimated/proxied. Even though to the best of our knowledge the use of a dataset with this feature is novel in scholarly work on pricing, such data is increasingly collected and maintained by firms. Other examples of where would-be sizes of failed deals are recorded or can be proxied are: Cloud Computing, B2B pay services such as Amazon Pay or PayPal, or Consulting Services.\footnote{For instance, many consulting companies use Salesforce software to maintain a dataset of their potential deals. The software helps record the intended size of a contract along different stages of the negotiation. The software also keeps record of ``pursuits'' that eventually did not lead to a deal.} In the specific case of this study, our results did lead to a change in LifeLabs' pricing strategy that is qualitatively in line with our recommendation, though not fully coinciding with it. More specifically, they reduced their prices as our study recommends; but did so only for larger deals.\footnote{Firms adjusting their pricing strategies in the direction of (but not exactly coinciding with) recommendations from academic studies is not unprecedented. An instance is \cite{dube2017scalable} and the impact of their study on Ziprecruiter's pricing.} See Figure \ref{fig: lifelab's reaction}. 

On the front of the model structure, the key assumption that may constrain the applications of the model is the functional form of the value function $V(\cdot)$, in particular, the assumptions that (i) the marginal utility is constant until consumption hits $\bar{q}$ and that (ii) it fully stagnates at $\bar{q}$. It was because of these assumptions that we were able to prove each customer $i$ would purchase either zero or $\bar{q}_i$ units. These assumptions imply that the best applications of our frameworks are those in which there is a strong notion of ``size of need'' imposed on the customer relatively independently of the price.\footnote{Another, more strict way of modeling such preferences would be to set $V(q):=\bar{q}\times v\times \textbf{1}_{q\geq \bar{q}}$. These preferences would yield the same outcomes as our main specification under any concave fee schedule $P(\cdot)$. They would, however, yield different results if, for instance, there is a small-business discount implemented in the form of lower fees for smaller number of workshops. Another example of where the results between the two models are different due to non-concavity is the analysis of incentives in section \ref{subsec: demand and incentives again}.} Once again, B2B applications are more likely to satisfy these conditions relative to such B2C applications as food, beverage, cellular services, etc (see \cite{kim2002modeling,iyengar2008conjoint,hortaccsu2018empirical} for instances of these markets). That said, even in those markets, the joint distribution between a notion of ``size'' and willingness to pay still matters. And frameworks that capture smoothness but ignore this distribution also have a shortcoming. As such, the full solution to this issue would be a model that captures both a smooth value function and a joint distribution between size and WTP. We provide further discussion of how to capture smoothness in $V(\cdot)$ in the following subsection. Also, in appendix \ref{appendix: smoothness}, we carry out additional sensitivity analysis to smooth $V(\cdot)$ in our specific context, and informally discuss  how to generalize the estimation procedure to the case of smooth $V(\cdot)$. We leave a full treatment of this problem, however, to future research.

\subsection{Limitations and Future Research}

We discuss some limitations of our research here. Of course this list is non-exhaustive.

\subsubsection{Piece-wise linear value function $V_i(\cdot)$}

Our formulation of gross value is $V_i(q)\equiv v_i\times \min(q,\Bar{q}_i)$. Though this captures the two key dimensions of heterogeneity that we are focused on in this paper (i.e., size and value,) it does abstract away from smoothly diminishing rate of return. That is, the marginal value for another unit of the product is  constant at $v_i$ until it hits the stagnation point and reduces to zero. A more general formulation would allow for smoothly diminishing returns to scale on the part of the buyer. One way to do so would be to assume $V_i(q)\equiv u(v_i,\min(q,\Bar{q}_i))$ where $u:\mathbb{R}^2\rightarrow\mathbb{R}$ is a function increasing in its both arguments and concave in the second one. An illustration of this structure is: $V_i(q)\equiv v_i\times (\min(q,\Bar{q}_i))^\alpha$ for some $\alpha\in(0,1)$. We do examine a form of such an extension in appendix \ref{appendix: smoothness}, and show that our results do not change drastically as a result of this extension. That said, in our extension in that appendix, we take the estimated distribution $f(\cdot,\cdot)$ over $\bar{q}_i$ and $v_i$ from our main analysis, and then add another parameter to the value functions in order to capture the degree of smoothness. 

A natural next step for future research would be to incorporate this smoothness parameter in the estimation stage rather than add it after the initial, non-smooth, model has been estimated. It can be expected  that in the smooth model, the purchase size need not coincide with $\bar{q}_i$ even under a concave contract. Similarly, it is less reasonable to assume that the recorded would-be size for unsuccessful deals corresponds to $\bar{q}_i$. As a result a new estimation process  would be necessary in order to incorporate smoothness. Such an estimation procedure should have a key feature: it needs to find a way to interpret the observed size for unsuccessful deals. In our current model, we assume that this observed would-be size is the satiation point $\bar{q}_i$ of the value function $V(\cdot)$. But if the value function no longer has a satiation point, our interpretation needs to adjust.\footnote{One possibility would be to assume this quantity represents the number of units that the buyer would expect to buy conditional on its uncertainty being resolved in a way that would make non-zero purchase optimal.} We leave this to future research but provide an informal discussion in appendix \ref{appendix: smoothness}.


\subsubsection{Exogeneity assumption on the set of potential customers}

Second, we assume that the set of deal-talks is exogenous. Put in the terminology of our timeline in Figure \ref{fig: timeline}, we assume that participating in a conversation with the seller, which leads to a record of $\bar{q}_i$, is costless and is, hence, done by any firm that could benefit from the product (i.e., any firm with $\bar{q}_i>0$). Assumptions similar to this have previously been made in the discrete choice literature (see \cite{cohen2016using} and \cite{dube2017scalable} for instance.) Nevertheless, this assumption is not ideal, neither in the discrete-choice setting nor in our setting. The reason is that the potential customers that reach out to the seller  are selected based on the likelihood to purchase. This means a change in price can change not only what deals succeed, but also who joins the pool of potential customers in the first place. This can be modeled by assuming that communicating with the seller in stage 2 of the timeline in Figure \ref{fig: timeline} is costly to any potential customer $i$, and that the any customer $i$ who expects to have low $v_i$ may avoid approaching the seller in order to save on this cost. The estimation of such an extension to the model would require additional assumptions. We skip this extension in this paper.

In spite of this issue, we believe our results are reasonable. This is because   potential customers who did not initiate a conversation are likely those with lower $v$ values. As a result, missing them from the analysis would \textit{overestimate} the overall optimal prices. Given that our current recommendation to the company is to lower their prices, our recommendation would only be strengthened if we were able to observe potential customers who did not start a conversation about a deal.

\subsubsection{The assumption that all potential customers would purchase at $p=0$}

Our analysis assumes that potential customers who engaged with LifeLabs would have been willing to purchase if there offered price was zero. We believe this is an acceptable assumption in our application, given that during 2020 and 2021, the only source of customer acquisition for LifeLabs was word of mouth. This implies that potential customers recorded in our data are those with an interest in the product because they were the ones who reached out to LifeLabs.

However, applying our framework to contexts with heavy marketing on the firm's side would require caution. To illustrate, if the deal acceptance rate is lower for larger deals only because the firm more heavily markets to larger users, this may be a source of endogeneity that would need to be addressed before our framework can be applied.

\subsubsection{Two-point identification}

Key in the identification of the price effects in the model is the variation between the observed demand under observed prices on the one hand and the assumed demand (i.e., total number of potential deals) under a price of zero. Although the within-schedule variation in the per-unit prices also plays a role in identification--see section \ref{sec: estimation} and appendix \ref{appendix: other specs} for the details--the key role played by the demand at zero price has important implications. Our use of this variation for making predictions on how demand would react to ``more localized'' changes in price levels would mean our assumed curvature of the demand curve (i.e., functional form) may be playing a role in shaping our results. Shortly, we will discuss a sensitivity analysis to one aspect of our functional form assumptions. Nevertheless, this issue of functional forms remains a limitation, even if we assume that the demand at $p=0$ is correctly measured by the sum of successful and unsuccessful deals.

\subsubsection{Functional form assumptions on the error term $\epsilon_{it}$}

We assume that the error term $\epsilon_{it}$, and hence value $v_{it}$ is a logistic random variable. This is limiting in at least two ways. First, it is a limitation by the virtue of being a functional form assumption. Second, this assumption implies that a strictly positive fraction of the potential customers may not purchase at $p=0$ of zero under the estimated parameters.

To deal with these issues, we estimate the model with normal (as opposed to logistic) error terms. The estimation results and the optimal price schedule remain similar to what our logistic model finds. See appendix \ref{apx: functional form} for more details. Also, the fraction not buying at $p=0$ remains minimal in both models, $0.133\%$ for under logistic error terms and $0.044\%$ under normal errors.

\subsubsection{Monopoly}

As long as competitors are not expected to respond to the changes in the firm's pricing, our framework is fully compatible with the presence of competition in the market. Competition would only imply that $v_i$ should be interpreted as the value relative to the next best option which can in principle be a competitor's product. However, if competitors are expected to respond to the firm's pricing, our framework falls short. This shortcoming can only be resolved if the framework is amended with an empirical strategy that helps to quantify how the overall price levels and shapes of nonlinear contracts offered by competitors  would respond to a given firm's pricing. This is beyond the scope of our paper.

\subsection{Additional analyses and robustness checks}

In addition to the appendices that have been mentioned throughout, we have carried out other supplemental analyses which are also presented in the appendix section of the paper. Appendix \ref{apx: predicting sizes} examines whether it would be possible to avoid using data on intended sizes of failed deals and, instead, predict those sizes by (i) estimating a regression  model of size $\bar{q}_{it}$ on observables \textit{using the successful deals data only}, and then (ii) using that model to project predicted $\bar{q}_{it}$ values for unsuccessful deals. Our analysis in that appendix shows that this alternative approach would fail to adequately capture the relationship between size and value, thereby delivering a less flexible joint distribution than our main approach.

Appendix \ref{apx: discontinuity} examines whether there is adequately strong inference can be made about willingnesses to pay by leveraging the observed discontinuities in the price schedule (as depicted in figure \ref{fig:pricesched}). If feasible, this would be important given that such an approach would no longer require a role for the zero price $p=0$ in creating the price variation. Our analysis in that appendix, however, shows that the data is not adequately powerful for making inference based solely on the discontinuity. We expect such power concerns to also hold in other B2B applications.


\section{Conclusion}\label{sec: conclusion}

This paper empirically analyzed optimal nonlinear pricing with an emphasis on capturing key insights from the mechanism design literature, and with a focus on B2B applications. We proposed and estimated a model of continuous-choice demand capturing a joint distribution of customer size and customer per-unit willingness to pay. We flexibly estimated this joint distribution by leveraging a novel dataset that records information (including prospective deal size) not only for successful deals but also for unsuccessful ones. We then used the estimated model to solve for the optimal nonlinear tariff.

We find that optimal nonlinear pricing improves upon the profitability of optimal linear pricing by at least 8.2\%. Nevertheless, this second-degree price discrimination method recovers only about 7.1\% of the profitability gap between linear pricing and first-degree price discrimination. We also find that second degree price discrimination improves consumer welfare by about 10.2\% and social welfare by about 8.9\%. 

We conducted further simulation analyses in order to generate general insights above and beyond our specific application. Among other analyses, we (i)  quantified the magnitude of the profit impact of incentive compatibility constraints, (ii) examined the role of cost side v.s. demand side factors in shaping the optimal contract, and (iii) studies the effects of using a fixed fee on the shape and profitability of the optimal contract.

We believe our analysis can be broadly applicable by firms that  seek to design a nonlinear pricing tariff, especially in B2B contexts such as cloud computing, pay systems, and SaaS. The usefulness of our framework for such applications stems from the fact that its empirical strategy design is cognizant of both the shortcomings and the advantages of these settings when it comes to data availability. Our method relies on detailed data on intended features (especially sizes) of unsuccessful transactions, which is increasingly available in B2B cases, and combines this data with identifying assumptions in order to compensate for the fact that direct price variation in many large B2B applications is commonly limited.


One major direction in which our analysis may be extended would be adding smoothness to the value functions by customers. Another potential future direction would be to extend the analysis from a monopolistic setting to an oligopolistic one.

\section*{Declaration}
This research was financially supported by the Yale Center for Customer Insights.

\bibliographystyle{chicago}
\bibliography{references}

\begin{thebibliography}{}

\bibitem[\protect\citeauthoryear{Anderson and Dana~Jr}{Anderson and Dana~Jr}{2009}]{anderson2009price}
Anderson, E.~T. and J.~D. Dana~Jr (2009).
\newblock When is price discrimination profitable?
\newblock {\em Management Science\/}~{\em 55\/}(6), 980--989.

\bibitem[\protect\citeauthoryear{Armstrong}{Armstrong}{1996}]{armstrong1996multiproduct}
Armstrong, M. (1996).
\newblock Multiproduct nonlinear pricing.
\newblock {\em Econometrica: Journal of the Econometric Society\/}, 51--75.

\bibitem[\protect\citeauthoryear{Aryal and Gabrielli}{Aryal and Gabrielli}{2020}]{aryal2020empirical}
Aryal, G. and M.~F. Gabrielli (2020).
\newblock An empirical analysis of competitive nonlinear pricing.
\newblock {\em International Journal of Industrial Organization\/}~{\em 68}, 102538.

\bibitem[\protect\citeauthoryear{Bodoh-Creed, Hickman, List, Muir, and Sun}{Bodoh-Creed et~al.}{2023}]{bodoh2023stress}
Bodoh-Creed, A.~L., B.~R. Hickman, J.~A. List, I.~Muir, and G.~K. Sun (2023).
\newblock Stress testing structural models of unobserved heterogeneity: Robust inference on optimal nonlinear pricing.
\newblock Technical report, National Bureau of Economic Research.

\bibitem[\protect\citeauthoryear{Carroll}{Carroll}{2017}]{carroll2017robustness}
Carroll, G. (2017).
\newblock Robustness and separation in multidimensional screening.
\newblock {\em Econometrica\/}~{\em 85\/}(2), 453--488.

\bibitem[\protect\citeauthoryear{Chan, Kadiyali, and Xiao}{Chan et~al.}{2009}]{chan2009structural}
Chan, T., V.~Kadiyali, and P.~Xiao (2009).
\newblock Structural models of pricing.
\newblock In {\em Handbook of pricing research in marketing}, pp.\  108--131. Edward Elgar Publishing.

\bibitem[\protect\citeauthoryear{Chan}{Chan}{2006}]{chan2006estimating}
Chan, T.~Y. (2006).
\newblock Estimating a continuous hedonic-choice model with an application to demand for soft drinks.
\newblock {\em The Rand journal of economics\/}~{\em 37\/}(2), 466--482.

\bibitem[\protect\citeauthoryear{Cohen, Hahn, Hall, Levitt, and Metcalfe}{Cohen et~al.}{2016}]{cohen2016using}
Cohen, P., R.~Hahn, J.~Hall, S.~Levitt, and R.~Metcalfe (2016).
\newblock Using big data to estimate consumer surplus: The case of uber.
\newblock Technical report, National Bureau of Economic Research.

\bibitem[\protect\citeauthoryear{Derdenger and Kumar}{Derdenger and Kumar}{2013}]{derdenger2013dynamic}
Derdenger, T. and V.~Kumar (2013).
\newblock The dynamic effects of bundling as a product strategy.
\newblock {\em Marketing Science\/}~{\em 32\/}(6), 827--859.

\bibitem[\protect\citeauthoryear{Devanur, Haghpanah, and Psomas}{Devanur et~al.}{2020}]{devanur2020optimal}
Devanur, N.~R., N.~Haghpanah, and A.~Psomas (2020).
\newblock Optimal multi-unit mechanisms with private demands.
\newblock {\em Games and Economic Behavior\/}~{\em 121}, 482--505.

\bibitem[\protect\citeauthoryear{Draganska and Jain}{Draganska and Jain}{2006}]{draganska2006consumer}
Draganska, M. and D.~C. Jain (2006).
\newblock Consumer preferences and product-line pricing strategies: An empirical analysis.
\newblock {\em Marketing science\/}~{\em 25\/}(2), 164--174.

\bibitem[\protect\citeauthoryear{Dub{\'e}}{Dub{\'e}}{2004}]{dube2004multiple}
Dub{\'e}, J.-P. (2004).
\newblock Multiple discreteness and product differentiation: Demand for carbonated soft drinks.
\newblock {\em Marketing Science\/}~{\em 23\/}(1), 66--81.

\bibitem[\protect\citeauthoryear{Dub{\'e} and Misra}{Dub{\'e} and Misra}{2017}]{dube2017scalable}
Dub{\'e}, J.-P. and S.~Misra (2017).
\newblock Scalable price targeting.
\newblock Technical report, National Bureau of Economic Research.

\bibitem[\protect\citeauthoryear{Frazier}{Frazier}{2018a}]{frazier2018bayesian}
Frazier, P.~I. (2018a).
\newblock Bayesian optimization.
\newblock In {\em Recent advances in optimization and modeling of contemporary problems}, pp.\  255--278. Informs.

\bibitem[\protect\citeauthoryear{Frazier}{Frazier}{2018b}]{frazier2018tutorial}
Frazier, P.~I. (2018b).
\newblock A tutorial on bayesian optimization.
\newblock {\em arXiv preprint arXiv:1807.02811\/}.

\bibitem[\protect\citeauthoryear{Ghili}{Ghili}{2022}]{ghili2022network}
Ghili, S. (2022).
\newblock Network formation and bargaining in vertical markets: The case of narrow networks in health insurance.
\newblock {\em Marketing Science\/}~{\em 41\/}(3), 501--527.

\bibitem[\protect\citeauthoryear{Ghili}{Ghili}{2023}]{ghili2022characterization}
Ghili, S. (2023).
\newblock A characterization for optimal bundling of products with nonadditive values.
\newblock {\em American Economic Review: Insights\/}~{\em 5\/}(3), 311--326.

\bibitem[\protect\citeauthoryear{Ghili and Schmitt}{Ghili and Schmitt}{2023}]{ghili2018risk}
Ghili, S. and M.~Schmitt (2023).
\newblock Risk aversion and double marginalization.
\newblock {\em Available at SSRN 3237732\/}.

\bibitem[\protect\citeauthoryear{Goldberg}{Goldberg}{2021}]{goldberg2021designing}
Goldberg, S. (2021).
\newblock Designing monitoring programs.

\bibitem[\protect\citeauthoryear{Gowrisankaran, Nevo, and Town}{Gowrisankaran et~al.}{2015}]{gowrisankaran2015mergers}
Gowrisankaran, G., A.~Nevo, and R.~Town (2015).
\newblock Mergers when prices are negotiated: Evidence from the hospital industry.
\newblock {\em American Economic Review\/}~{\em 105\/}(1), 172--203.

\bibitem[\protect\citeauthoryear{Haghpanah and Hartline}{Haghpanah and Hartline}{2021}]{haghpanah2019pure}
Haghpanah, N. and J.~Hartline (2021).
\newblock When is pure bundling optimal?
\newblock {\em The Review of Economic Studies\/}~{\em 88\/}(3), 1127--1156.

\bibitem[\protect\citeauthoryear{Hendel}{Hendel}{1999}]{hendel1999estimating}
Hendel, I. (1999).
\newblock Estimating multiple-discrete choice models: An application to computerization returns.
\newblock {\em The Review of Economic Studies\/}~{\em 66\/}(2), 423--446.

\bibitem[\protect\citeauthoryear{Hendel and Nevo}{Hendel and Nevo}{2013}]{hendel2013intertemporal}
Hendel, I. and A.~Nevo (2013).
\newblock Intertemporal price discrimination in storable goods markets.
\newblock {\em American Economic Review\/}~{\em 103\/}(7), 2722--2751.

\bibitem[\protect\citeauthoryear{Horta{\c{c}}su and McAdams}{Horta{\c{c}}su and McAdams}{2018}]{hortaccsu2018empirical}
Horta{\c{c}}su, A. and D.~McAdams (2018).
\newblock Empirical work on auctions of multiple objects.
\newblock {\em Journal of Economic Literature\/}~{\em 56\/}(1), 157--184.

\bibitem[\protect\citeauthoryear{Howell, Lee, and Allenby}{Howell et~al.}{2016}]{howell2016price}
Howell, J.~R., S.~Lee, and G.~M. Allenby (2016).
\newblock Price promotions in choice models.
\newblock {\em Marketing Science\/}~{\em 35\/}(2), 319--334.

\bibitem[\protect\citeauthoryear{Iyengar and Gupta}{Iyengar and Gupta}{2009}]{iyengar2009nonlinear}
Iyengar, R. and S.~Gupta (2009).
\newblock Nonlinear pricing.
\newblock In {\em Handbook of pricing research in marketing}, pp.\  355--383. Edward Elgar Publishing.

\bibitem[\protect\citeauthoryear{Iyengar and Jedidi}{Iyengar and Jedidi}{2012}]{iyengar2012conjoint}
Iyengar, R. and K.~Jedidi (2012).
\newblock A conjoint model of quantity discounts.
\newblock {\em Marketing Science\/}~{\em 31\/}(2), 334--350.

\bibitem[\protect\citeauthoryear{Iyengar, Jedidi, and Kohli}{Iyengar et~al.}{2008}]{iyengar2008conjoint}
Iyengar, R., K.~Jedidi, and R.~Kohli (2008).
\newblock A conjoint approach to multipart pricing.
\newblock {\em Journal of Marketing Research\/}~{\em 45\/}(2), 195--210.

\bibitem[\protect\citeauthoryear{Iyer and Villas-Boas}{Iyer and Villas-Boas}{2003}]{iyer2003bargaining}
Iyer, G. and J.~M. Villas-Boas (2003).
\newblock A bargaining theory of distribution channels.
\newblock {\em Journal of marketing research\/}~{\em 40\/}(1), 80--100.

\bibitem[\protect\citeauthoryear{Jeuland and Shugan}{Jeuland and Shugan}{1983}]{jeuland1983managing}
Jeuland, A.~P. and S.~M. Shugan (1983).
\newblock Managing channel profits.
\newblock {\em Marketing science\/}~{\em 2\/}(3), 239--272.

\bibitem[\protect\citeauthoryear{Kadiyali, Vilcassim, and Chintagunta}{Kadiyali et~al.}{1996}]{kadiyali1996empirical}
Kadiyali, V., N.~J. Vilcassim, and P.~K. Chintagunta (1996).
\newblock Empirical analysis of competitive product line pricing decisions: lead, follow, or move together?
\newblock {\em Journal of Business\/}, 459--487.

\bibitem[\protect\citeauthoryear{Kim, Allenby, and Rossi}{Kim et~al.}{2002}]{kim2002modeling}
Kim, J., G.~M. Allenby, and P.~E. Rossi (2002).
\newblock Modeling consumer demand for variety.
\newblock {\em Marketing Science\/}~{\em 21\/}(3), 229--250.

\bibitem[\protect\citeauthoryear{Laffont, Maskin, and Rochet}{Laffont et~al.}{1987}]{laffont1987optimal}
Laffont, J.-J., E.~Maskin, and J.-C. Rochet (1987).
\newblock Optimal nonlinear pricing with two-dimensional characteristics.
\newblock {\em Information, Incentives and Economic Mechanisms\/}, 256--266.

\bibitem[\protect\citeauthoryear{Lambrecht, Seim, Vilcassim, Cheema, Chen, Crawford, Hosanagar, Iyengar, Koenigsberg, Lee, et~al.}{Lambrecht et~al.}{2012}]{lambrecht2012price}
Lambrecht, A., K.~Seim, N.~Vilcassim, A.~Cheema, Y.~Chen, G.~S. Crawford, K.~Hosanagar, R.~Iyengar, O.~Koenigsberg, R.~Lee, et~al. (2012).
\newblock Price discrimination in service industries.
\newblock {\em Marketing Letters\/}~{\em 23}, 423--438.

\bibitem[\protect\citeauthoryear{Leslie}{Leslie}{2004}]{leslie2004price}
Leslie, P. (2004).
\newblock Price discrimination in broadway theater.
\newblock {\em RAND Journal of Economics\/}, 520--541.

\bibitem[\protect\citeauthoryear{Liu, Brazell, and Allenby}{Liu et~al.}{2022}]{liu2022non}
Liu, Y.~M., J.~D. Brazell, and G.~M. Allenby (2022).
\newblock Non-linear pricing effects in conjoint analysis.
\newblock {\em Quantitative Marketing and Economics\/}~{\em 20\/}(4), 397--430.

\bibitem[\protect\citeauthoryear{Luo, Perrigne, and Vuong}{Luo et~al.}{2018}]{luo2018structural}
Luo, Y., I.~Perrigne, and Q.~Vuong (2018).
\newblock Structural analysis of nonlinear pricing.
\newblock {\em Journal of Political Economy\/}~{\em 126\/}(6), 2523--2568.

\bibitem[\protect\citeauthoryear{Maskin and Riley}{Maskin and Riley}{1984}]{maskin1984monopoly}
Maskin, E. and J.~Riley (1984).
\newblock Monopoly with incomplete information.
\newblock {\em The RAND Journal of Economics\/}~{\em 15\/}(2), 171--196.

\bibitem[\protect\citeauthoryear{McManus}{McManus}{2007}]{mcmanus2007nonlinear}
McManus, B. (2007).
\newblock Nonlinear pricing in an oligopoly market: The case of specialty coffee.
\newblock {\em The RAND Journal of Economics\/}~{\em 38\/}(2), 512--532.

\bibitem[\protect\citeauthoryear{Mussa and Rosen}{Mussa and Rosen}{1978}]{mussa1978monopoly}
Mussa, M. and S.~Rosen (1978).
\newblock Monopoly and product quality.
\newblock {\em Journal of Economic theory\/}~{\em 18\/}(2), 301--317.

\bibitem[\protect\citeauthoryear{Narayanan, Chintagunta, and Miravete}{Narayanan et~al.}{2007}]{narayanan2007role}
Narayanan, S., P.~K. Chintagunta, and E.~J. Miravete (2007).
\newblock The role of self selection, usage uncertainty and learning in the demand for local telephone service.
\newblock {\em Quantitative Marketing and economics\/}~{\em 5}, 1--34.

\bibitem[\protect\citeauthoryear{Nelder and Mead}{Nelder and Mead}{1965}]{nelder1965simplex}
Nelder, J.~A. and R.~Mead (1965).
\newblock A simplex method for function minimization.
\newblock {\em The computer journal\/}~{\em 7\/}(4), 308--313.

\bibitem[\protect\citeauthoryear{Nevo, Turner, and Williams}{Nevo et~al.}{2016}]{nevo2016usage}
Nevo, A., J.~L. Turner, and J.~W. Williams (2016).
\newblock Usage-based pricing and demand for residential broadband.
\newblock {\em Econometrica\/}~{\em 84\/}(2), 411--443.

\bibitem[\protect\citeauthoryear{Oi}{Oi}{1971}]{oi1971}
Oi, W.~Y. (1971).
\newblock A {Disneyland} dilemma: two-part tariffs for a {Mickey Mouse} monopoly.
\newblock {\em Quarterly Journal of Economics\/}~{\em 85\/}(1), 77--96.

\bibitem[\protect\citeauthoryear{Reiss and White}{Reiss and White}{2005}]{reiss2005household}
Reiss, P.~C. and M.~W. White (2005).
\newblock Household electricity demand, revisited.
\newblock {\em The Review of Economic Studies\/}~{\em 72\/}(3), 853--883.

\bibitem[\protect\citeauthoryear{Rey and Tirole}{Rey and Tirole}{1986}]{rey1986logic}
Rey, P. and J.~Tirole (1986).
\newblock The logic of vertical restraints.
\newblock {\em The American economic review\/}, 921--939.

\bibitem[\protect\citeauthoryear{Rey and Verg{\'e}}{Rey and Verg{\'e}}{2008}]{rey2008economics}
Rey, P. and T.~Verg{\'e} (2008).
\newblock Economics of vertical restraints.
\newblock {\em Handbook of antitrust economics\/}~{\em 353}, 390.

\bibitem[\protect\citeauthoryear{Rochet and Stole}{Rochet and Stole}{2002}]{rochet2002nonlinear}
Rochet, J.-C. and L.~A. Stole (2002).
\newblock Nonlinear pricing with random participation.
\newblock {\em The Review of Economic Studies\/}~{\em 69\/}(1), 277--311.

\bibitem[\protect\citeauthoryear{Rochet and Stole}{Rochet and Stole}{2003}]{rochet2003economics}
Rochet, J.-C. and L.~A. Stole (2003).
\newblock The economics of multidimensional screening.
\newblock {\em Econometric Society Monographs\/}~{\em 35}, 150--197.

\bibitem[\protect\citeauthoryear{Schmalensee}{Schmalensee}{1981}]{schmalensee1981output}
Schmalensee, R. (1981).
\newblock Output and welfare implications of monopolistic third-degree price discrimination.
\newblock {\em The American Economic Review\/}~{\em 71\/}(1), 242--247.

\bibitem[\protect\citeauthoryear{Song and Chintagunta}{Song and Chintagunta}{2007}]{song2007discrete}
Song, I. and P.~K. Chintagunta (2007).
\newblock A discrete--continuous model for multicategory purchase behavior of households.
\newblock {\em Journal of marketing Research\/}~{\em 44\/}(4), 595--612.

\bibitem[\protect\citeauthoryear{Varian}{Varian}{1985}]{varian1985price}
Varian, H.~R. (1985).
\newblock Price discrimination and social welfare.
\newblock {\em The American Economic Review\/}~{\em 75\/}(4), 870--875.

\bibitem[\protect\citeauthoryear{Verboven}{Verboven}{2002}]{verboven2002quality}
Verboven, F. (2002).
\newblock Quality-based price discrimination and tax incidence: evidence from gasoline and diesel cars.
\newblock {\em RAND Journal of Economics\/}, 275--297.

\bibitem[\protect\citeauthoryear{Wilson}{Wilson}{1993}]{wilson1993nonlinear}
Wilson, R.~B. (1993).
\newblock {\em Nonlinear pricing}.
\newblock Oxford University Press on Demand.

\bibitem[\protect\citeauthoryear{Yang}{Yang}{2021}]{yang2021costly}
Yang, F. (2021).
\newblock Costly multidimensional screening.
\newblock {\em arXiv preprint arXiv:2109.00487\/}.

\bibitem[\protect\citeauthoryear{Yin, Zhang, Zhou, Yuan, Zhao, Li, and Liu}{Yin et~al.}{2020}]{yin2020kaml}
Yin, L., H.~Zhang, X.~Zhou, X.~Yuan, S.~Zhao, X.~Li, and X.~Liu (2020).
\newblock Kaml: improving genomic prediction accuracy of complex traits using machine learning determined parameters.
\newblock {\em Genome biology\/}~{\em 21\/}(1), 1--22.

\end{thebibliography}

\newpage
\appendix
\section*{Appendix}

This appendix provides multiple complementary analyses to the paper.  Section \ref{appendix: lemma proof} provides the proof to lemma \ref{claim: concavity and estimation of q bar}. Section \ref{appendix: concave} analyzes the robustness of the results to our approximation whereby we treated the observed price schedule by LifeLabs as concave. In another robustness check, section \ref{appendix: smoothness} studies the optimal price schedule under the assumption that the gross valuation functions $V_i(\cdot)$ are smooth rather than piecewise linear. Section \ref{appendix: other specs} demonstrates the robustness of our results to model specification. Section   \ref{appendix: optimization method} sets out our grid-bisection optimization method, compares it to other alternatives, and provides recommendations on algorithm choice for researchers studying similar problems. Section \ref{appendix: data simulation} describes the process by which our counterfactual data (visually depicted in figure \ref{fig: incentive compatibility demand CF} of section \ref{sec: further CF analysis}) are generated. Section \ref{apx: 3rd degree} carries out  a  number of counterfactual analyses in which we study the impact of 3rd-degree price discrimination as well as a combination of 2nd and 3rd degree discrimination strategies. Section \ref{sec:alt estimation} provides an alternative estimation method that one could implement if exogenous (experimental) variation in prices were indeed available. Section \ref{apx: predicting sizes} examines whether it would be possible to avoid using data on intended sizes of failed deals and, instead, predict those sizes by (i) estimating a regression  model of size $\bar{q}_{it}$ on observables \textit{using the successful deals data only}, and then (ii) using that model to project predicted $\bar{q}_{it}$ values for unsuccessful deals. Section  \ref{apx: discontinuity} examines whether there is adequately strong inference can be made about willingnesses to pay by leveraging the observed discontinuities in the price schedule (as depicted in figure \ref{fig:pricesched}). Section \ref{apx: functional form} studies the implications  of our functional form assumption on the error terms $\epsilon_{it}$ in the value regression equation \ref{eq: value regression}. Section \ref{apx: alternative cost model} provides sensitivity analysis to using an alternative cost model. 

\section{Proof of Lemma \ref{claim: concavity and estimation of q bar}}\label{appendix: lemma proof}
As a reminder:

$$q^*(P|\Bar{q},v):=\arg\max_{q\geq0} \big[V(q|\Bar{q},v)-P(q)\big]$$

where $V(q|\Bar{q},v):=\min(q,\Bar{q})\times v$. 

Note that $V(q|\Bar{q},v)$ is constant in $q$ for $q\geq\Bar{q}$ whereas $P(q)$ is strictly increasing. Thus, no quantity $q>\Bar{q}$ can be in the arg max. As a result, we can rewrite: 

$$q^*(P|\Bar{q},v):=\arg\max_{q\in[0,\Bar{q}]} \big[V(q|\Bar{q},v)-P(q)\big]$$

But within the $[0,\Bar{q}]$ interval, the value function can be written as: $V(q|\Bar{q},v):=q\times v$ which yields:

$$q^*(P|\Bar{q},v):=\arg\max_{q\in[0,\Bar{q}]} \big[q\times v-P(q)\big]$$

Note that by strict concavity of $P(q)$ and linearity of $q\times v$, the function $q\times v-P(q)$ is strictly convex. This implies that its global maximum on the interval has to be an extreme point of the interval. That is: $q^*(P|\Bar{q},v)\subset\{0,\Bar{q}\}$. \textbf{Q.E.D.}

\section{Robustness analysis and discussion of the treatment of the observed schedule as increasing and concave}\label{appendix: concave}

In estimating the model, we leveraged Lemma \ref{claim: concavity and estimation of q bar} which stated that the observed $q$ for each customer $i$ would have to equate $\bar{q}_i$ if the observed price schedule $P(\cdot)$ charged by the firm is weakly concave. As Figure \ref{fig:pricesched} depicts, however, the observed price schedule is not concave. It does have a decreasing slope but it also have some discontinuous downward jumps. In this appendix, we accomplish two goals. First, we carry out a robustness check. We show that the schedule is indeed approximately concave. That is, these jumps are sufficiently small for the result to be robust to them. Second, we provide a discussion of how our estimation method would need to be modified if, loosely speaking, the degree of non-concavity were substantial.

\subsection{Robustness Analysis}\label{apx: robustness concavity}

More precisely, we re-estimate the model and the optimal pricing analysis without using the data points $i$ for whom the observed $q_i$ may be different from $\bar{q}_i$ due to non-concavities in the price schedule. Those are basically all data points that fall on the ``dips'' of the observed schedule, i.e., those observations $i$ with $q_i$ such that $\exists q<q_i$ with $P(q)>P(q_i)$, where $P(\cdot)$ is the observed price schedule.

\begin{figure}[H]
\begin{subfigure}{0.475\textwidth}
    \includegraphics[width=\textwidth]{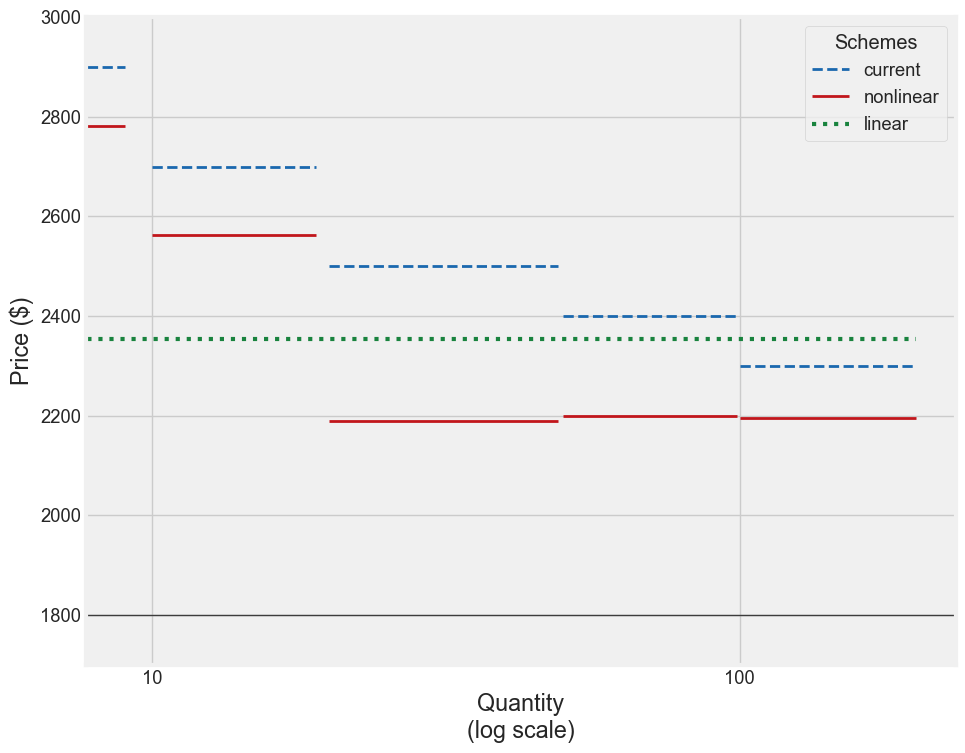}
    \caption{Marginal Prices (restricted sample)}
    \label{fig:marginalprice} 
\end{subfigure}
\begin{subfigure}{0.475\textwidth}
    \includegraphics[width=\textwidth]{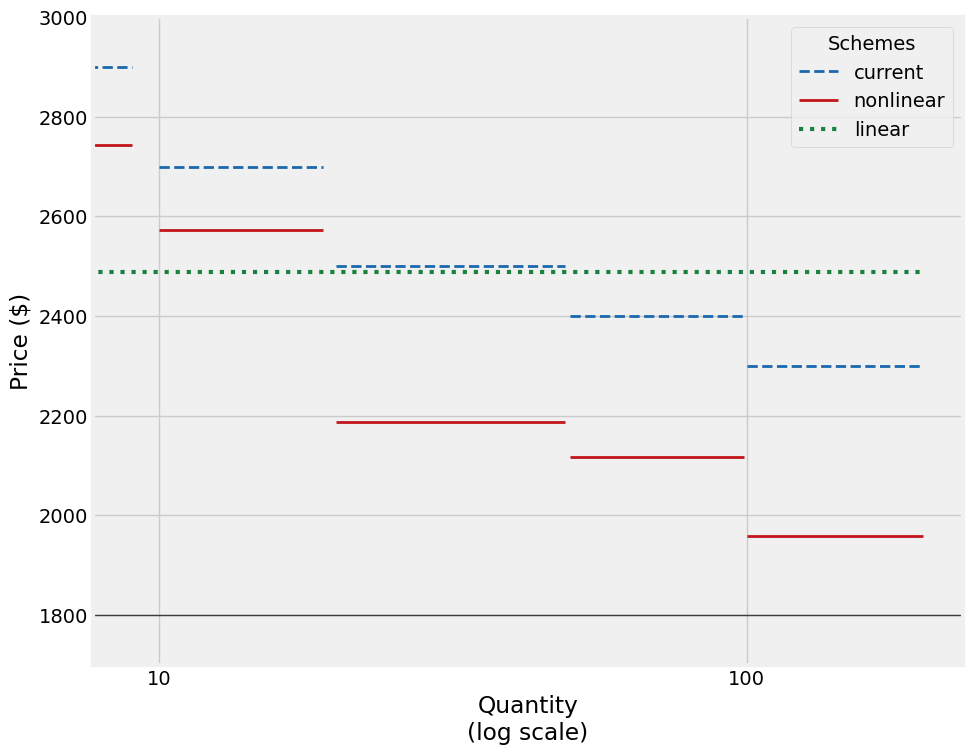 }
    \caption{Marginal Prices (full sample)}
    \label{fig:marginalprice} 
\end{subfigure}
    \caption{Optimal price schedules under restricted and full sample. As can be seen from this figure, differences between the two are negligible.}
    \label{fig:schemes_concavity}
\end{figure}

This new sample is slightly smaller than the original one (2,465 data points instead of 2,566). Based on this sample, the optimal price schedule and the profitability measures are re-computed and depicted in figure \ref{fig:schemes_concavity} and table \ref{tab:profitandwelfare_concavity}.

\begin{table}[H]
\begin{center}
{\small
\begin{tabular}{ l r l r l r l r l }
\toprule
 &  &  &  &  & \textit{Consumer} &  & \textit{Social} & \\
\hspace{5mm}\textit{Scheme} & \textit{Revenue} & change & \textit{Profit} & change & \textit{Welfare} & change & \textit{Welfare} & change \\
 & \textit{(\$M)} & (\%) & \textit{(\$M)} & (\%) & \textit{(\$M)} & (\%) & \textit{(\$M)} & (\%)\\
 \midrule
\hspace{4mm}\textit{current}  
& 27.54 & \multicolumn{1}{c}{-} & 15.07 & \multicolumn{1}{c}{-} & \ 6.84 & \multicolumn{1}{c}{-} & 21.91 & \multicolumn{1}{c}{-} \\
\cdashline{2-9}[1pt/1pt]
\textit{1$^\text{st}$ degree} 
& 52.89 & +92.05\% & 31.39 & +108.29\% &	\ 0 & $-$100.00\% & 31.39 & +43.27\%\\
\hspace{6mm}\textit{linear} 
& 31.41 & +14.05\% & 14.57 & -\ \ 3.32\% & 10.54 & +\ 54.09\% & 25.11 & +14.61\% \\
\textit{nonlinear} 
& 30.41 & +10.42\% & 15.8 & +\ \ 4.84\% & 8.81 & +\ 28.80\%	& 24.61 & +12.32\%\\
\bottomrule
\end{tabular}}
\end{center}
\caption{Profit and Welfare Analysis}
\label{tab:profitandwelfare_concavity}
\end{table}

As can be seen from comparing these results to the corresponding ones from the main text, there is little difference caused by the few points for the observed purchase size and customer size differ and the results are more or less robust.\footnote{Perhaps the most notable difference is that profit and revenue levels are generally slightly lower here compared to the results from the main text. This should not be surprising given that by removing some buyers from the sample, we are also removing the profit and revenue they generate for the seller.} 

\subsection{Discussion}\label{apx: discussion concavity}

In spite of the overall robustness of the results to the concavity assumption in our case, a broader question still stands: how should we estimate the model in other settings if the observed schedule is not concave, and the ``degree of non-concavity'' is substantial? In this subsection, we provide an informal discussion but do not fully implement the method.

When an observed price schedule $P(\cdot)$ is non-concave, some observed purchase quantities $q_i$ may have been chosen to benefit from the effective ``low-volume discount'' implied by the convex part of the schedule, as opposed to being chosen to satisfy the need of the customer for its size of use of $\bar{q}_i$ products. As an example, suppose a price schedule $P(\cdot)$ is given by $P(q)=2000\times q$ for $q\leq 100$ and $P(q)=2500\times q$ for $q>100$. In this case, a customer $i$ with $\bar{q}_i=120$ and $v_i=2200$ would optimally choose $q_i^*=100\neq \bar{q}_i$. Thus, for a researcher who does not observe the true size $\bar{q}_i=120$, it is no longer feasible (unlike the case of a concave tariff) to infer $\bar{q}_i$ with full precision. The researcher will only learn $\bar{q}_i\geq 100$.

To sum up the insight from the illustration in the previous paragraph, lack of concavity in the observed schedule $P(\cdot)$ implies that for some observations $i$, instead of point-identifying $\bar{q}_i$, we will only be able to set-identify $\bar{Q}_i$ such that $\bar{q}_i\in\bar{Q}_i$. As a result, an extra step would need to be added to our estimation procedure: we would need to get from sets $\bar{Q}_i$ to point estimates $\hat{\bar{q}}_i\in\bar{Q}_i$ for sizes $\bar{q}_i$.  In other words, we need to choose a proper $\hat{\bar{q}}_i$ from each, possibly singleton,\footnote{Note that some of the sets $\bar{Q}_i$ are indeed singletons for those observations that still allow for point identification. For instance, in the example in the previous paragraph, if we observe $q_i=50$, we can infer $\bar{q}_i=50$.} set $\bar{Q}_i$. One possible approach to doing so would be by choosing $\hat{\bar{q}}_i$ values that provide the best fit to a linear characterization based on some observables. Formally:

\begin{equation}
    \big\{\{\hat{\bar{q}}_i\}_i,\hat{\Gamma}\big\}=\arg\min_{\{\tilde{q}_i\}_i,\Gamma}\Sigma_i\big(\tilde{q}_i-X_i\times \Gamma\big)^2 \quad \text{s.t.}\, \forall i: \, \tilde{q}_i\in\bar{Q}_i
\end{equation}

A problem formulation of this format has previously been used to estimate costs to insurers of including hospitals into their networks of providers (in these environments, sometimes only inequalities about such costs are available but point estimates are needed, which is similar to our problem here). For examples of this formulation, see \cite{gowrisankaran2015mergers,ghili2022network}. In addition, \cite{ghili2022network} provides a fast algorithm called ``Regression Fixed Point'' which can be used to solve this problem.

Once we solve this problem and obtain point estimates for all $\bar{q}_i$, we can move on to the next step and estimate a model of values $v_i$.

\section{Robustness analysis and discussion of value functions with smoothly diminishing returns}\label{appendix: smoothness}

As mentioned in the main text of the paper, we make the simplifying assumption that value functions take the form $V_i(q)\equiv v_i\times\min(q,\bar{q}_i)$. In this appendix, we explore smoothing these value functions. More specifically, this appendix has two objectives. Subsection \ref{apx: smoothness robustness} carries out a robustness analysis \textit{in our specific context}, and examines the effects of such smoothness on the shape of the optimal schedule as well as the welfare effects. Subsection \ref{apx: smoothness discussion}, then, provides an informal discussion on how our estimation procedure should be modified if we are to estimate a smooth value function in more general contexts.

\subsection{Robustness analysis}\label{apx: smoothness robustness}

To formally capture smoother value functions, we modify the original  from $V_i(q)\equiv v_i\times\min(q,\bar{q}_i)$ to:

\begin{equation}\label{eq: smooth values}
    V_i(q)\equiv v_i \times [\zeta  \min(q,\bar{q}_i)^\alpha]
\end{equation}

In this formulation, $\alpha\in(0,1]$ allows the value to exhibit diminishing return as $q$ increases. The multiplier $\zeta$ does not carry any economic meaning, as  it could be absorbed in to $v_i$ and we would still have the same general formulation. We incorporate it, however, to make the above value function and the original piece-wise-linear one ``comparable''. More specifically, $\zeta$ is chosen so that the smoothed and original value functions yield the same value when evaluated at $q=\Bar{q}_i$:

$$v_i\times \zeta \Bar{q}_i^\alpha=v_i\Bar{q}_i\Leftrightarrow \zeta=\Bar{q}_i^{1-\alpha}$$

\begin{figure}[H]
    \centering
    \includegraphics[width=.38\textwidth]{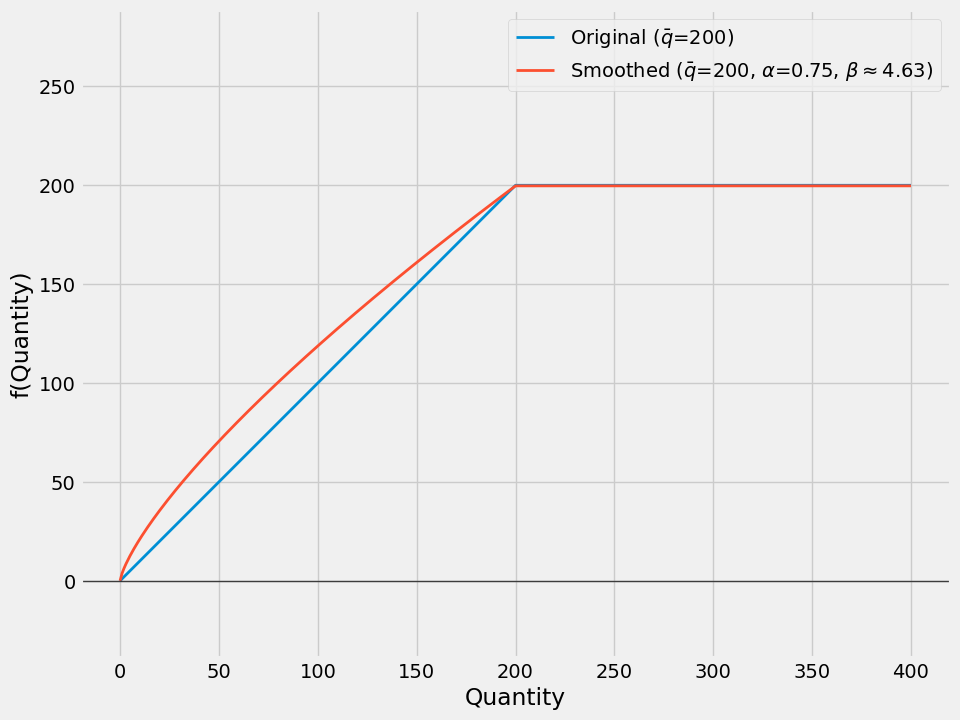}
    \caption{Original and smoothed utility comparison}
    \label{fig:smnooth_illustration}
\end{figure}

Figure \ref{fig:smnooth_illustration} illustrates one such smoothed value function when $\Bar{q}=200, v=1,$ and $\alpha=0.75$. 

\begin{figure}[H]
\begin{center}
\begin{subfigure}{0.4\linewidth}
    \includegraphics[width=\linewidth]{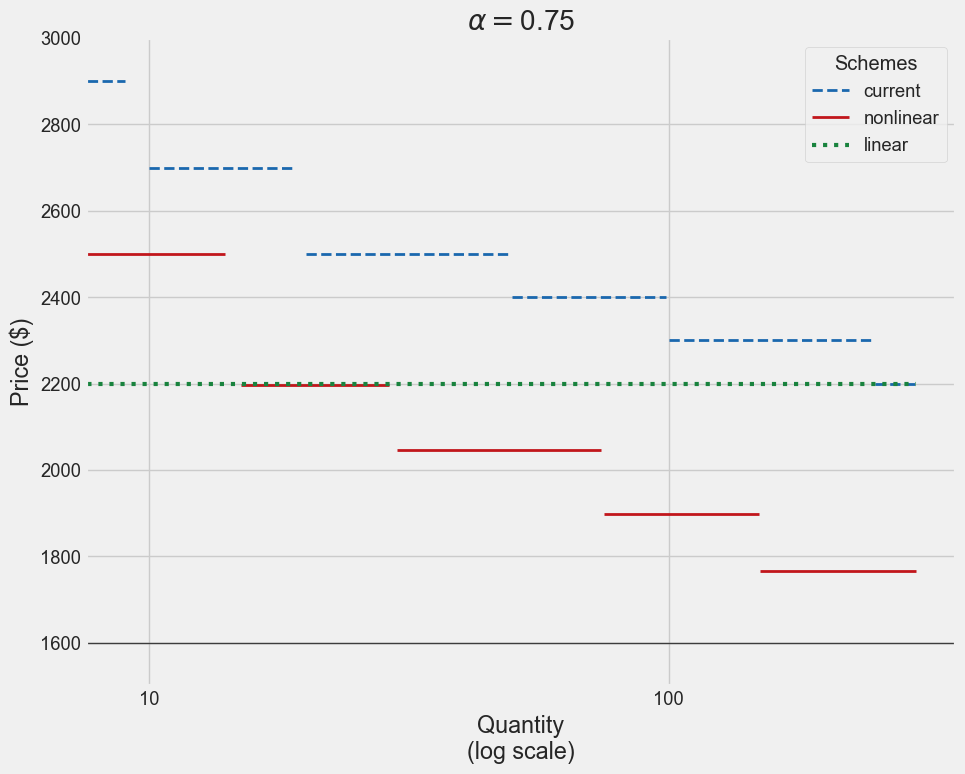}
    \caption{$\alpha=0.75$}
    \label{fig:marginalprice} 
\end{subfigure}
\begin{subfigure}{0.4\linewidth}
    \includegraphics[width=\linewidth]{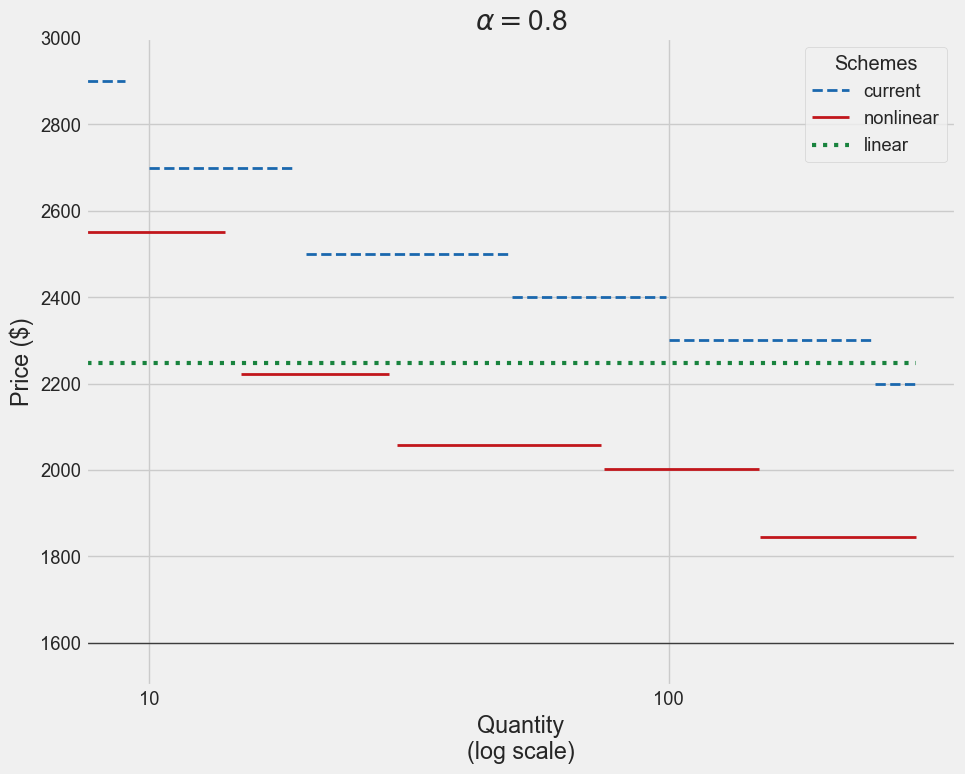}
    \caption{$\alpha=0.8$}
    \label{fig:marginalprice} 
\end{subfigure}
\begin{subfigure}{0.4\linewidth}
    \includegraphics[width=\linewidth]{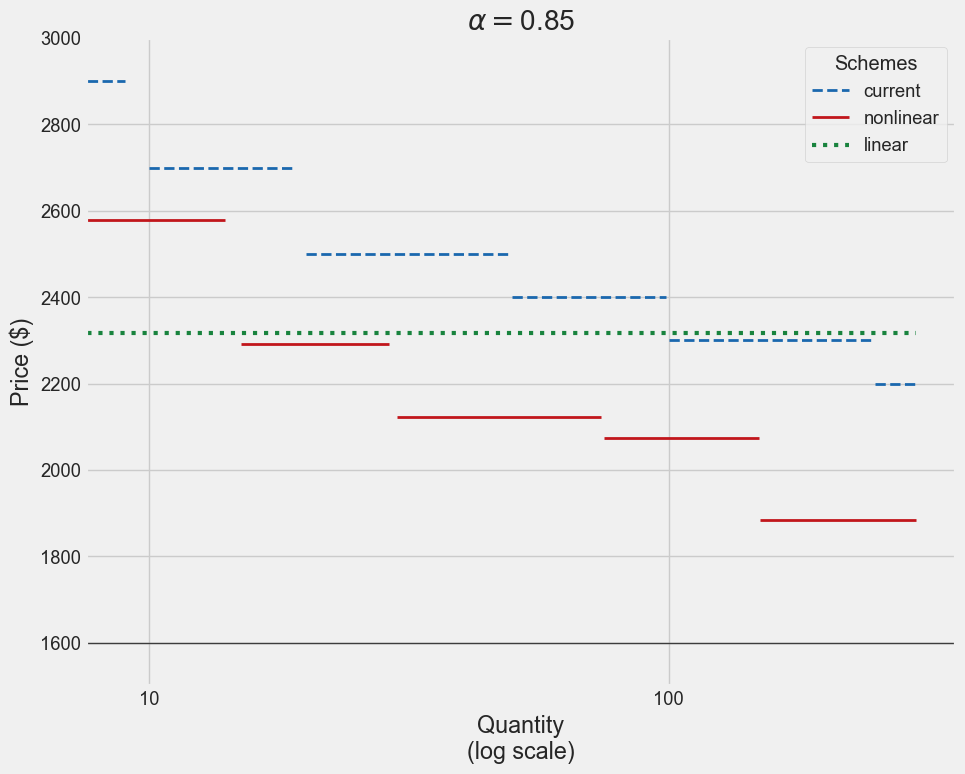}
    \caption{$\alpha=0.85$}
    \label{fig:marginalprice} 
\end{subfigure}
\begin{subfigure}{0.4\linewidth}
    \includegraphics[width=\linewidth]{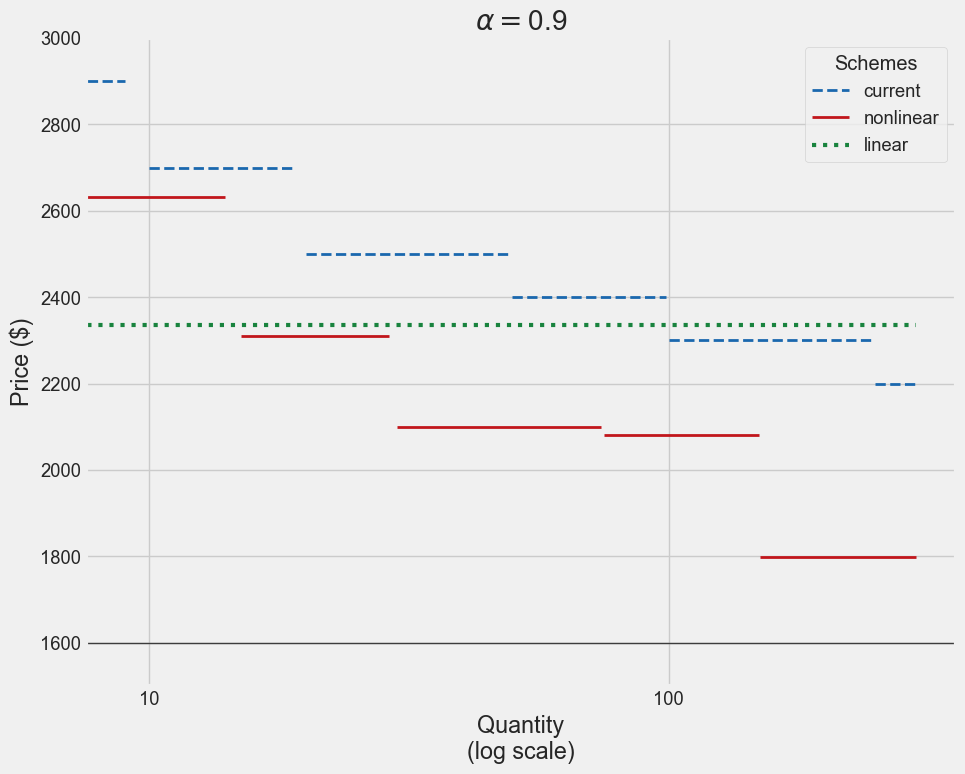}
    \caption{$\alpha=0.9$}
    \label{fig:marginalprice} 
\end{subfigure}
\begin{subfigure}{0.4\linewidth}
    \includegraphics[width=\linewidth]{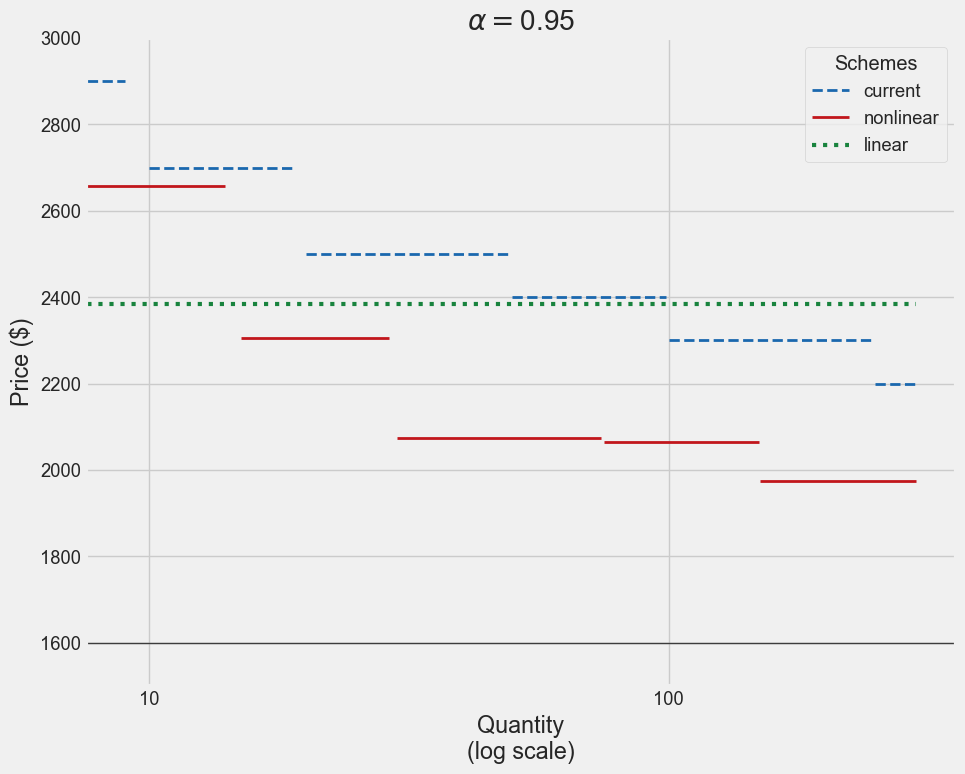}
    \caption{$\alpha=0.95$}
    \label{fig:marginalprice} 
\end{subfigure}
\end{center}
    \caption{Robustness analysis to the rate of diminishing return $\alpha$. The results on optimal linear and nonlinear pricing schedules seem robust.}
    \label{fig:smoothness robustness}
\end{figure}

We next turn to an analysis of how sensitive our results will be to $\alpha$. Figure \ref{fig:smoothness robustness} shows the optimal linear and nonlinear price schedules as a function of the smoothness factor $\alpha$. As can be seen from this figure, the results are robust to the smoothness factor in the range $[0.75,1]$. Note that even though $\alpha=1$ has not been analyzed in a separate subfigure here, the results from this case have already been examined in the paper. This is because the $\alpha=1$ special case coincides with our original, piece-wise linear, utility function.

As these figures show, the optimal pricing schedule seems robust to the level of the smoothness parameter $\alpha$. Note, however, that this robustness check was performed based upon an estimated value function $V(\cdot)$. A more complete analysis should, instead, allow for robustness \textit{in the estimation phase} and estimate the parameter $\alpha$ in the process. Fully estimating a smooth value function is beyond the scope of this paper. But the next subsection in this appendix provides an informal discussion of what such an estimation process would look like.


\subsection{Discussion}\label{apx: smoothness discussion}

The robustness checks in the previous subsections took as given the estimated parameters of the model,  added a smoothness parameter, and examined the robustness of the main result to different values that parameter could take. In this section, we ask how should the estimation procedure be modified if we are to incorporate $\alpha$ in equation \ref{eq: smooth values} as a parameter to be estimated?\footnote{The other parameter in the equation, i.e., $\xi$, need not be estimated and can be set to $1$. The reason we had to incorporate it into the robustness analysis in section \ref{apx: smoothness robustness} was to make the comparison against the main specification apples to apples.} We provide informal discussions on two fronts: how the estimation routine needs to change, and whether/how additional data can be helpful. Note that this discussion is informal and has been composed without actually estimating the smooth model. As a result, the process, once rigorously implemented, may differ from what comes below.

\noindent\textbf{Modifications to the estimation procedure:} We imagine at least three ways in which the estimation procedure should be modified if one is to estimate a smooth model with $\alpha$ in equation \ref{eq: smooth values} as a parameter.

\begin{enumerate}
    \item The two-step estimation procedure in the main text will no longer be feasible to maintain. In the main model, due to (i) the assumed functional form for $V(\cdot)$ and (ii) the (almost) concavity of the observed schedule, we were able to show (see Lemma \ref{claim: concavity and estimation of q bar}) that each customer $i$ chooses to purchase the quantity $q_i\in\{0,\bar{q}_i\}$. This result was what allowed us to separate the estimation into two steps. First, for anyone who purchases, the observed purchase quantity was assumed equal to that customer's $\bar{q}_i$, which helped estimate the marginal distribution of $\bar{q}_i$. Next, a discrete-choice model with binary dependent variable was fitted to a deal-success dummy variable to infer the  distribution of $v_i$ conditional on $\bar{q}_i$.
    
    With a smooth $V(\cdot)$ the lemma no longer holds, since the observed purchase quantity $q_i$ is no longer equal to size $\bar{q}_i$. Thus, the above two-stage process may no longer be utilized. We expect that the full distribution should be estimated ``all at once'' based on a single loss function that measures the distance between  the optimal $q^*_i$ as a function of the parameters and the observed $q_i$ across observations $i$. Though the simplest possible formulation for such a loss function would be $\Sigma_i (q^*_i-q_i)^2$, we expect that more complex forms may need to be adopted. For instance, one might consider adding a term that punishes for incorrectly predicting the deal status (i.e., $|s^*_i-s_i|:=|\textbf{1}_{q^*_i>0}-\textbf{1}_{q_i>0}|$). 

    \item Adding an extra parameter to the model and breaking Lemma \ref{claim: concavity and estimation of q bar} will likely imply that additional assumptions need to be imposed in order to maintain the estimation feasibility. We expect that some functional form assumption should be imposed on the distribution of customer size $\bar{q}_i$, given that it can no longer be directly read off the data.

    \item Finally, and perhaps most importantly, with a smooth model of $V(\cdot)$, one may need to find an alternative way to interpret the observed ``intended'' size for deals that eventually fail. Although one could still directly assign those numbers as sizes $\bar{q}_i$, such an approach might be inconsistent: if we assume $q_i$ is not necessarily equal to $\bar{q}_i$ for successful deals, why should we assume so for unsuccessful ones? As a result, more careful work would need to be done in order to incorporate the observed $q_i$ data for unsuccessful deals into the estimation procedure. This may indeed require further data, a topic to which we turn next.
 \end{enumerate}

\noindent\textbf{Useful additional data:} To estimate a model with smooth value functions, as mentioned in the last  bullet point above, a key consideration would need to be how to interpret the data on the intended sizes $q_i$ of deals that eventually failed. Especially helpful on this front, we believe, would be more granular data on how the intended deal size \textit{evolves} over the course of the pursuit until eventual success or failure. Such data is increasingly collected and maintained by firms. For instance, many B2B firms use versions of the Salesforce software for keeping track of their pursuits (also called ``opportunities'').  This software allows to record the intended size of the deal across different stages (as opposed to only once), which helps quantify how the deal size evolves. This goes beyond what we observe with the LifeLabs data which is a static picture of the intended deal size.

Such deal-size evolution data should be helpful in inferring size information for failed deals with more accuracy. A formal analysis of this is left to future research utilizing such data; here, we provide an informal discussion. Data can be leveraged to quantify time trends (over the course of stages) of both successful and unsuccessful deals. Relative to failed deals, successful deals have ``continuation trends'' as they last for more stages until closure (typically stage 6 in many settings). These continuation trends may be used to project what the purchase size would have been for those deals that  did not ultimately materialize.

\section{Robustness to other model specifications}\label{appendix: other specs}

In this section, we examine the sensitivity of the results to some model specifications. The number of alternative specifications one can potentially estimate is prohibitively large. As a result, we focus on a small set of alternatives. In particular, we examine robustness along two dimensions. First, we study how the model estimates and simulated optimal pricing policy would change if we changed the set of behavioral features or incorporated number of employees as another independent variable. Second, we check robustness to a specification in which instead of three distinct size groups, we have five size groups. This latter specification is formally implemented by changing the fixed effects model for sizes. Table \ref{tab:robustness mle} summarizes the estimation results across these alternative specifications. Note that the first column is the same as the default specification in the model, and has been presented for ease of comparison.

\newpage

\makeatletter
\def\adl@drawiv#1#2#3{%
        \hskip.5\tabcolsep
        \xleaders#3{#2.5\@tempdimb #1{1}#2.5\@tempdimb}%
                #2\z@ plus1fil minus1fil\relax
        \hskip.5\tabcolsep}
\newcommand{\cdashlinelr}[1]{%
  \noalign{\vskip\aboverulesep
           \global\let\@dashdrawstore\adl@draw
           \global\let\adl@draw\adl@drawiv}
  \cdashline{#1}
  \noalign{\global\let\adl@draw\@dashdrawstore
           \vskip\belowrulesep}}
\makeatother

\begin{figure}[H]
\begin{subfigure}{0.45\linewidth}
\centering
\includegraphics[width=0.9\linewidth]{Figures/2023joint1og.png}
\caption{Joint PDF of Model (1)}
\label{fig:subim2}
\end{subfigure}
\begin{subfigure}{0.45\linewidth}
\centering
\includegraphics[width=0.9\linewidth]{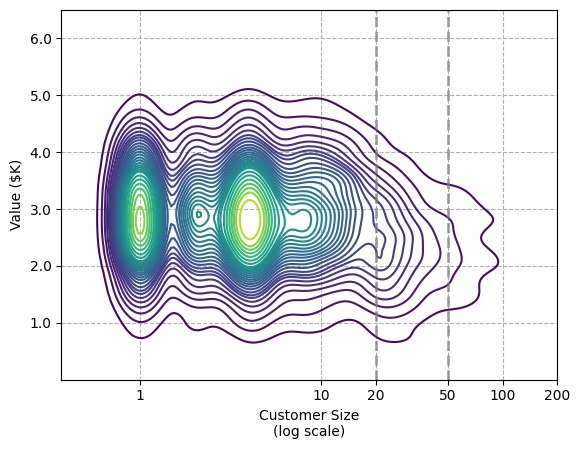} 
\caption{Joint PDF of Model (2)}
\label{fig:subim1}
\end{subfigure}
\begin{subfigure}{0.45\linewidth}
\centering
\includegraphics[width=0.9\linewidth]{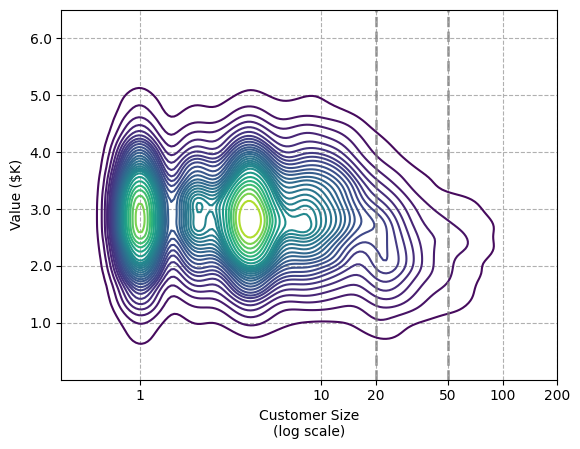}
\caption{Joint PDF of Model (3)}
\label{fig:subim2}
\end{subfigure}
\begin{subfigure}{0.45\linewidth}
\centering
\includegraphics[width=0.9\linewidth]{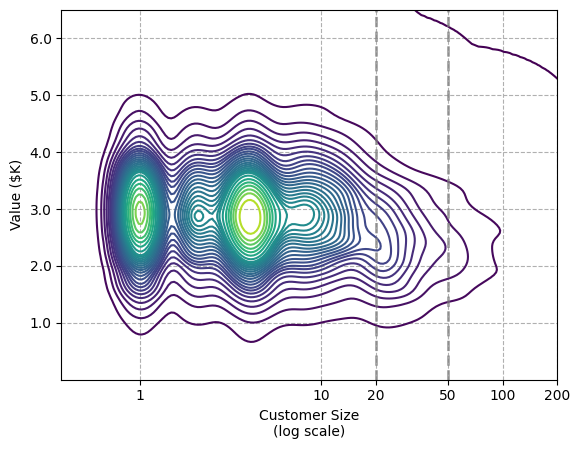} 
\caption{Joint PDF of Model (4)}
\label{fig:subim1}
\end{subfigure}
\begin{subfigure}{0.45\linewidth}
\centering
\includegraphics[width=0.9\linewidth]{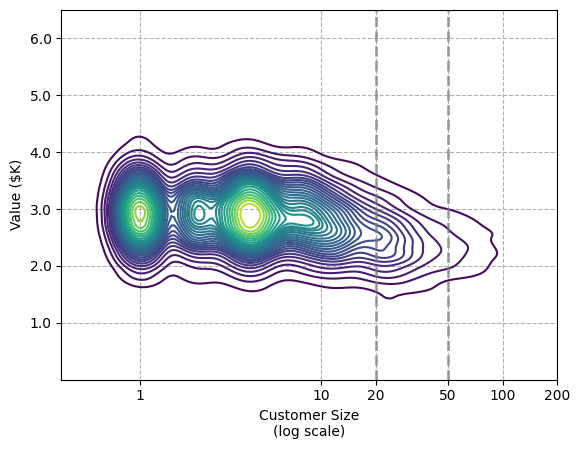} 
\caption{Joint PDF of Model (5)}
\label{fig:subim1}
\end{subfigure}
\caption{Estimated joint distributions $f(\cdot,\cdot)$ visualized for the five different model specifications labeled by columns of Table \ref{tab:robustness mle}}\label{fig:robustness joint dist}
\end{figure}

Figure \ref{fig:robustness joint dist} is complementary to Table \ref{tab:robustness mle} in that instead of  regression results, it visually compares the estimated joint distribution $f(\cdot,\cdot)$ across model specifications.

\begin{table}[h]
\begin{center}
{
\begin{tabular}{l c c c c c c}
\Xhline{1pt}
\thead[l]{Model} & \thead[c]{(1)} & \thead[c]{(2)} & \thead[c]{(3)} & \thead[c]{(4)} & \thead[c]{(5)}  \\ \hline\\[-2ex]
\textit{Intercept} 
 & 2,260 & 2,314 & 2,253 & 2,435  & 2,531\\ [-.5ex] 
 & (56) & (58) & (59) & (57) & (33) \\
 \cdashlinelr{2-6}
\textit{Size} $\in$ [20,50)
    & -657 & -638 & -659 & -535 & \\
    [-.5ex] & (44) & (41) & (43) & (41) & \\
\textit{Size} $\in$ [50,$\infty$)
    & -835 & -797 & -840 & -610  &\\
    [-.5ex] & (68) & (69) & (70) & (74) & \\
\textit{Size} $\in$ [10,20)
    & & & &  & -270 \\
    [-.5ex] &  &  &  &  & (23) \\
\textit{Size} $\in$ [20,50)
    & & & &  & -583\\
    [-.5ex] &  &  &  &  & (26)\\
\textit{Size} $\in$ [50,100)
    & & & & & -711\\
    [-.5ex] &  &  &  &  & (53) \\
\textit{Size} $\in$ [100,$\infty$)
    & & & & & -832\\
    [-.5ex] &  &  &  &  & (72) \\
 \cdashlinelr{2-6}
\textit{Log Firm Age} 
    & -40 & -38 & -42 & -46  & -22\\
    [-.5ex] & (16) & (18) & (18) & (17) & (10) \\
\textit{Time}
    & 70 & 118 & 71 & 123  & 47\\
    [-.5ex] & (26) & (25) & (25) & (24) & (15) \\
\textit{Computer Software}
    & 39	& 40 & 38 & 44  & 25 \\
    [-.5ex] & (29) & (29) & (28)  & (27) & (17) \\
\textit{Marketing and Advertisement}
    & -224	& -177 & -222 & -144  & -130\\
    [-.5ex] & (67) & (64) & (66) & (58) & (39) \\
 \cdashlinelr{2-6}
\textit{Log Number of Employees}
    & & -8 & 1 & 2  & \\
    [-.5ex] &  & (9)  & (9) & (8) & \\
\textit{Log Feature 1}
    & 133 & 144 & 133 & 151  & 79\\
    [-.5ex] & (9) & (9) & (9)  & (9) & (6) \\
\textit{Feature 2}
    & 686 & 680 & 687 &  & 417\\
    [-.5ex] & (30) & (29) & (29)  &  & (25) \\
\textit{Feature 3}
    & & -462 & & -441 &\\
    [-.5ex] &  & (34)  &  & (31)  & \\
 \cdashlinelr{2-6}
\textit{Scale}
    & 385 & 374 & 386 & 383   & 228\\
    [-.5ex] & (8) & (7) & (7) & (8) & (13)\\
 \hline
Negative Log Likelihood & 2649 & 2557 & 2649 & 2865  &  2613 \\[-0.3ex]
Number of Obs. & 4,468  & 4,468 & 4,468  & 4,468   & 4,468  \\
\Xhline{1pt}
\end{tabular}}
\end{center}
\begin{tablenotes}
\small
\linespread{1}\small
\item \textit{Notes:} \textit{Firm Age}, which is the current year minus founded year was log transformed and  \textit{Time} is time fixed effect coded $1$: 2021 and $0$: 2020. \textit{Size}, \textit{Industry}, \textit{Feature 2} and \textit{3} are all indicator variables. The values in parentheses represent the bootstrapped standard errors using 1,000 samples.
\end{tablenotes}
\caption{MLE estimates table}
\label{tab:robustness mle}
\end{table}

\newpage

Table \ref{tab:robustness mle} and Figure \ref{fig:robustness joint dist} suggest that our estimation results are robust across specifications. In order to complete the robustness analysis, however, it would add value to examine the sensitivity of our key counterfactual analysis result (i.e., the shape of the optimal nonlinear price schedule) to the specification. Table \ref{tab:robustness price schedule} accomplishes this.

\begin{table}[H]
\begin{center}
{\footnotesize
\begin{tabular}{ c c c c c c}
\Xhline{1pt}
\thead{Model Specification} & $p^*_1$ & $p^*_2$ & $p^*_3$ & $p^*_4$ & $p^*_5$ \\
\hline 
(1) & 2776 & 2617 & 2113 & 2113 & 1948\\
(2) & 2749 & 2642 & 2177 & 2114 & 1957\\
(3) & 2791 & 2420 & 2187 & 2147 & 2136\\
(4) & 2744 & 2572 & 2188 & 2117 & 1959\\
(5) & 2688 & 2299 & 2069 & 1979 & 1787\\

\Xhline{1pt}
\end{tabular}}
\end{center}
\caption{Optimal nonlinear pricing schedule $P^*=(p^*_1,...,p^*_5)$  remains robust to model specification.}\label{tab:robustness price schedule}
\end{table}

As this table shows, the shape of the optimal contract remains robust to the model specification.

\section{Optimization Method}\label{appendix: optimization method}

Though mechanism design theory does provide tractable methods to compute optimal nonlinear tariffs, all of those methods are devised under assumptions that seldom hold empirically.\footnote{For instance, methods provided in \cite{mussa1978monopoly,maskin1984monopoly} and similar papers all rely on a ``single-crossing'' condition which make the customer heterogeneity one-dimensional, an assumption that is clearly violated in our context (for instance in the left panel of Figure \ref{fig: illustration}).} As a result, we turn to numerical approaches to find the optimal schedule.

\textbf{Our method: ``Grid-Bisection''.} Our problem has some features that are crucial in determining what method is used for optimization. First, the number of dimensions (i.e., five) is non-trivial but not too large. Second, we seek to find the global maximum as opposed to a local one. Third, each instance of computing the objective function (i.e., the profit) is costly. Fourth, the objective function is expected to behave non-smoothly, and there are no guarantees on concavity/linearity features that one could leverage.

Under the above circumstances, the optimization literature recommends the use of Bayesian methods (see \cite{frazier2018bayesian,frazier2018tutorial} for an overview). We found, however, that a multi-dimensional variant of bisection search outperforms a Bayesian approach in the sense of delivering a higher objective-function value in a shorter amount of time. In this appendix, we briefly describe our method. We also comparatively analyze multiple alternatives (a variant of Bayesian Optimization and multiple versions of gradient descent)  and provide a summary on which approach we think is the most appropriate one under different problem settings.

In a nutshell, our method starts with a 5-dimensional grid of all the five possible prices $(p_1,...,p_5)$ described in the previous subsection. Each grid is of size $d$, which implies that we initially evaluate the profit function $\pi$ under $d^5$ possible values for vector $P$. The values for each $p_k$ ($k\in\{1,...,5\}$) in this initial grid are chosen to be equi-distant points on the interval $[0,5000]$. In other words, the lower and upper bounds in iteration-1 of the algorithm for each $p_k$ are given by $\underline{p}^1_k=0$ and $\bar{p}^1_k=5000$. Denote the length of the interval on the $k$-th dimension by $l_k^1=\bar{p}^1_k-\underline{p}^1_k$.\footnote{Note that by our choices of the $\underline{p}^1_k$ and $\bar{p}^1_k$ values, all $l^1_k$ are equal to each other; but this need not be the case generally.}

We then proceed with an iterative process. In each iteration $t$, we first find the optimal price vector $P^{t*}=(p^{t*}_1,...,p^{t*}_5)$  among the $d^5$ candidates in our price grid. Next, we form a new grid (for the next iteration) by ``zooming in'' on $P^{t*}$. More precisely, the new grid is constructed based on $P^{t*}$ and is a zoom factor $z\in(0,1)$ in the following manner:

\begin{equation}\label{eq: grid binary iteration}
    (\underline{p}^{t+1}_k,\bar{p}^{t+1}_k)=\begin{cases} (\underline{p}^{t}_k,\underline{p}^{t}_k+l^t_k), & \text{if   }   p^{t*}_k-z\frac{l^t_k}{2}< \underline{p}^{t+1}_k 
        \\
        (\bar{p}^{t}_k-l^t_k,\bar{p}^{t}_k),   & \text{if   }
        p^{t*}_k+z\frac{l^t_k}{2}> \bar{p}^{t+1}_k \\
        (p^{t*}_k-z\frac{l^t_k}{2},p^{t*}_k+z\frac{l^t_k}{2}),   & \text{otherwise}
    \end{cases}
\end{equation}

In words, the new bounds $\underline{p}^{t+1}_k$ and $\bar{p}^{t+1}_k$ form an interval of length $z\times l^t_k$ centered around $p^{t*}_k$, unless this interval itself falls partially outside of the old bounds $\underline{p}^{t}_k$ and $\bar{p}^{t}_k$ (in which case the new interval is moved until it falls just within the old one). We then iterate and construct smaller and smaller intervals until we reach $t$ such that all $l^t_k$ are smaller than a pre-determined threshold $\varepsilon$, at which point the algorithm stops. 

\begin{figure}[H]
    \centering
    \begin{center}
    \includegraphics[width=0.4\linewidth]{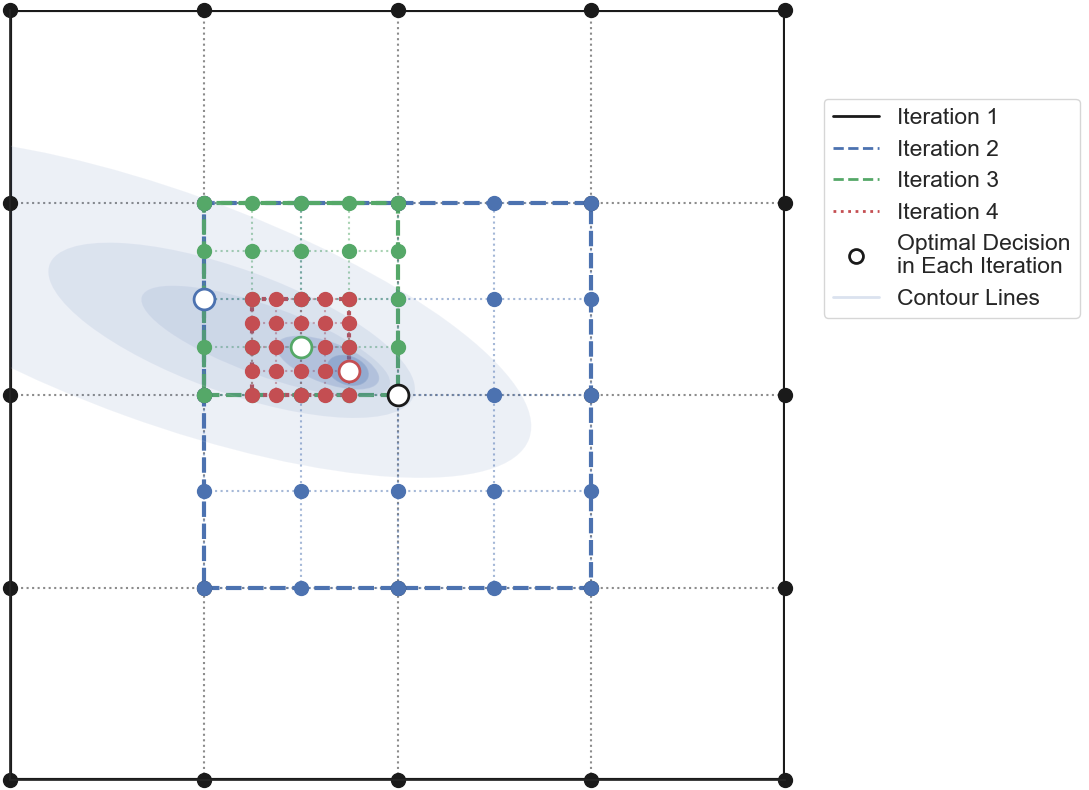}
    \end{center}
    \footnotesize  \emph{Notes:} (1) We are using $d = 2$ for demonstration. In the paper, $d = 5$. (2) In each loop $t$, the length of the binary search interval for each dimension $k$ shrinks by $z = 0.5$.
    \caption{Visual illustration of the grid-bisection optimization method.}
    \label{fig:grid-binary}
\end{figure}

Though we are not aware of a systematic analysis of the properties and performance of this grid-bisection approach, we are aware that variants of it have been previously used in different fields. One example is \cite{yin2020kaml}, who also provide a visual illustration of their variation. Following their illustration, Figure \ref{fig:grid-binary} schematically describes our grid-binary optimization procedure.

The algorithm involves $d^5$ instances of computing the profit function $\pi$ in each iteration. It approximately takes $\frac{\log(l^1_k)-\log(\varepsilon)}{-\log(z)}$ iterations for the algorithm to stop. We choose $d=5$ and $z=\frac{1}{2}$ for our application. 

\textbf{Comparison Against Alternative Methods.} With this grid-bisection optimization method in hand, we next turn to  a comparison between it and other alternatives we considered. In particular, we examine the following three approaches: 1) Bayesian Optimization a la \cite{frazier2018bayesian,frazier2018tutorial}, 2) Gradient method with the initial candidate for $P^*$ chosen from an initial grid of $5^d$ possible $P\in[0,3000]^d$ vectors, and 3) A simplex-based direct search method a la \cite{nelder1965simplex} with the initial candidate for $P^*$ chosen from an initial grid of $10^d$ possible $P\in[0,3000]^d$ vectors. We do this comparison in three scenarios: (i) solving for the optimal linear price (d=1), (ii) solving for the optimal price when we have three marginal costs for small (i.e., $\Bar{q}_i\in I_1 \cup I_2$), medium-size (i.e., $\Bar{q}_i\in I_3$), and large (i.e., $\Bar{q}_i\in I_4 \cup I_5$)customers, i.e., $d=3$, and finally (iii) the optimal price $P^*$  in our main 2nd-degree price discrimination setting which was five-dimensional $d=5$. Table \ref{tab: comparison among computation algoirthms} presents the results. It presents both the objective value (i.e., profit) and runtime of each algorithm. Note that in addition to the methods whose performances are reported in this table, we examined a number of Gradient-method alternatives, such as ``conjugate gradient”, ``BFGS”, ``Truncated newton algorithm”, and ``Powell algorithm”. However, due to weak performance, we do not report them here.

\begin{table}[H]
\begin{center}
{\footnotesize
\begin{tabular}{ l r l c r}
\Xhline{1pt}
\thead{Method\hspace{2cm}} & \hspace{3cm} & \thead{Marginal Price (\$)} & \thead{Profit (\$M)} & \thead{Time (sec)} \\
\hline 
\multirow{3}{*}{Grid-Bisection} 
& linear & [2392] & 18.71 & 7\\
& nonlinear 3 & [2584 2132 2136] & 20.07 & 276\\
& nonlinear 5 & [2682 2477 2133 2125 2133] & 20.19 & 1243\\
\Xhline{0.05pt}
\multirow{3}{*}{Bayesian Optimization} 
& linear & [2392] & 18.71 & 25\\
& nonlinear 3 & [2578 2134 2132] & 20.06 & 342\\
& nonlinear 5 & [2696 2456 2098 2079 1810] & 20.07 & 1194\\
\Xhline{0.05pt}
\multirow{3}{*}{Nelder-Mead\,5} 
& linear & [2390] & 18.71 & 3\\
& nonlinear 3 & [2590 2077 2074] & 20.08 & 14\\
& nonlinear 5 & [2694 2456 2072 2034 1835] & 20.13 & 157\\
\Xhline{0.05pt}
\multirow{3}{*}{Nelder-Mead\,10} 
& linear & [2390] & 18.71 & 3\\
& nonlinear 3 & [2585 2133 2130] & 20.08 & 55\\
& nonlinear 5 & [2656 2443 2077 2076 2075]  & 20.16 & 3006\\
\Xhline{1pt}
\end{tabular}}
\end{center}
\caption{Comparative analysis of the performances of different optimization algorithms}

\label{tab: comparison among computation algoirthms}
\end{table}

Note that this analysis has been performed on an earlier specification of the model. This is why the optimal schedule is slightly different from what can be found in the main specification presented in the paper.

\textbf{Discussion and Recommendations.} Based on the results  presented in table \ref{tab: comparison among computation algoirthms}, the grid-bisection approach consistently outperforms other candidates with respect to the objective function. It also outperforms the Bayesian method on the front of the algorithm run-time. The comparison between Grid-Bisection and the Nelder Mead simplex approach \citep{nelder1965simplex} is more nuanced. In the case of $d=5$, which is the main case we analyze, Grid-Binary takes shorter to run than Nelder-Mead with an initial grid of $10^5$ but takes much longer if compared instead to Nelder-Mead with an initial grid of size $5^5$. 

Given the above summary, we recommend that Grid-Bisection be used for similar empirical multi-dimensional-screening problems if (i) precision is of utmost importance, and (ii) the number of dimensions is not too high (e.g., $d=5$). For higher numbers of dimensions (e.g., $d=10$ or larger), we conjecture that the Nelder-Mead approach with an initial grid that is not too large will make the best trade-off between accuracy and runtime.

\section{Details on data simulation}\label{appendix: data simulation}

This section provides details on how the counterfactual datasets for the middle and right-most columns of Figure \ref{fig: incentive compatibility demand CF} were produced.

To produce the counterfactual data in the middle column, we alter the original data in a way that increases the deal acceptance rate in the for mid-size customers (i.e., $\Bar{q}_i\in I_3$) while decreasing the acceptance rates for smaller ($\Bar{q}_i\in I_1\cup I_2$) and larger ($\Bar{q}_i\in I_4\cup I_5$) ones. More specifically, for each row of the data (i.e., each $it$ combination), we alter the observed value $s_{it}$ for deal success status with probability $p$ (i.e., according to a random draw from a Bernoulli distribution with parameter $p$). If the Bernoulli draw is 1, (that is, if we are set to alter $s_{it}$), then we alter it to $1$ if customer $i$ is mid-size and to 0 otherwise. If the Bernoulli draw is 0, however, we do not alter $s_{it}$.

The procedure for generating the counterfactual data for the right-most column of the figure is similar, except that the $s_{it}$ values for mid-size deals are altered to 0 and those for small and large deals are altered to 1. Another difference between the generating processes for the middle and right-most columns of Figure \ref{fig: incentive compatibility demand CF} is that we set $p=0.7$ for the former and $p=0.6$ for the latter.

The algorithm below formally describes the process.

\begin{center}
\begin{algorithm}[H]
	\SetAlgoNoLine
	\KwIn{Demand data with the number of samples $N$ and for some predetermined probability $p\in(0,1)$}
	\KwOut{Demand data with modified deal outcomes}
        \emph{\textbf{Note}} $Group_i$: size group(small/medium/large) of sample $i$\\
        \hspace{7mm}$Deal\,Success_i$: binary deal success variable of sample $i$\\
        \textbf{Obtain} $N$ draws from Bernoulli distribution with $p$ and let it $\mathbf{d}\in \{0,1\}^N$\\
            \eIf {$\mathbf{d}_i==1$}{
                \eIf{$Group_i==medium$}{
                    $Deal\,Success_i=1$
                }{ $Deal\,Success_i=0$
                }
            }{\textbf{pass}}
        \caption{Counterfactual data simulation process}
	\label{alg: data simulation}
\end{algorithm}
\end{center}

\section{Third Degree Price Discrimination}\label{apx: 3rd degree}
In this appendix, we study third-degree price discrimination. In particular, we consider two kinds: (i) third-degree price discrimination based on customer sizes, and (ii) combining second-degree price discrimination (based on size) with third-degree discrimination (based on other observables).


\subsection{Third-degree discrimination based on customer size}

In the discussion of incentive compatibility constraints,  we introduced the ``individually optimized'' contract $\Tilde{P}$. We argued this price schedule is ``naive'' in that it fails to anticipate the ability by each customer of size $\Bar{q}_i\in I_k$ to tune its purchase size in order to take advantage of other, lower, prices $p_{k'}$. In this section, we ask ``what if such anticipation is indeed correct?" In other words, what if the seller is able to force customers of each size group $k$ to only pay $p_k$ per unit no matter how many units $q$ they purchase? This means the firm can third-degree price discriminate based on size $\Bar{q}$ and need not worry about incentive compatibility. Under these conditions, $\Tilde{P}_k$ would indeed become the true optimal price schedule. The profit here, will be equal to $\Sigma_k \pi_k (\Tilde{p}_k)$ where ``local'' profit functions  $\pi_k(\cdot)$ are defined as in equation \ref{eq: profit local}.

Note that due to the relaxation of the incentive constraints, the third-degree-discrimination profit $\Sigma_k \pi_k (\Tilde{p}_k)$ is larger than $\pi(\Tilde{P})$ and, likely, also than the second-degree-discrimination profit $\pi(P^*)$. Table \ref{tab: 3dpdbysize} empirically analyzes these profits under the three data scenarios described in Figure \ref{fig: incentive compatibility demand CF} (recall that one scenario is the original data and the other two are counterfactual datasets, modifying the demand system).

\begin{table}[H]
    \centering
    \small
    \begin{center}
        \begin{tabular}{lccc}
            \toprule
             & \textbf{Original} & \textbf{(i)} & \textbf{(ii)} \\
            \midrule
            Jointly Optimized Pricing & 17.54 & 15.55 & 22.02 \\
            \cdashline{1-4}[1pt/1pt]
            Individually Optimized Pricing & 17.53 & 14.22 & 20.19 \\
                          & ($-$0.06\%) & ($-$8.55\%) & ($-$8.31\%) \\
            Third-Degree PD (by Size) &    17.58 &    15.94 &    22.77 \\
                          & (+0.23\%) &  (+2.51\%) & (+3.41\%) \\
            \bottomrule
        \end{tabular}
    \end{center}
        \footnotesize \emph{Notes:} Columns (i) \& (ii) indicate profits for Figure \ref{fig: incentive compatibility demand CF} middle and right columns. (i) is produced from the scenario with the highest success rate in the median group (20\%/78\%/16\%) while (ii) is from the scenario with the lowest success rate in the median group (80\%/21\%/77\%).
    \caption{Profits from Jointly and Individually Optimized Schedules  and Third-Degree PD by Size}
    \label{tab: 3dpdbysize}
\end{table}

As the left column of this table shows, size-based third degree price discrimination delivers little to no extra profitability above second-degree discrimination (only by 0.23\%). This should not be too surprising given that the main advantage of size-based third-degree discrimination is the relaxation of incentive-compatibility constraints. But as first reported in figure \ref{fig: incentive compatibility demand CF} and repeated in the middle row of table \ref{tab: 3dpdbysize}, in the original dataset the profit impact of incentive compatibility constraints was too slim (less than 0.1\%). Looking at alternative data scenarios (mid and right columns in Table \ref{tab: 3dpdbysize}) confirm our intuition: the benefit of third-degree price discrimination based on size is large (small) if the loss from charging the individually optimized contract instead of the jointly optimized one is large (small). Formally, there seems to be a strong and positive association between $\pi(P^*)-\pi(\Tilde{P})$ on the one hand and $\big(\Sigma_k\pi_k(\Tilde{p}_k)\big)-\pi(P^*)$ on the other.

\subsection{Combining second and third degree discrimination}

\begin{figure}[H]
    \centering
    \makebox[\textwidth]{
    \includegraphics[height=2.25in]{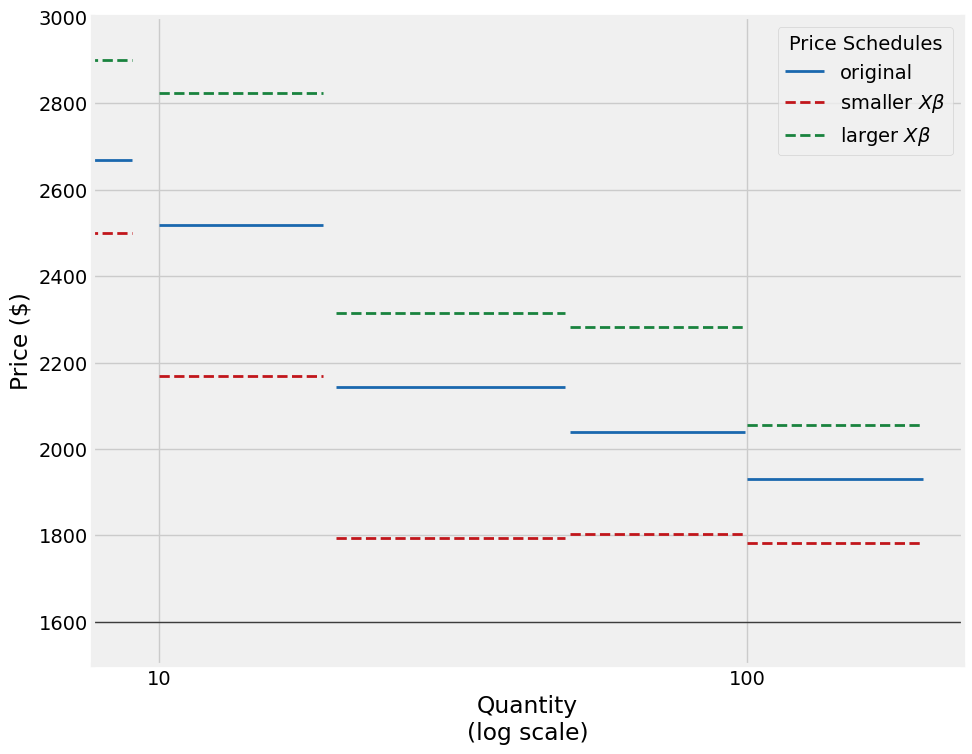}}
    \caption{Optimal price schedules $P^{*1}$ and $P^{*2}$ when customers $i$ are divided into $J=2$ groups based on ranked $X_i\beta$ values}
    \label{fig: 3pd-pricing}
\end{figure}

Another approach to third degree price discrimination would be to combine it with second degree. More precisely, the firm can rank customers based on their $\beta X_i$ from equation \ref{eq: value regression} in a descending order and create equally-sized groups $j\in\{1,...,J\}$. The firm can then offer each group $j$ a separate optimal price schedule $P^{*j}(\cdot)$. Given the specification of $\beta$, this means customers who were established earlier, or those with higher amounts of behavioral feature 1 will be put in higher-ranked bins and face higher prices.\footnote{Of course many other customer characteristics can be incorporated into this. But as mentioned before, in our specific contexts, many potentially relelvant features turned out to be of little impact empirically.}. Figure \ref{fig: 3pd-pricing} shows the optimal strategy if the firm divides all the customers into $J=2$ groups based on ranked $X_i\beta$ values. The original (i.e., one-group) optimal schedule $P^*$ has also been plotted for comparison. As can be seen from the figure, the higher-willingness-to-pay group $j=1$ gets charged an additional \$200  or more per workshop (across different sizes) relative to the lower-willingness-to-pay group $j=2$.

\begin{figure}[H]
    \centering
    \makebox[\textwidth]{
    \includegraphics[height=2in]{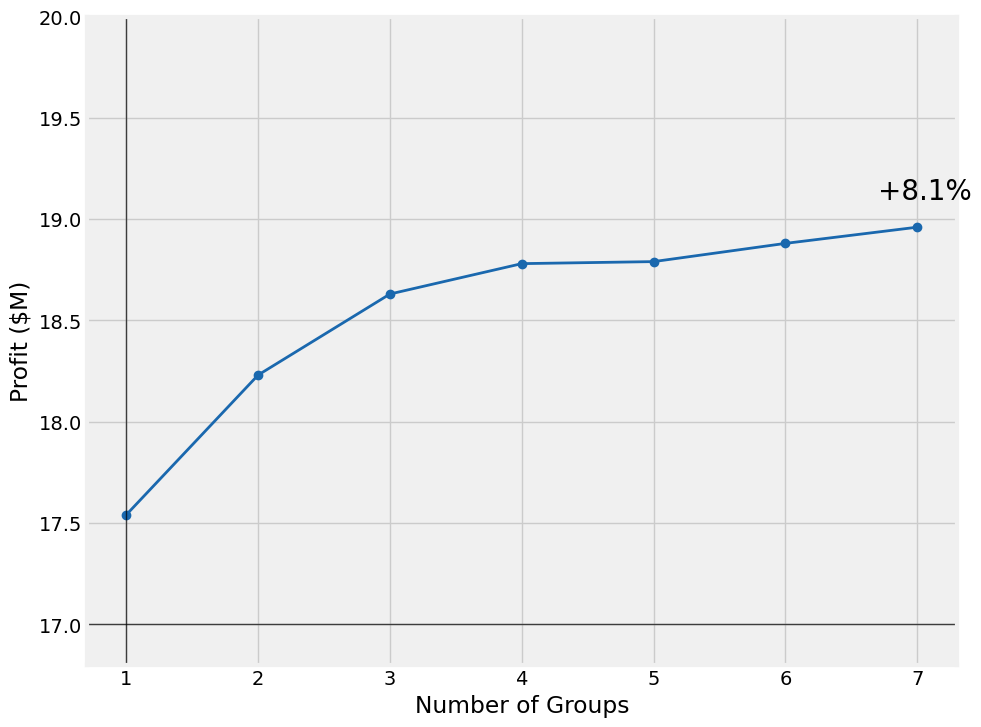}}
    \caption{Total Profit from optimal 2nd+3rd degree price discrimination as a function of the number of groups $J$}
    \label{fig: 3pd-groups}
\end{figure}

We close this section by a discussion of the profitability of third-degree price discrimination. Figure \ref{fig: 3pd-groups} plots the profitability of the optimal pricing as a function of the ``extent of third degree discrimination $J$''. Under purely second-degree discrimination ($J=1$), total profit is $\$17.54/y$ whereas under substantial third-degree discrimination $J=7$, the profitability is around $\$18.96M/y$. This profit surpasses the second-degree-discrimination profit by only 8.1\%. As a result, third-degree discrimination, though non-trivially useful when deployed alone or in conjunction with second-degree, does not seem to generate substantial extra profitability above sole second-degree discrimination.

We do not find the less-than-stellar performance of third-degree price discrimination in our context surprising. This is because as mentioned before in the process of variable selection for the modeling of $v$ (see equation \ref{eq: value regression}), we found that many potentially relevant variables (especially industry fixed effects) have little explanatory power. We do not expect this empirical observation to be generalizable. In other contexts where observable customer characteristics are highly predictive of purchasing power (see \cite{dube2017scalable} for instance), third degree price discrimination may even closely approximate the profitability of first-degree discrimination.

\section{Alternative Estimation Method for Experimental Data}\label{sec:alt estimation}

Here, we discuss an alternative estimation procedure that one could use if exogenous variation on prices were available. There are multiple reasons why such an analysis is worth carrying out. First, with exogenous price variation, one  would no longer require data on the sizes of unsuccessful deals for estimation. Given that this change of data leads to changes in the estimation methodology in a non-trivial way, it is useful to set out this alternative estimation routine  so that it can be used by firms/researchers that do have access to experimental data. Second, our analysis here will provide a more formal sense of how taxing the requirements will be for firms of running experiments that can allow for inference of the joint distribution between size and value, which helps further motivate our original approach of leveraging data on intended sizes of failed deals.

Our analysis here has two components. First, based on a ``\textit{ground truth}'' joint distribution $g(v,\bar{q},X)$ over customer sizes, values, and customer observable characteristics, we construct a simulated dataset of customers' purchase decisions under experimental price variation. Next, we exposit a method that  takes as inputs  the simulated data and recovers an estimate of $g(\cdot,\cdot,\cdot)$.

\textbf{Simulated Data:} We construct our ``ground truth'' distribution $g$ from the previously estimated distribution $f(\cdot,\cdot)$ over sizes and values  (visualized in Figure \ref{fig: joint dist}) and its co-variation with observable characteristics $X$. To this end, we start from  our original dataset\footnote{By original, we mean the version before carrying out the augmentation described previously.} and implement the following changes:

\begin{enumerate}
    \item Concatenate multiple copies of the dataset to achieve a desired number of rows (e.g., 1 Million)
    \item For each row $i$ of the resulting dataset:
    \begin{enumerate}
        \item Create a value column  $v_i$ by taking a draw from the estimated distribution from equation \ref{eq: value regression}.
        \item Create a price column $p_i$  by a randomly selecting a price in the set \{0,\$1K,\$2K,\$3K,\$4K,\$5K\}.
        \item Replace the deal success column $d_i$ by $\textbf{1}_{v_i\geq p_i}$.
        \item Remove from the dataset the entire column $v_i$. Also remove (i.e., replace by \textit{NA}) the value of $\bar{q}_i$ wherever $d_i=0$.
    \end{enumerate}
\end{enumerate}

Once we have this dataset, the objective is to recover $g(\cdot,\cdot,\cdot)$ \textit{without} using data on $\Bar{q}_i$ for the cases of non-purchase $d_i=0$, and \textit{without} using any data on $v_i$. Step (d) above in the construction of the dataset ensures we indeed do not have access to those data elements.

\textbf{Estimation Method:} We start by noting that under $p_i=0$, almost every customer purchases. Hence, the marginal distribution $g_{\bar{q},X}(\cdot,\cdot)$ over $\Bar{q}$ and $X$ can be approximated by the empirical distribution over 
$X$ and $\Bar{q}$ when $p_i=0$ and $d_i=1$. Denoted   $\tilde{g}_{\bar{q},X}(\cdot,\cdot)$, this empirical distribution is by construction observable. We then use the law of total probability to recover conditional distributions $g_{\bar{q},X}(\cdot,\cdot|v_i<\Bar{p})$ for $\Bar{p}\in\{\$1K,\$2K,\$3K,\$4K,\$5K\}$. For each such $\Bar{p}$, we have:

\begin{equation}\label{eq: law of total prob}
    g_{\bar{q},X}(\cdot,\cdot) = g_{\bar{q},X}(\cdot,\cdot|v_i\geq\Bar{p}) \times \text{Prob}(v_i\geq\Bar{p}) +g_{\bar{q},X}(\cdot,\cdot|v_i<\Bar{p})\times \text{Prob}(v_i<\Bar{p})
\end{equation}

Observe that in the above equation, $g_{\bar{q},X}(\cdot,\cdot|v_i<\Bar{p})$ is the only unknown: $g_{\bar{q},X}(\cdot,\cdot)$ we constructed an estimate for from the empirical distribution under zero price; $\text{Prob}(v_i\geq\Bar{p})$ and $\text{Prob}(v_i<\Bar{p})$ can be respectively estimated by directly measuring the fractions of $d_i=1$ and $d_i=0$ once the data is filtered for the $\Bar{p}$ treatment group: $p_i=\Bar{p}$; and, in a similar fashion, $g_{\bar{q},X}(\cdot,\cdot|v_i\geq\Bar{p})$ can be estimated by taking the empirical distribution over $X,\Bar{q}$ when $p_i=\Bar{p}$ and $d_i=1$. Plugging all of those into the equation, we can come up with estimate $\tilde{g}_{\bar{q},X}(\cdot,\cdot|v_i<\Bar{p})$ for $g_{\bar{q},X}(\cdot,\cdot|v_i<\Bar{p})$.

To summarize: even though in our analysis we do not have data on $\Bar{q}_i$ for individual unsuccessful deals, we can still estimate the distribution of $\Bar{q}_i$ (joint with observables) over all unsuccessful deals, conditional on each treatment price. This will be sufficient for the purpose of estimating the full joint distribution $g(\cdot,\cdot,\cdot)$ and, by extension, the optimal nonlinear pricing schedule, tasks which we turn to now.

Once equipped with estimates $\tilde{g}_{\bar{q},X}(\cdot,\cdot)$ and $\tilde{g}_{\bar{q},X}(\cdot,\cdot|v_i<\Bar{p})$ for $g_{\bar{q},X}(\cdot,\cdot)$ and $g_{\bar{q},X}(\cdot,\cdot|v_i<\Bar{p})$, we can use them to ``fill in'' the missing $\Bar{q}_i$ data for unsuccessful deals. To this end,  for each row with $p_i=\Bar{p}$ and $d_i=0$ and observable characteristics $X$, one can construct the conditional probability distribution $\tilde{g}_{\bar{q}|X}(\cdot|v_i<\Bar{p},X=X_i)$, take a draw of $\bar{q}$, and use it to fill the missing value for $\Bar{q}_i$.\footnote{To construct the conditional probability $\tilde{g}_{\bar{q}|X}(\cdot|v_i<\Bar{p},X)$, one would first need to ``smooth''  $\tilde{g}_{\bar{q},X}(\cdot,\cdot|v_i<\Bar{p})$ using functional form assumptions. To avoid this, another approach (which is what we take) would be to take a joint draw of $\bar{q},X$ from the original distribution $\tilde{g}_{\bar{q},X}(\cdot,\cdot|v_i<\Bar{p})$ and replace them for the missing $\Bar{q}_i$ as well as the non-missing $X_i$. Asymptotically, they should lead to the same outcome.}

With the last step finished, we have a dataset akin to the one that was used to estimate a model of $v_i$ according to equation \ref{eq: value regression}. As a result, we can repeat the same analysis here and estimate this equation, which would complete our estimation process.

\begin{figure}[H]
\begin{center}
\begin{subfigure}{0.32\textwidth}
    \includegraphics[width=\textwidth]{Figures/2023joint1og.png}
    \caption{Original distribution}
\end{subfigure}
\begin{subfigure}{0.32\textwidth}
    \includegraphics[width=\textwidth]{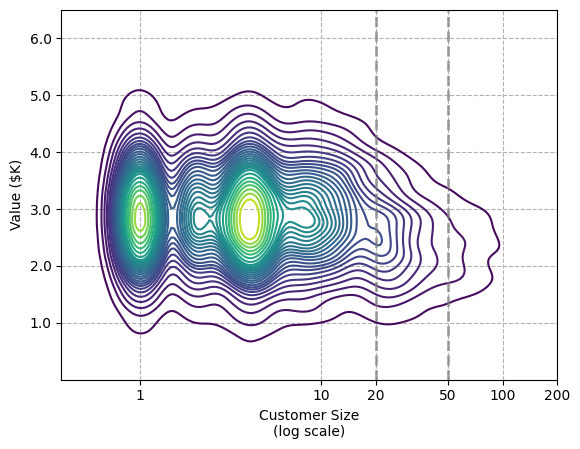}
    \caption{Recovered distribution}
\end{subfigure}
\begin{subfigure}{0.32\textwidth}
    \includegraphics[width=\textwidth]{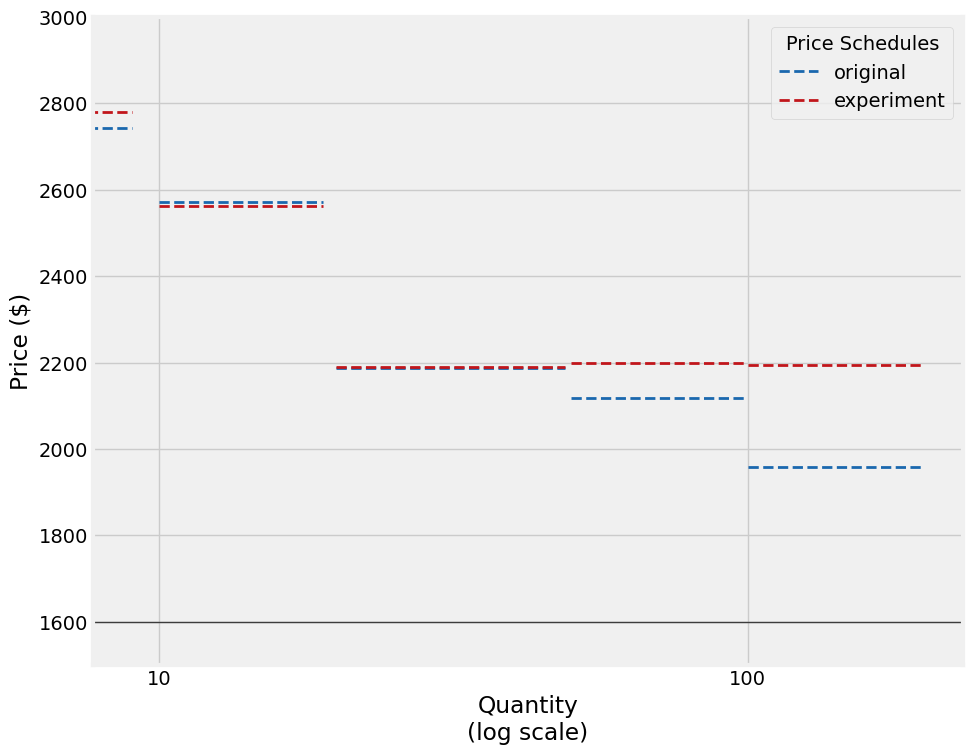}
    \caption{Price schedules}
\end{subfigure}
\end{center}
\caption{Comparing the recovered distribution $f$ in panel (b) with the assumed ``ground truth'' in panel (a). The optimal price schedules from the two distributions are shown in panel (c)}
\label{fig: experimental results}
\end{figure}

\textbf{Results:} We carried out the procedure above, using a simulated dataset with one million rows. We estimated the joint distribution $g$ over $\Bar{q},v,X$. Distribution $g$ is difficult to visualize. But the most critical (and yet easy to visualize) object of estimation is the less-general joint distribution $f$ over $\Bar{q},v$. Thus, we use $f$ to present our results. See Figure \ref{fig: experimental results} for a comparison between the assumed ground truth $f$ and our recovered $\hat{f}$ from the procedure outlined above. As the figure shows, the two distributions are similar. The implied optimal price schedules from them are also similar but there is some divergence for large deals.

We finish this section by reviewing some positives and negatives about the experimental approach. As for positives, there are at least two. First, it can help us estimate the model and come up with the optimal \textit{nonlinear} pricing schedule even if the only variation we have is a set of \textit{linear} prices. Second, if the range of prices experimented with is wide, we can estimate the $v_i$ equation without the functional form assumptions that we imposed in \ref{eq: value regression}.

On the front of drawbacks to the experimental approach, the first issue is that it is expensive. As our analysis shows, we need a wide range of experimental prices. Especially, we need to experiment with very low prices in order to get an estimate for the unconditional marginal distribution $\tilde{g}_{\bar{q},X}(\cdot,\cdot)$ on $\Bar{q},X$. In what we proposed above, we set one of the experimental prices as zero for this purpose. This is expensive  to firms. One can still estimate $\tilde{g}_{\bar{q},X}(\cdot,\cdot)$ through extrapolation if the lowest experimental price is larger than zero. But such extrapolation leads to some accuracy loss; and the higher the smallest experimental price, the larger the accuracy loss. Another reason why the experimental approach is expensive is how many data points it needs: we need representative samples across all size groups and under each price experiment. As the illustrative analysis above suggests, this requisite sample size can be substantial. This is far from reality in most B2B markets where the number of potential customers is small in the first place, and firms are at best willing to run experiments on a small subset of their markets. Additionally, a drawback of our main proposed method (i.e., using data on unsuccessful deals) is also the case for price experiments. More specifically, one worry about inference based on unsuccessful deals is that it ignores those potential customers who did not even enter a conversation with the firm because their $v_i$ was low but would have purchased under lower-than-observed prices. The same issue exists with experimental price variation. This is because under an experiment, customers decide whether to approach the seller based on the publicly posted price schedules, and not based on the price-treatment-group they would be assigned to.

\section{Predicting sizes for unsuccessful deals based on observed sizes for successful deals}\label{apx: predicting sizes}

In our main analysis, we assume $\bar{q}_{it}$ for any unsuccessful deal is the size information that we observe in the data corresponding to that deal. One possible way to avoid using that data would be through a prediction model that learns the relationship between $\bar{q}_{it}$ and some observable characteristics \textit{by training on the successful deals data}, and then uses that learned relationship to project predictions of $\bar{q}_{it}$ for $it$ observations where the deal is unsuccessful. We implement this approach in this appendix to find whether it can help us study this market without a need to use data on intended sizes of failed deals. 

The first step (i.e., learning), is implemented through the following regression equation:

\begin{equation}\label{eq: learning q bar from observables}
    \ln \bar{q}_{it}=X_{it}\times \psi+\nu_{it}
\end{equation}

where $X_{it}$ is the same vector of characteristics as what was used in the values regression equation \ref{eq: value regression}. We estimate the regression above using data for successful deals only. Resuls are reported in table \ref{tab: learning q bar from observables}. The key point to take from this table is that the explanatory power of these characteristics seems rather low, with an $R^2$ of less than 0.1. 

\begin{table}[H]
    \centering
{
\begin{tabular}{lrrrrrr}
\toprule
 & $\gamma$ & SE & $t$ & $P>|t|$ & $[0.025$ & $0.975]$ \\
\midrule
\textit{Intercept}
&    -1.3489 &     0.558 &    -2.418 &  0.016 &    -2.443 &    -0.255 \\
\textit{Log Feature 1}
&     0.2043 &     0.052 &     3.959 &  0.000 &     0.103 &     0.306 \\
$\text{\textit{Log Feature 1}}^2$ 
&    -0.0268 &     0.007 &    -3.766 &  0.000 &    -0.041 &    -0.013 \\
\textit{Feature 2}
&     0.3513 &     0.037 &     9.619 &  0.000 &     0.280 &     0.423 \\
\textit{Log \# Employees}
&     0.5041 &     0.090 &     5.628 &  0.000 &     0.328 &     0.680 \\
\textit{Log Revenue}
&     0.1021 &     0.033 &     3.114 &  0.002 &     0.038 &     0.166 \\
\textit{Log \# Employees}$\times$\textit{Log Revenue}
&    -0.0234 &     0.005 &    -4.670 &  0.000 &    -0.033 &    -0.014 \\
\textit{Time}
&     0.2790 &     0.037 &     7.501 &  0.000 &     0.206 &     0.352 \\
\midrule
& & & \textit{s} & 0.703 & \textit{N.Obs.} & 2,120 \\
& \multicolumn{3}{r}{\textit{Negative Log-Likelihood}} & 2631 & $R^2$ & 0.099 \\
\bottomrule
\end{tabular}
}
    \caption{Regression Summary Table}
    \label{tab: learning q bar from observables}
\end{table}

Once equipped with estimate $\hat{\psi}$ of $\psi$, we can use it to project values of $\bar{q}_{it}$ for unsuccessful deals $it$. With these projections at hand, one can form an estimation for the marginal distribution of $\bar{q}$ without using the intended-size-of-use data. In forming this distribution, for the \textit{successful deals}, there are two options. One can use projections for those observations as well, or alternatively, one can use directly observed purchase sizes $\bar{q}_{it}$ in the case of successful deals. We decided to go with the former method (i.e., using projections). This is because the projected values for $\bar{q}_{it}$ based on regression \ref{eq: learning q bar from observables} do not have any values above 15.\footnote{This, in turn, is because there are much fewer deals of size 15 or above in the data compared to deals of smaller sizes. As a result, our best fitting model seems to have essentially treated the large sizes as noise.} This means if for successful deals we use the original $\bar{q}_{it}$ data, we will end up with a dataset where the deal success rate for large deals is always 100\%, which we find inaccurate. As a result, we use projections $e^{X_{it}\hat{\psi}}$ for all observations regardless of deal success status.

With these projections at hand, we can then implement the estimation method described in the main text of the paper, and obtain results on the joint distribution of size and value.  The results are provided below. To make side by side comparison possible, we also provide results from the main analysis.

\begin{table}[h]
\begin{center}
{
\begin{tabular}{ l r r r}
\toprule
&  &  \textit{\textbf{Alternative Model}} & \textit{\textbf{Original Model}} \\
\textit{\textbf{Coefficient}} &  &  \textit{\textbf{Estimates}} & \textit{\textbf{Estimates}} \\
\midrule
\textit{Intercept} & $\beta_{0}$ & 2129.43 & 2260.56 \\[-0.3ex]
\textit{Log Feature 1} & $\beta_{1}$ & 54.39 & 133.79 \\[-0.3ex]
\textit{Feature 2} & $\beta_{2}$ & 811.34 & 686.64\\[-0.3ex]
\textit{Computer Software} & $\beta_{cs}$ & 26.29 & 39.18 \\[-0.3ex]
\textit{Marketing and Advertisement} & $\beta_{ma}$ & -286.7 & -224.29\\[-0.3ex]
\textit{Log Firm Age} & $\beta_{age}$ & -69.72 & -40.99 \\[-0.3ex]
\textit{Time} & $\alpha_{2021}$ & 51.98 & 70.20 \\[-0.3ex]
\textit{Mid Size} & $\gamma_{med}$ & -48.06 & -657.40 \\[-0.3ex]
\textit{Large Size} & $\gamma_{big}$ & 54.39 & -835.73\\[-0.3ex]
\textit{Scale} & $\sigma$ & 495.17 & 385.44\\[-0.3ex]
\midrule
\textit{Negative Log-Likelihood} & & 0.3075 & 0.2951\\
\bottomrule
\end{tabular}}
\end{center}
\caption{Maximum Likelihood Estimates for the parameters describing $f_{V|\bar{Q}}(\cdot)$. Results provided for the main specification as well as the model variant in this appendix. Results suggest predicting $\bar{q}_i$ based on observables (as opposed to using the data directly) compromises the quality of the estimation.}
\label{tab:mleestimates robustness to predicting sizes}
\end{table}

\begin{figure}[H]
\begin{center}
\begin{subfigure}{0.45\textwidth}
    \includegraphics[width=\textwidth]{Figures/2023joint1og.png}
    \caption{Estimated $f(\cdot,\cdot)$ under the main specification}
\end{subfigure}
\begin{subfigure}{0.45\textwidth}
    \includegraphics[width=\textwidth]{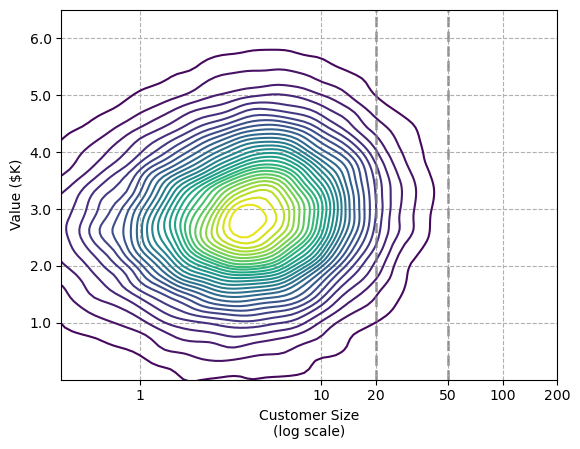}
    \caption{Estimated $f(\cdot,\cdot)$ under the specification in this appendix}
\end{subfigure}
\end{center}
\caption{As the results suggest, using predicting sizes based on observables (instead of using direct data on them) compromises the flexibility with which we can capture the joint distribution between size and value.}
\label{fig: comparison btwn main model and predicting sizes}
\end{figure}

Table \ref{tab:mleestimates robustness to predicting sizes} and figure \ref{fig: comparison btwn main model and predicting sizes} summarize these comparisons. The key takeaway is that the projection approach fails to ``realize'' the gap in deal acceptance rate across different deal sizes. This can be seen by observing that, in table \ref{tab:mleestimates robustness to predicting sizes},  the coefficients $\gamma_{med}$ and $\gamma_{big}$ are much smaller in magnitude under the model in this appendix relative to the main analysis. Also, figure \ref{fig: comparison btwn main model and predicting sizes}, which visualizes the estimated joint distributions between sizes and values for the analysis in this appendix as well as the main analysis of the paper. This figure too can show that the projection approach fails to capture key aspects of the co-variation between size and valuation.

\begin{figure}[H]
\begin{subfigure}{0.475\textwidth}
    \includegraphics[width=\textwidth]{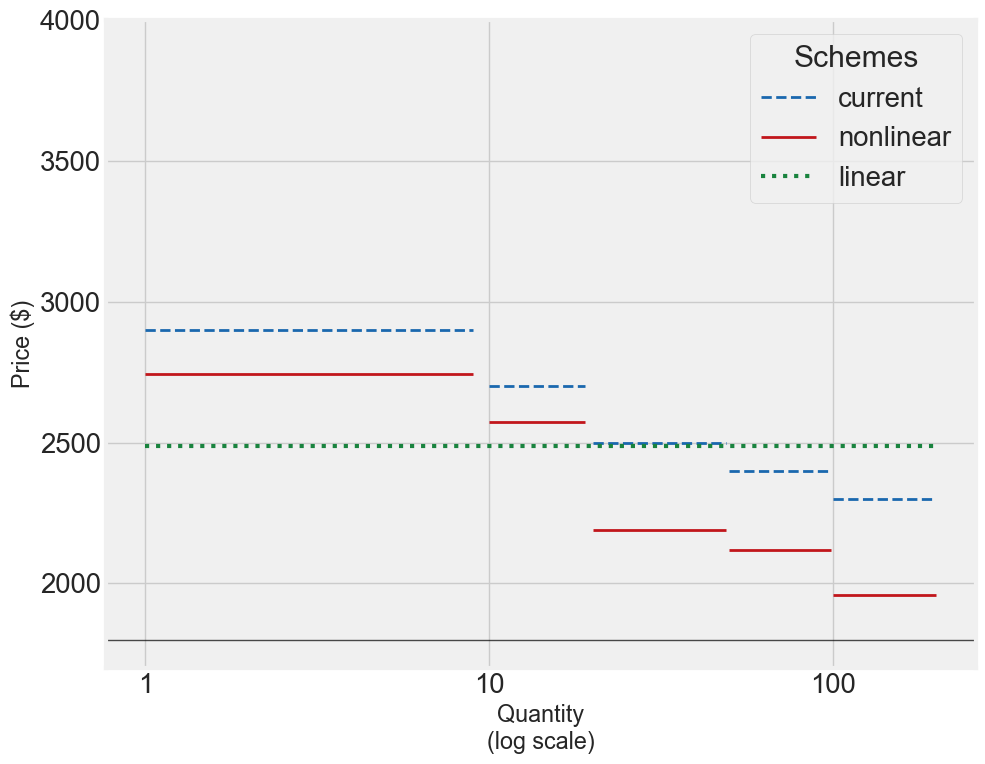}
    \caption{Marginal Prices (original)}
    \label{fig:marginalprice_original} 
\end{subfigure}
\begin{subfigure}{0.475\textwidth}
    \includegraphics[width=\textwidth]{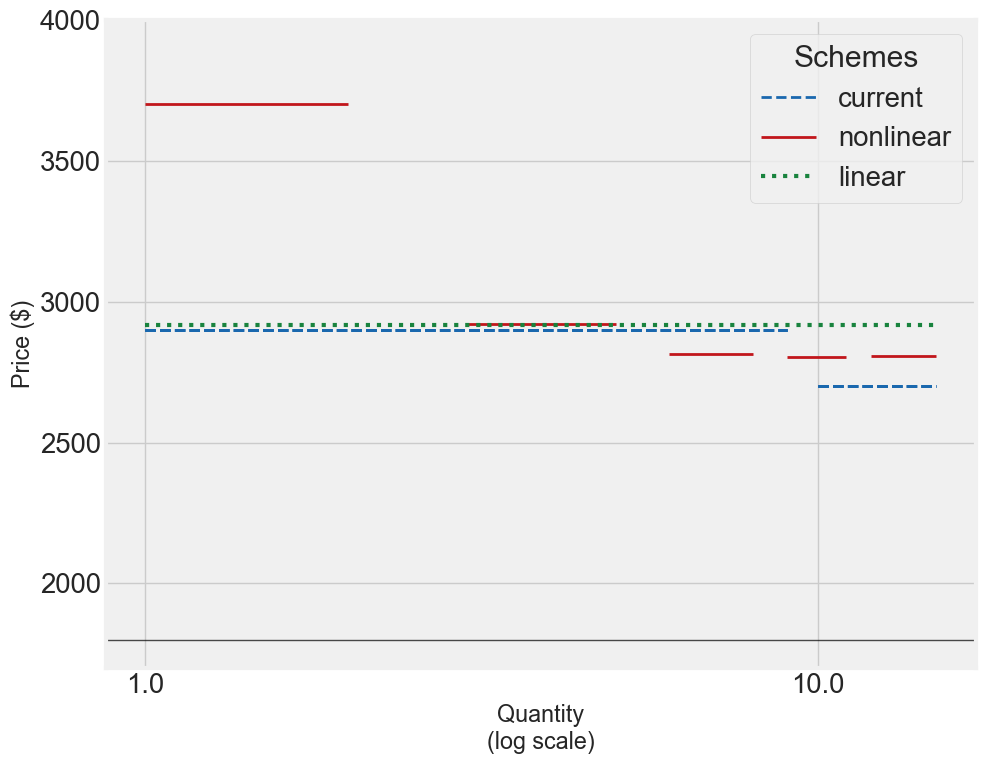}
    \caption{Marginal Prices (estimated $\bar q_{it}$)}
    \label{fig:marginalprice_qestimated} 
\end{subfigure}
    \caption{Optimal price schedules under original model and estimated $\bar q$ model. Note that size ranges from 1 to 15 for the right panel: [1-15] for linear and [1-3, 4-6, 7-9, 10-12, 13-15] for nonlinear scheme.}
    \label{fig:schemes_qestimated}
\end{figure}

Finally, figure \ref{fig:schemes_qestimated} below compares the optimal pricing schedule under the projection approach used in this appendix against the optimal schedule from the main analysis of the paper. As can be seen from this figure, our projection approach does not have recommendations on how to price for larger deal sizes. Even the recommendations on the smaller deal sizes do not seem reasonable.

\section{Leveraging discontinuities in the observed price schedule for identification}\label{apx: discontinuity}

As mentioned in the main text, our estimation approach makes a clear functional form assumption by making inference based on a comparison between the would be demand at very low price to the observed demand at observed prices. In this appendix, we study whether we can avoid making this functional form assumption by turning to a different source of variation for estimating price sensitivity? That source of variation is the observed discontinuities in the price schedule. As figure \ref{fig:pricesched} in the main text of the paper shows, the price per unit abruptly changes a number of thresholds: 10, 20, 50, and 100. In this appendix, we ask whether such discontinuities can make it possible to estimate price sensitivities for different deal sizes? We do not carry out a full estimation analysis based on the idea but show results that suggest the answer is no. We will then finish by a discussion of whether such an approach could work in other B2B contexts.

Table \ref{tab: discontinuity} provides an overview of the data points that we have around these thresholds. Specifically, we have multiple ``window sizes''. To illustrate, for a window size of 1 and a threshold of 10, we only look at deals with an (intended/realized) size of 9 and 10, respectively to represent ``just below'' and ``just above'' the threshold. As another illustration, for a window size of 3 and a threshold of 50, the just-below window is 47-49 and the just-above is 50-52. For different threshold/window-size combinations, the table reports the counts of deals (successful and unsuccessful) observed within the just-below and just-above windows, as well as the percent success rates.

\begin{table}[H]
\centering
{\small
\begin{tabular}{ccrcrcrcr}
\toprule
 \textbf{\textit{size group}}&
     \multicolumn{2}{c}{\textbf{\textit{window size 1}}} & \multicolumn{2}{c}{\textbf{\textit{window size 3}}} & \multicolumn{2}{c}{\textbf{\textit{window size 5}}} & \multicolumn{2}{c}{\textbf{\textit{window size 7}}} \\
 \cmidrule(lr){2-3} \cmidrule(lr){4-5} \cmidrule(lr){6-7} \cmidrule(lr){8-9}
 group & \% success & count & \% success & count & \% success & count & \% success & count \\
\midrule
\textbf{\textit{10$-$}} & 0.3571 & 28 & 0.4580 & 441 & 0.4287 & 856 & 0.4430 & 2,212 \\
\textbf{\textit{10$+$}} & 0.5697 & 165 & 0.5374 & 294 & 0.4874 & 476 & 0.4962 & 524 \\
\textbf{\textit{20$-$}} & - & 0 & 0.3590 & 39 & 0.4828 & 87 & 0.3590 & 39 \\
\textbf{\textit{20$+$}} & 0.5426 & 129 & 0.5584 & 154 & 0.5220 & 182 & 0.4980 & 247 \\
\textbf{\textit{50$-$}} & - & 0 & 0.0000 & 1 & 0.1250 & 8 & 0.0833 & 12 \\
\textbf{\textit{50$+$}} & 0.3750 & 16 & 0.3750 & 24 & 0.4000 & 30 & 0.4722 & 36 \\
\textbf{\textit{100$-$}} & - & 0 & - & 0 & - & 0 & 1.0000 & 1 \\
\textbf{\textit{100$+$}} & 0.6000 & 5 & 0.6000 & 5 & 0.5000 & 6 & 0.5000 & 6 \\
\textbf{\textit{250$-$}} & - & 0 & - & 0 & - & 0 & - & 0 \\
\textbf{\textit{250$+$}} & - & 0 & - & 0 & - & 0 & - & 0 \\
\bottomrule
\end{tabular}
}
\caption{Number of observations and deal success rates immediately below and above the discontinuities in the observed price schedule. The below-above comparison in deal success rates is intuitive, but the number of observations is too small for inference to be possible based solely on these observations.}
\label{tab: discontinuity}
\end{table}

There are two key insights from table \ref{tab: discontinuity}, which are suggestive that the discontinuity method alone may be insufficient for inferring price sensitivities. First, the number of data points in the just below and just above intervals are too small too allow for meaningful statistical  inference. For example, even with a window size of 7 which may be considered too large for regression discontinuity,\footnote{The reason why a window size of 7 may be considered too large for the discontinuity analysis is that it will have total prices in the just-above window exceed the total prices in the just below.} we only have one observation for the just-below-100 window and 6 observations for the just-above.

The second key observation from table \ref{tab: discontinuity} is that customers may be a bit strategic with the price schedule, trying to take advantage of the discontinuity. To see this, observe that with a window size of one, the number of customers in the just-above window is substantially higher than the number of customers in the just-below window. This makes the inference based on this selected sample difficult.\footnote{Note that this strategic issue could in principle create problems with our main estimation approach as well. That said, as shown in Appendix \ref{appendix: concave}, our results are not sensitive to it given that these windows only comprise a small portion of our data. But for a discontinuity method, these windows would be the entire data.}

As a result of the above observations, we opted for the estimation approach outlined in the main text, although it involves an implicit functional form assumption. Note, however, that although not in the form of regression discontinuity, the within-schedule variation in prices do have a role in our estimation routine. See the identification discussion in the main text for a more detailed discussion.

We finish this appendix by a brief discussion what the requirements would be for the discontinuity approach to work in similar contexts. Following our analysis of table \ref{tab: discontinuity}, we believe there would be at least two key conditions: 1) a sufficiently high number of data points within just-below and just-above windows for all thresholds, and 2) negligible strategic positioning of purchase size. If these conditions were met, the discontinuity approach could be fruitful given that the comparisons between deal acceptance rates just below and above make directional sense: In all cases where a comparison is feasible, the acceptance rate just above (which gets a lower price) is higher.

\section{Sensitivity analysis to functional form assumptions on the error term}\label{apx: functional form}

In this appendix, we dive deeper into the role of functional forms and examine how the results would change if we assumed alternative forms for the error term in regression equation \ref{eq: value regression} for value $v_{it}$ per unit. The  functional form examined in the main text of the paper is logistic. In this appendix, we also estimate the model assuming the error term is a normal random variable.

\begin{table}[H]
\begin{center}
{\small
\begin{tabular}{l c c }
\Xhline{1pt}
\thead[l]{Model} & \thead[c]{Logistic} & \thead[c]{Normal}  \\ \hline\\[-2ex]
\textit{Intercept}
& 2,261 & 2,170 \\
\textit{Log Feature 1}
& 134 & 150 \\
\textit{Feature 2}
& 687 & 782 \\
\textit{Computer Software}
& 39 & 45 \\
\textit{Marketing and Advertisement}
& -224 & -240 \\
\textit{Log Firm Age}
& -41 & -43 \\
\textit{Time}
& 70 & 79 \\
\textit{Mid Size}
& -657 & -693 \\
\textit{Large Size}
& -836 & -873 \\
\textit{Scale}
& 386 & 731 \\
 \hline
Average NLL & 0.2951 & 0.2946 \\[-0.3ex]
\Xhline{1pt}
\end{tabular}}
\end{center}
\caption{MLE estimation results under logistic (main specification) and normal error terms. Results appear to be robust.}
\label{tab: logistic vs normal, estimates}
\end{table}

Table \ref{tab: logistic vs normal, estimates} presents compares the estimation results from these two functional form assumptions. As the table shows, the results are by-and-large similar.

\begin{figure}[H]
\begin{subfigure}{0.475\textwidth}
    \includegraphics[width=\textwidth]{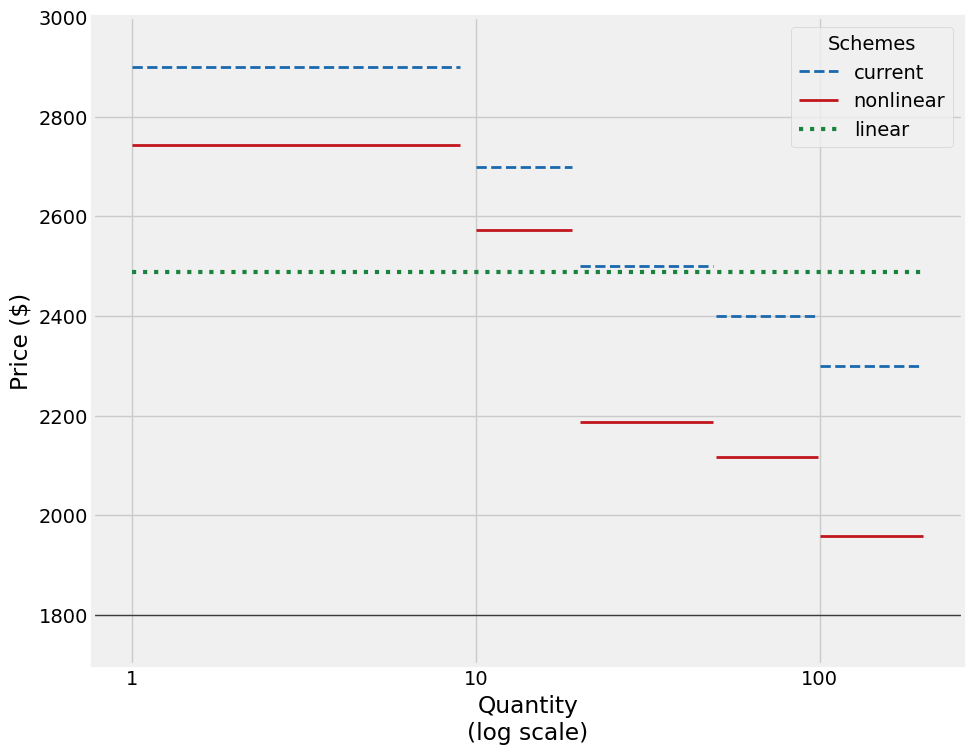}
    \caption{Marginal Prices (logistic error)}
    \label{fig:marginalprice_logistic} 
\end{subfigure}
\begin{subfigure}{0.475\textwidth}
    \includegraphics[width=\textwidth]{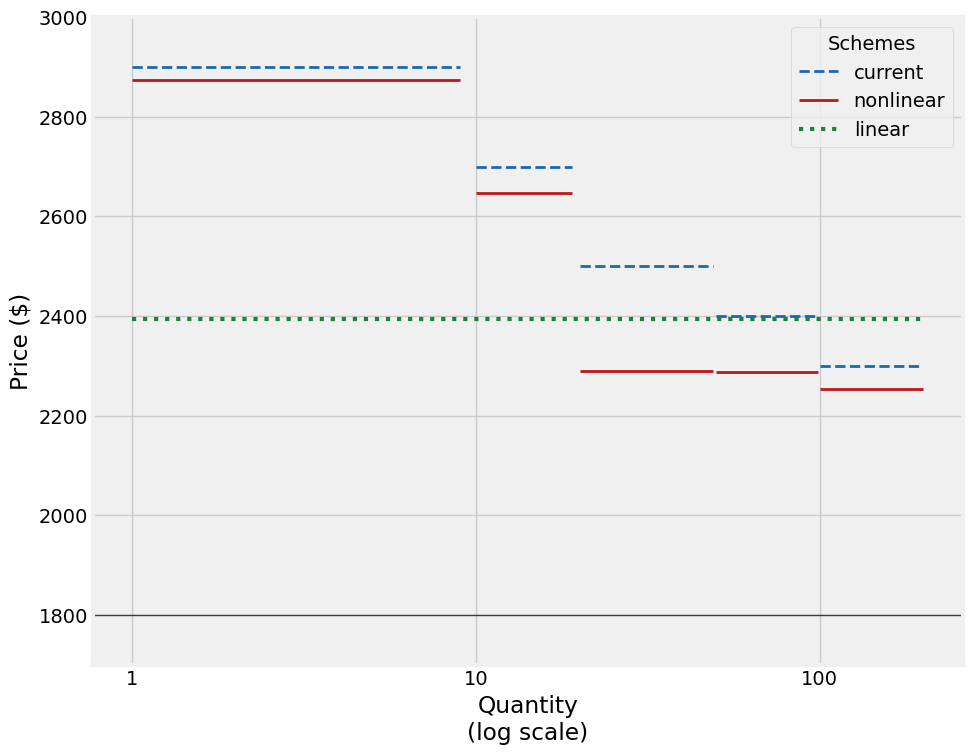}
    \caption{Marginal Prices (normal error)}
    \label{fig:marginalprice_gaussian} 
\end{subfigure}
    \caption{Optimal price schedules under logistic and normal error distributions.}
    \label{fig:schemes_errorterm}
\end{figure}

Also, figure \ref{fig:schemes_errorterm} presents the optimal price schedules arising from these two different demand estimates. As the figure shows, the optimal schedule, too, is relatively robust to the functional form assumption.

\section{Alternative cost modeling}\label{apx: alternative cost model}

The cost model used in our main analysis assumed that the Service and Consulting (SNC) costs take only one of the three possible values recorded by the firm: 1868, 4670, or 9340. This three-tier recording of SNC costs was due to convenience. In reality, the SNC costs should vary across deals in a more smooth fashion. This appendix studies whether smoothing SNC costs before using them in our cost model can significantly alter the results.

To this end, we carry out a two-step process. In the first step, we predict a smooth version of the SNC costs which we denote $\widehat{\*{SNC}}$. More specifically, we focus on successful deals $it$ with $d_{it}=1$ and  estimate the following linear regression model:

\begin{equation}\label{eq: SNC regression}
    SNC_{it}=\zeta \times Z_{it}+\varepsilon_{it}
\end{equation}

Using this regression, we estimate coefficients $\hat{\zeta}$; and then we set $\widehat{\*{SNC}}_{it}:=\hat{\zeta}\times Z_{it}$ for all data points $it$. In this regression, $Z_{it}$ is a set of observable characteristics. Table \ref{tab:sncregestimates} presents the estimation results for the SNC cost model.

\begin{table}[H]
\begin{center}
{\small
\begin{tabular}{ l c  }
\toprule
\textit{\textbf{Coefficient}}  &  \textit{\textbf{Estimate}}\\
\midrule
\textit{Intercept}  & 166.31 \\[-0.3ex]
& (225.67)\\[1ex]
\textit{Log Feature 1} & 515.16\\[-0.3ex]
& (37.11)\\[1ex]
\textit{Feature 2} & 517.04\\[-0.3ex]
& (102.03)\\[1ex]
\textit{Log Firm Age} & 344.87 \\[-0.3ex]
& (66.04)\\[1ex]
\textit{Size} & 63.90 \\[-0.3ex]
& (3.81)\\[1ex]
\midrule
\textit{Observations} & 2,233\\
\textit{R-squared} & 0.220\\
\bottomrule
\end{tabular}}
\end{center}
\caption{Results for the estimation of Service and Consulting (SNC) costs on observable characteristics}
\label{tab:sncregestimates}
\end{table}

The second step of our process is that we use similar cost-calibration formulas for $c_1$ and $c_2$ (i.e., equations \ref{eq: fixed cost calibration} and \ref{eq: var cost calibration} from the main text) but this time plug in the smoothed $\widehat{\*{SNC}}_{it}$ instead of $SNC_{it}$. Formally, we set:

\begin{equation}\label{eq: fixed cost calibration smooth}
    c_{1t}= 1253 + 0.65 \times \frac{\sum_{i\,\text{s.t. } d_{it}=1}\widehat{\*{SNC}}_{it}}{\sum_{i\,\text{s.t. } d_{it}=1}1}
\end{equation}
and 
\begin{equation}\label{eq: var cost calibration smooth}
    c_{2t} = 601 + 0.35 \times \frac{\sum_{i\,\text{s.t. } d_{it}=1}\widehat{\*{SNC}}_{it}}{\sum_{i\,\text{s.t. } d_{it}=1}\bar{q}_{it}}
\end{equation}

\begin{figure}[H]
\begin{center}
\begin{subfigure}{0.475\textwidth}
    \includegraphics[width=\textwidth]{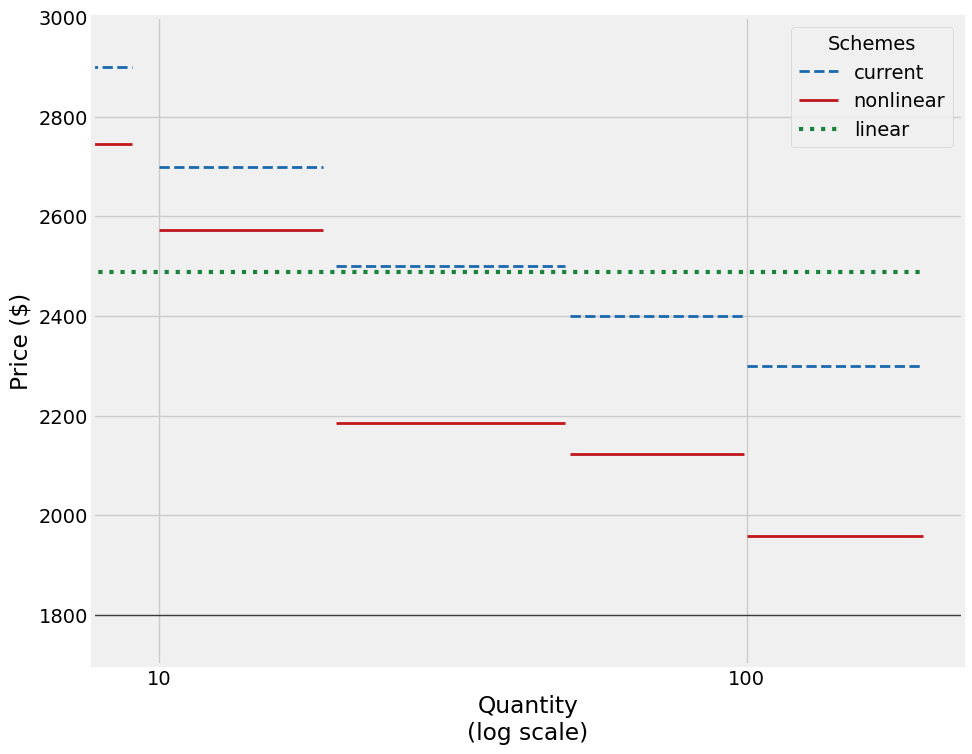}
    \caption{Marginal Prices (smoothed cost)}
    \label{fig:marginalprice} 
\end{subfigure}
\begin{subfigure}{0.475\textwidth}
    \includegraphics[width=\textwidth]{Figures/rnr2024fig4b.png }
    \caption{Marginal Prices (observed cost)}
    \label{fig:marginalprice} 
\end{subfigure}
\end{center}
    \caption{Optimal price schedules under smoothed and observed SNC cost. As can be seen from this figure, differences between the two are negligible.}
    \label{fig:schemes_smoothcost}
\end{figure}

 Based on this analysis, we arrive at the following figures: $\hat c_1=$\$2,782/customer and $\hat c_2 =$\$738/workshop. Figure \ref{fig:schemes_smoothcost} compares the optimal pricing schedules under these costs estimates to the schedule under the cost estimates in the main model. The results appear robust. Additionally, table \ref{tab:profitandwelfare_cost} examines profit and welfare results which also appear robust to this change in the cost model specification.
 
\begin{table}[H]
\begin{center}
{\small
\begin{tabular}{ l r l r l r l r l }
\toprule
 &  &  &  &  & \textit{Consumer} &  & \textit{Social} & \\
\hspace{5mm}\textit{Scheme} & \textit{Revenue} & change & \textit{Profit} & change & \textit{Welfare} & change & \textit{Welfare} & change \\
 & \textit{(\$M)} & (\%) & \textit{(\$M)} & (\%) & \textit{(\$M)} & (\%) & \textit{(\$M)} & (\%)\\
 \midrule
\hspace{4mm}\textit{current}  
& 30.08 & \multicolumn{1}{c}{-} & 16.98 & \multicolumn{1}{c}{-} & \ 7.35 & \multicolumn{1}{c}{-} & 24.33 & \multicolumn{1}{c}{-} \\
\cdashline{2-9}[1pt/1pt]
\textit{1$^\text{st}$ degree} 
& 58.64 & +94.95\% & 35.47 & +112.65\% &	\ 0 & $-$100.00\% & 35.47 & +47.61\%\\
\hspace{6mm}\textit{linear} 
& 32.38 & +\ 7.65\% & 16.55 & -\ \ 0.78\% & 9.33 & +\ 26.94\% & 25.88 & +\ 7.70\% \\
\textit{nonlinear} 
& 33.91 & +12.73\% & 17.90 & +\ \ 7.31\% & 10.26 & +\ 39.59\% & 28.16 & +17.19\%\\
\bottomrule
\end{tabular}}
\end{center}
\caption{Profit and Welfare analysis under the smooth cost model. Results appear similar to those under the model in the main manuscript.}
\label{tab:profitandwelfare_cost}
\end{table}

\end{document}